\documentclass[11pt]{article}
\usepackage[utf8]{inputenc} 
\usepackage[T1]{fontenc}    
\usepackage[margin=1in]{geometry}
\usepackage{graphicx} 
\usepackage{amsmath,amsfonts,amssymb,amsthm, bm}
\usepackage{color}
\usepackage{hyperref,cleveref}
\usepackage{algorithm}
\usepackage{algpseudocode}
\usepackage{tikz}
\usetikzlibrary{calc,angles,quotes,arrows.meta, intersections,decorations.pathreplacing}
\usepackage{mathtools}
\usepackage{thm-restate}

\usepackage[normalem]{ulem}
\usepackage{cancel}

\theoremstyle{plain}
\newtheorem{theorem}{Theorem}[section]

\newtheorem{lemma}[theorem]{Lemma}

\theoremstyle{definition}
\newtheorem{definition}[theorem]{Definition}

\theoremstyle{remark}

\DeclareMathOperator{\arccosh}{arccosh}
\DeclareMathOperator{\arcsinh}{arcsinh}
\DeclareMathOperator{\arctanh}{arctanh}

\DeclareMathOperator{\volume}{Volume}
\DeclareMathOperator{\lvl}{lvl}

\DeclareMathOperator{\Cl}{Cl}
\DeclareMathOperator{\low}{low}
\DeclareMathOperator{\CAP}{Cap}

\def\eps{\varepsilon}

\DeclareMathAlphabet{\mathui}{T1}{cmr}{m}{ui}

\def\dim{\mathui{d}}

\title{Navigable Graphs for Low-Dimensional Hyperbolic Spaces}
\author{%
Chengyuan Deng \quad Jie Gao \quad Kuijun Liu \quad Kevin Lu\\
Feng Luo \quad Puheng Weng \quad Cheng Xin\\[0.5em]
{\small Rutgers University}%
\thanks{Email addresses, in author order:
\texttt{cd751@rutgers.edu};
\texttt{jg1555@cs.rutgers.edu};
\texttt{kuijun.liu@rutgers.edu};
\texttt{kll160@scarletmail.rutgers.edu};
\texttt{fluo@math.rutgers.edu};
\texttt{pw344@rutgers.edu};
\texttt{cx122@cs.rutgers.edu}.}
}

\begin{document}

\maketitle

\begin{abstract}
    A directed graph on a point set $P$ in a finite metric space $(X, d)$ is navigable, if for every distinct ordered pair of distinct points $s,t \in P$, there is an out-neighbor $v$ of $s$ satisfying $d(v,t) < d(s,t)$. This condition makes routing inherently local: at every step, the current vertex chooses a neighbor that is strictly closer to the destination. By repeatedly making such local moves, greedy routing is guaranteed to reach the destination. Navigable graphs have found applications in small-world phenomena, geographic routing, and recently, approximate nearest neighbor search. The central goal is to construct, for a given metric space, sparse navigable graphs that admit short greedy routing paths.

    Existing work on sparse navigable graphs mostly assumes bounded doubling dimensions. In this paper, we show that constant-dimensional hyperbolic spaces that do not have bounded doubling dimension also admit sparse navigable graphs. 
    Specifically, for an $n$-point set in $\mathbb H^k$ with a constant $k$, we construct a navigable graph of size $O_k(n)$ with Steiner points, where each node has constant out-degree. We also show that Steiner points can be removed: every $n$-point set in $\mathbb H^k$ admits a navigable graph on the input points alone with $O_k(n)$ edges, though a single vertex may have out-degree $\Omega(n)$. 
    Furthermore, with a refined construction with $O_k(n\log n)$ edges and an accompanying greedy routing algorithm, we can obtain routing paths of expected $O(\log ^2 n)$ in both settings with or without Steiner points. With Steiner points, the maximum out-degree is bounded by $O_k(\log n)$. 
    
    We complement these upper bounds with matching lower bound results. 
    In the hyperbolic plane, there are point sets for which every navigable graph requires maximum out-degree $\Omega(n)$, so the maximum linear out-degree above is unavoidable and constant out-degree is only possible with Steiner points. Next, we show that for every constant $\alpha>1$, an $\alpha$-navigable graph may require $\Omega(n^2)$ edges, which rules out the standard route of using $\alpha$-navigability to obtain greedy paths of bounded hops. Last, once the dimension $k$ reaches $\Omega(\log n)$, there are point sets for which every navigable graph may require $n^{3/2-o(1)}$ edges, matching the upper bounds for general metrics.
\end{abstract}

\newpage
\setcounter{page}{1}

\section{Introduction}

Given a finite point set $P$ in a metric space $(X,d)$, a directed graph
with $P$ as the vertex set is \emph{navigable} if, for every ordered pair of distinct vertices $s,t\in P$, there is an out-neighbor $v$ of $s$ such that $d(v,t)<d(s,t)$. 
In a navigable graph, from any source $s$ to any destination $t$, there is a distance-decreasing path $s=x_0,x_1,\ldots,t$ such that
$d(x_{i+1},t)<d(x_i,t)$ for every $i$. Further, such a path can be discovered locally, where at each node we select a neighbor closer to the destination. 
One can also ask for an $\alpha$-navigable graph, for a constant $\alpha>1$, where for any source $s$ and destination $t$, there is a neighbor $x$ of $s$ such that $d(x, t)<1/\alpha \cdot d(s, t)$. This requirement is stronger, as each node has a neighbor that is substantially closer to the destination. 

The topic of greedy routing and navigable graphs have gained attention in many application areas, including small-world navigation~\cite{milgram_small-world_1967,Kleinberg2000-qa}, routing in wireless and sensor networks~\cite{karp2000gpsr}, and more recently in approximate nearest neighbor search~\cite{subramanya2019diskann,malkov2020efficient,Fu19MRNG}. For approximate nearest neighbor search, a point set $X$ is given together with a distance function $d$ and the goal is to construct a graph $G$ that is navigable for $(X, d)$. For an approximate nearest neighbor query of $t$, we may start from any point $s$ and iterate through the neighbor deemed closer to the destination $t$ until a good candidate is identified.  The cost of such a search depends on two parameters: the max out-degree of the navigable graph $G$, which bounds the cost of looking for a closer neighbor to $t$ in each step, and the number of hops until termination. Of course, the complete graph is always navigable, but nearest-neighbor search on a complete graph degenerates to a trivial linear search. Ideally, a low out-degree graph is preferred, but a sparse graph (of $\tilde{O}(n)$ edges \footnote{The $\tilde{O}$ may hide logarithmic factors.}) is also nice, as it provides a low degree on average.

Greedy routing on a navigable graph, despite guaranteed delivery, may not guarantee a small number of hops -- consider a long path of points on a line. 
On an $\alpha$-navigable graph with $\alpha>1$, for each step one can choose a neighbor whose distance to the destination shrinks by a factor of $1/\alpha$ -- thus making a significant jump towards the destination. 
$k$ such jumps will shrink the distance to the destination by a factor of $1/\alpha^k$. 
As shown 
in~\cite{pmlr-v267-gollapudi25a}, in $O(\log_{\alpha}\frac{\Delta}{(\alpha-1)\epsilon})$ steps a greedy search returns an $(\frac{\alpha}{\alpha-1})$-approximate nearest neighbor to the destination $t$,  where $\Delta$ is the aspect ratio of the input metric.

These ideas help to explain the popularity and wide adoption of graph-based approximate nearest neighbor search methods in industry: DiskANN~\cite{subramanya2019diskann}, Hierarchical Navigable Small World (HNSW) method~\cite{malkov2020efficient}, and the Navigating Spreading-out Graph (NSG) or Monotonic Relative Neighborhood Graph (MRNG)~\cite{Fu19MRNG}. 
These methods often show remarkable empirical performance, which has motivated both empirical optimization methods~\cite{zhang2026prunedonstrebuild,pmlr-v267-gollapudi25a} and theoretical studies of constructing sparse navigable graphs for a given input metric in the past few years. We quickly review the state-of-the-art bounds below. 

For any given metric where the pairwise distances are distinct\footnote{We may assume with perturbation that all distances are distinct. A uniform metric, trivially, requires all pairs of vertices to be connected in a navigable graph.}, there is a navigable graph of total size of $O(n^{3/2})$~\cite{diwan24navigable}. For a general metric setting this is almost the best one can do. Indeed, it is shown that for points in $\Omega(\log n)$ dimensional Euclidean space, a navigable graph may require $\Omega(n^{3/2-\delta})$ edges in total for any $\delta>0$~\cite{diwan24navigable}. On the other hand, a number of recent work~\cite{conway2026efficiently,khanna2026sparse} have considered the ``instance-optimal'' scenario, where they try to find a navigable graph whose size is comparable to the sparsest navigable graph possible for the input metric. They show that an $O(\log n)$ approximation (in terms of the total number of edges) is possible with a running time of $O(n^2)$, and improving beyond this approximation factor is NP-hard. Thus, if an input metric indeed supports a sparse navigable graph, these algorithms can find a decent one. 
Therefore, an interesting question is which metrics support a sparse, navigable graph and what properties are required for it. 

To find sparse navigable graphs, it appears that the exclusively used property so far is the constant \emph{doubling dimension} -- that any disk of size $r$ can be covered by at most a constant number of disks of radius $r/2$~\cite{heinonen2001lectures}. A $\dim$-dimensional Euclidean space $\mathbb{R}^{\dim}$ equipped with the standard distance metric has doubling dimension of $\Theta(\dim)$~\cite{gupta03bounded}. A metric of constant doubling dimension exhibits the same packing property as constant dimensional Euclidean spaces and thus supports nice hierarchical nets and other useful data structures~\cite{harpeled2006fastSIAM}. For points in $\dim$-dimensional Euclidean space, one can find a navigable graph of out-degree $O(2^{\dim}\log n)$ such that the greedy navigation has a bounded hop length of $O(\log^2n)$ and a total query time of $O(\log^3n)$. Har-Peled et.al.~\cite{Har-Peled2026-zm} showed several graph constructions for a metric of constant doubling dimension $\dim$, with graph size of $O(n/\eps^{\dim})$ for $(1+\eps)$-approximate nearest neighbor search in query time of $O(\eps^{-\dim}\log \Delta)$. All of these constructions extensively use the packing property.



Inspired by these results, one natural question that arises is, 
\begin{quote}
    Are there any metrics that do not have a constant doubling dimension yet support a sparse navigable graph?
\end{quote}

The answer to this question is yes. First, an easy proof shows that for a tree metric, the minimum spanning tree is a navigable graph (\Cref{thm:tree}). A tree metric does not necessarily have a constant bounded doubling dimension. Indeed, on a simple binary tree the volume of a ball of radius $r$ grows as an exponential function in the radius. A tree metric is one with a ``hyperbolic'' property and specifically has $\delta$-hyperbolicity with $\delta=0$ in the notion of Gromov coarse geometry~\cite{gromov1987hyperbolic}. Naturally one can ask, 
\begin{quote}
Are there other hyperbolic metrics/spaces that admit sparse navigable graphs? 
\end{quote}

The answer is again yes for points in the hyperbolic plane. Specifically, in both the Euclidean space and hyperbolic space~\cite{Nielsen2010-vg,Bogdanov2013-sl}, the Delaunay triangulation of a set of input points is a navigable graph. This is because the Delaunay triangulation is the dual graph of the Voronoi diagram. Thus, if the destination $t$ is different from the source $s$, it falls outside the Voronoi cell of $s$. In other words, $t$ falls on the different side of a bisector of $s$ and a point $x$ that defines the boundary of the Voronoi cell of $s$. By definition, the neighboring vertex $x$ of $s$ is closer to $t$ than $s$ is.  This immediately implies that for both Euclidean plane and hyperbolic plane, the Delaunay triangulation, which is a planar graph and thus of linear size, is a navigable graph. Unfortunately, for 3D or higher, the Delaunay triangulation may have a quadratic size and thus is not super interesting as a navigable graph. Enlightened by this observation, we modify our question as
\begin{quote}
    Are there linear-size navigable graphs for points in hyperbolic space of dimension $\dim \ge 3$?
\end{quote}
This paper mainly answers this question affirmatively when the dimension $\dim$ is a constant and negatively when the dimension $\dim$ is $\Omega(\log n)$.

Before we present our main results, we note that hyperbolic geometry has received considerable attention in modern data analysis and machine learning. Non-Euclidean data representation has also been studied through multidimensional scaling with symmetric bilinear forms~\cite{deng24NeucMDS} and extensions of the Johnson--Lindenstrauss lemma~\cite{deng2025jl}. 
Hyperbolic geometry drops the parallel postulate of Euclidean geometry. In hyperbolic geometry, for any line and a point not on that line, there are at least two distinct lines through the point that do not intersect the original line. Unlike Euclidean spaces of low dimensions, where the volume of a ball grows as a polynomial function of its radius, balls in hyperbolic spaces of low dimensions have exponential volume growth in radius. This naturally accommodates hierarchical, tree-like, and exponentially branching structures that arise in many real-world data sets. 
For this reason, hyperbolic embeddings have been adopted in network representation learning, hierarchical clustering, and discrete geometric data analysis~\cite{nickel17poincare,gulcehre2018hyperbolic,Ganea2018hyperbolic,tifrea2018poincare,De-Sa2018-cc,Chami2020-ug,Chami2019-ie, Wilson2014-qx, Karliga2004-fr, Tabaghi2024-bn}. 
More generally, as pointed out by Gromov’s hyperbolization theorem \cite{gromov1987hyperbolic}, spaces with complicated topology tend to resemble spaces with negative curvature, such as hyperbolic spaces. This partly explains the importance of hyperbolic geometry in representing the negative curvature phenomena arising from complex geometric data~\cite{krioukov2010hyperbolic,Shavitt2004-au}.  Graph-based nearest neighbor search has been applied in hyperbolic
space~\cite{prokhorenkova2022graph}, but with guarantees only under the assumption that the data is uniformly distributed in a hyperbolic ball.


\paragraph{Our Results.}
We start with some negative results (\Cref{sec:lower}). A well known example in a hyperbolic plane is to take a set $X$ of $n$ points uniformly on a circle of radius $r=O(\frac{1}{\eps} \cdot \ln n)$. 
All pairs of points have distance to each other between $2r(1-\eps)$ and $2r$~\cite{Kisfaludi-Bak_vanWordragen_2025}. Therefore, if we ask for an $\alpha$-navigable graph for any fixed $\alpha>1$, by choosing $\eps$ properly for any pair of points $p, q\in X$, none of the other points can be a candidate neighbor (i.e., being substantially closer to $q$ than $p$). We have to add all pairs of edges, forcing a complete graph.  Therefore, although a lot of prior work focuses on $\alpha$-navigable graphs and hop bound, such graphs are meaningless in the hyperbolic setting. This also means we do not immediately have the nice hop count bound arising from the $\alpha$-navigability. 
Second, if we add the center $o$ of the circle to the point set, which has distance $r$ to all the points in $X$, any navigable graph must connect edges from the center to all other $n$ points, as none of the other points is closer to the destination than $o$ is. This means that we have to have max out-degree of $\Omega(n)$ for navigable graphs in the hyperbolic plane. Both lower bound examples extend to high dimensions as well.  
Finally, because sufficiently small regions of hyperbolic space are nearly Euclidean, the lower bound example for Euclidean space of dimension $\Omega(\log n)$~\cite{diwan24navigable} can be reconstructed in the hyperbolic space of dimension  
$\Omega(\log n)$. Thus, any navigable graph for hyperbolic space of dimension $\Omega(\log n)$ requires at least $n^{3/2-o(1)}$ edges, matching the upper bound for any generic metric space~\cite{diwan24navigable}. 

Despite the pessimism conveyed by the negative results, we show positive results (\Cref{sec:positive}) for low-dimensional hyperbolic spaces $\mathbb{H}^{\dim}$ with $\dim=O(1)$.
For a $\dim$-dimensional hyperbolic space, we have a navigable graph of size $O_{\dim}(n)$ (\Cref{thm:nosteiner}), where the constant factor may be exponential in $\dim$ (which is necessary). Although the worst case maximum out-degree may be forced to be $\Omega(n)$, the average out-degree is still a constant. Further, we show that if we are allowed additional \emph{Steiner points}, which are additional points in the hyperbolic space, we can build a navigable graph of size $O_{\dim}(n)$ such that the maximum out-degree is a constant $O_{\dim}(1)$. If we allow the max degree to be $O_{\dim}(\log n)$, we can bound the hop count of greedy navigation by $O_{\dim}(\log^2 n)$. Finally, our results can be extended to the case when the destination $q$ is not necessarily a point in the input set $X$, but can be anywhere in $\mathbb{H}^{\dim}$. Using a revised navigable graph of size $O_{\dim}(n)$ we can ensure to arrive at a point in $X$ which is an $(1+\eps)$-approximate nearest neighbor of $q$. 

Our results show that greedy navigation and approximate nearest-neighbor search in low dimensional hyperbolic spaces can still be quite effective and practical. This is of particular interest, since hyperbolic embedding has been a popular approach for representing and navigating in large, complex networks with heterogeneous degree distributions and strong community structures~\cite{krioukov2010hyperbolic,boguna2010sustaining,Papadopoulos2010-fi}. Typically, such work only considers constant-dimensional hyperbolic space.  

\paragraph{Overview of Main Techniques.} We will explain the main intuition using the hyperbolic plane $\mathbb{H}^2$, and the same idea works for other constant dimensions. 

Euclidean distances and Euclidean point sets enjoy a \emph{scale invariant} property. If we scale the entire point set by the same factor, this typically does not affect how the algorithms for these points should be designed. Hyperbolic distances, due to their negative curvature, lack such scale invariance. For small distance scales, the behavior of hyperbolic distances is ``nearly Euclidean'' -- our lower bound actually uses this property to approximately embed a Euclidean point set in hyperbolic spaces at tiny scales. At a constant distance scale, we also enjoy a constant doubling dimension. On the other hand, for large distance scales, the constant doubling dimension assumption no longer holds. Yet we can approximate the hyperbolic distance, up to an additive constant, by a \emph{binary tiling}, which has the structure of a set of infinitely sized binary trees with ``horizontal edges'' forming a chain connecting all tiles on the same level. These tiles are isometric. And the hyperbolic distance of two tiles (and any two points in these tiles) can be approximated by the (scaled) hop length of a path going upward from both tiles along the tree edges until reaching a pair of horizontally adjacent tiles. We call this hop length the $d_H$ distance. 
See \Cref{fig:binary-tiling} for an example. This additive approximation of $d_H$ to the real hyperbolic distance $d_{\mathbb{H}}$ has been made concrete by Park and Vigneron~\cite{Park2025-ge}.

Given the vastly different structures of hyperbolic distances at the small and large distance scales, we design two types of edges to enable greedy navigation. We fix a constant threshold
$D$ and use \emph{local routing} for pairs at distance at most $D$, via a
construction inspired by low-dimensional Euclidean geometry. For pairs at
distance greater than $D$, we use \emph{global routing}, whose edges are guided by the binary tiling. By merging the two sets of edges, we support greedy navigation for all pairs. When we are at a distance more than $D$ away from the destination, global routing enables jumping to a neighbor closer to the destination until we arrive at a node with a distance smaller than $D$, at which the edges by local routing kick in. 

For local routing, we use the Yao graph. For each vertex $p$, consider dividing the hyperbolic plane by cones of a fixed angle $\delta$ centered at $p$. We connect a directed edge from $p$ to the closest vertex inside each cone. When $\delta$ is small enough (as a function of $D$), for any destination $t$ falling in the same cone, the neighbor inside this cone is closer to $t$ than $p$ is. Since we make $\delta$ a constant, the number of edges added in the local phase is a constant per node. 

For global routing, we can keep only one vertex per tile (the representative) and only need to worry about hopping from tile to tile. For all non-empty tiles, we also keep the lowest bridge connecting their ancestor tiles, and the first junction of the ancestor chains reaching the lowest ancestor.  Any tiles that take these ``junction'' positions are given Steiner points. Due to the nice structure of $d_H$ and its alignment with the tree structure, we add edges for all vertices in one tile with the representatives in neighboring tiles that are either non-empty or contain Steiner points. The total number of such outgoing edges per vertex is a constant.  These edges immediately enable greedy navigation among the tiles by using the $d_H$ distance. 

If we wish to allow navigation with the real hyperbolic distance $d_{\mathbb{H}}$ instead of the discrete approximation $d_H$, we can add additional edges to ``correct'' the errors. Since there is at most a constant additive error, if we add edges for each node to the tiles within a sufficiently large (yet a constant) 
radius in terms of $d_H$ distance, this allows us to jump sufficiently far ahead in $d_H$ which ensures monotonic reduction in hyperbolic distance to the destination.  
Combining local and global routing, we have a navigable graph with only $O(n)$ Steiner points and constant out-degree per node. 

Next, greedy navigation in this graph may not have a bounded hop length. To address this, we need to re-examine local and global routing to ``shortcut'' long chains. For local routing, we adapt the construction in the Euclidean setting~\cite{arya93approximate} to the hyperbolic setting. The construction is fairly simple -- take a random permutation for all points, and for each cone, instead of connecting to the closest vertex, connect to the closest point whose index is a new record low. This adds around $O(\log n)$ outgoing edges per node, by the analysis of record setters in a random permutation. To show that the number of hops is reduced to $O(\log^2n)$, we need both a new greedy routing algorithm (that specifies which neighbor to take, when there are multiple candidates with smaller distance to the destination) as well as a few non-trivial geometric structures in hyperbolic geometry so that we can quantify ``sufficient progress'' towards the destination. For global routing, to avoid long ancestor chains, we use the heavy-light decomposition of tree edges and add ``shortcut'' edges to the heavy paths, a standard technique in hopset construction to reduce the graph diameter. These shortcut edges increase the out-degree to be $O(\log n)$, and the total number of hops along any ancestor chain is reduced to $O(\log^2 n)$.

We show that the Steiner points can be removed. Local routing does not need Steiner points, so the only problem is with the global routing. These Steiner points hold the ``junction'' positions. If we get rid of these Steiner points, we need to reconnect the other pairs that initially navigate through them. The challenge is that we cannot naively connect all tiles that were initially connected via a Steiner node, as there can be quadratically many such edges. 
In~\Cref{subsec:global-nosteiner}, we show how to properly rank the tiles by their distances to a Steiner tile, such that if we connect them in a proper order, we can limit the total number of edges created to be still linear in $n$. The only caveat is that non-Steiner global routing is greedy with respect to the discrete distance $d_H$ (the discrete approximation) rather than the hyperbolic distance. Nevertheless, navigation using $d_H$ is equally practical -- $d_H$ can be computed in constant time and is a completely local measure that requires only the coordinates of the current point and the destination.   

Finally, we show how to use such graph navigation idea for approximate nearest neighbor search (ANNS) (\Cref{subsec:ANNS}). Essentially when the destination $q$ is not necessarily a point in $X$, we cannot really arrive at $q$. But we hope to land on a point in $X$ which is an approximate nearest neighbor. This requires edits to both the global and local routing section. For global routing, we adopt the approximate Voronoi diagram as in~\cite{Park2025-ge} that lands at a point of $X$ with a constant additive error to the distance between $q$ and its true nearest neighbor. Then we modify the local routing algorithm (with the same graph) to ensure that we arrive at a $(1+\eps)$-ANN in $O(\log^2 n)$ hops.

\section{Related Work}

\paragraph{Navigable Graphs and Spanners}  In a spanner graph (of spanning ratio or stretch factor of $\alpha\ge 1$), there is a path between any source $s$ and destination $t$ with total length at most $\alpha\cdot d(s, t)$. Note that the paths supported in navigable graphs and spanners can be different. A distance-decreasing path can have a total length much longer than $d(s, t)$ while an $\alpha$-approximate path in a spanner can have vertices that are further away from the destination than their predecessor. 

There has been a lot of work on sparse spanners in (low-dimensional) Euclidean settings and, more recently, in (low-dimensional) hyperbolic space settings~\cite {Kisfaludi-Bak_vanWordragen_2025,kisfaludibak2025nearoptimaldynamicsteinerspanners}. While these ideas are inspirational for the design of navigable graphs, it is not clear how they immediately translate into navigable graphs, for the reasons mentioned above. In particular, it is shown in~~\cite{Kisfaludi-Bak_vanWordragen_2025,kisfaludibak2025nearoptimaldynamicsteinerspanners} that for points in the hyperbolic plane one cannot find a spanner graph of subquadratic size with stretch factor $\alpha<2$. This is a sharp contrast to spanners in low-dimensional Euclidean spaces, where a $(1+\eps)$-spanner of $O_{\dim, \eps}(n)$ size is possible for any $\eps>0$. Here, the big-O notation hides factors that are exponential in $d$. In addition, a linear size $(1+\eps)$-spanner is possible if Steiner points are allowed. 

\paragraph{Greedy Embedding} A related topic is greedy embedding~\cite{Papadimitriou2005-cm}, where the goal is to find the coordinates of vertices of a given graph in a geometric space/metric $d$ such that $G$ is navigable using the metric $d$. Greedy embedding does not always exist -- for example, a star of more than 6 edges does not admit a greedy embedding in the Euclidean plane. But any planar 3-connected graph has a greedy embedding in the Euclidean plane~\cite{Leighton2010-qa,Angelini2010Greedy}. Greedy embedding in the hyperbolic space is actually very easy. As shown in~\cite{kleinberg2007geographic}, any tree has a greedy embedding in the hyperbolic plane $\mathbb{H}^2$. The embedding can be made to have small-sized coordinates~\cite{Eppstein2009-ai} or as the Delaunay graph of the vertices~\cite{Sarkar2012-tr}.
Thus, any graph has a greedy embedding since one can take any spanning tree and embed it. 

\paragraph{Approximate Nearest Neighbor Search (ANNS)}
In low-dimensional Euclidean space, approximate nearest neighbor search can be done by standard range query structures~\cite{AryaMNSW98}.
For high-dimensional Euclidean spaces, exact nearest neighbor search cannot be substantially better than linear scan~\cite{harpeled2012approximate}, and approximate solutions can be done by using locality sensitive search with sublinear query time~\cite{indyk1998approximate}. Deng et al.~\cite{deng2026localitysensitivehashinghyperbolic} develop locality-sensitive hashing for approximate near-neighbor search in hyperbolic space, including the high-dimensional setting. Approximate nearest neighbor search has also been studied in low-dimensional hyperbolic spaces with linear storage and $O_{\dim}(\log n)$ query time for a data set of $n$ points in $\mathbb{H}^{\dim}$, with multiplicative approximation factor~\cite{kisfaludibak2025nearoptimaldynamicsteinerspanners,Kisfaludi-Bak_vanWordragen_2025} or additive approximation error~\cite{Park2025-ge}. Last, Krauthgamer and Lee
\cite{Krauthgamer2006-zz} studied approximate nearest neighbor search for $\delta$-hyperbolic spaces with local doubling dimension (i.e., bounded doubling dimension for distances no greater than $\delta/3$), and presented data structures of $O(n^2)$ storage and logarithmic query time with a $O(\delta)$ additive error.  



\section{Preliminaries}
\label{sec:pre}
\subsection{Navigable Graphs}

\begin{definition}\label{def:navigable}
    An $\alpha$-navigable graph on a finite metric space $M = (X, d)$ is a directed graph $G(X, E)$ with $X$ as its set of vertices and the set of edges $E$ satisfies the following property: for every pair of distinct vertices $s, t \in X$, there exists an edge $(s,v) \in E$ such that $d(s, t) > \alpha \cdot d(v, t)$.
\end{definition}

When $\alpha=1$, we say that the graph is simply navigable. For a uniform metric space, any navigable graph has to be a complete graph. Throughout the paper we assume that the distances of pairs of vertices in $X$ are distinct. In practice this can be ensured by perturbing the distances slightly. 

On a navigable graph, greedy routing, i.e., going to a vertex closer to the destination, will monotonically get closer to the destination and eventually arrive at $t$ on any finite-sized graph. The maximum out-degree of such a graph is directly related to the cost of greedy routing at each step. To ensure additional bounds on hop length of such greedy routing path, we need specific algorithms that decide which neighbor to take.  We will elaborate such algorithmic designs when needed.

We may allow a navigable graph with additional \emph{Steiner} points $S$, and define the graph on $X\cup S$. 
We typically want to use only $O(n)$ Steiner vertices, with $n=|X|$.
\begin{definition}
    A Steiner navigable graph on a finite metric space $M = (X, d)$ is a graph $G(X \cup S,E)$ with $X \cup S$ as its set of vertices and $E$ as its set of edges that satisfy the following: there is a greedy routing algorithm $\mathcal{A}$ that, for every pair of distinct vertices $s, t$ with $s, t\in X$, uses only the information of the current vertex  and destination $t$ and produces a path $P$ ending at $t$; for any vertex $u$ on the path, the next hop $v$ on the path is a neighbor of $u$ such that $d'(u, t) > d'(v, t)$ where $d'$ is a metric on $X \cup S$ extending $d$ on $X$. 
\end{definition}

This definition preserves the property that points in the given set can be navigated to, but does not require the Steiner points, $S$ to have the navigable property with all other vertices. Further, the definition allows for a metric $d'$ which is different from $d$, although in many concrete settings we may just choose the Steiner points from the same ambient space as $X$ and use $d'=d$.

\subsection{Hyperbolic Spaces}

We will use the Poincar\'e half-space model to represent hyperbolic space $\mathbb{H}^{\dim}$. Each point $p$ in $\mathbb{H}^{\dim}$ is represented by a pair $(z, x)$ with $z\in \mathbb{R}^+$ and $x\in \mathbb{R}^{\dim-1}$. The hyperbolic distance between two points $p_1=(z_1, x_1)$ and $p_2=(z_2, x_2)$ is defined by the following function
\[d_{\mathbb{H}}(p_1, p_2)=\arccosh \left (1+\frac{\|x_1-x_2\|^2+(z_1-z_2)^2}{2z_1z_2}\right ),\] where $\|x_1-x_2\|$ is the Euclidean distance between $x_1, x_2$. The geodesic between two points $p_1$ and $p_2$ follows an arc that goes through both points and is orthogonal to the plane $z=0$.

\begin{figure}
\centering\includegraphics[width=0.35\linewidth]{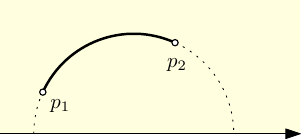} \hspace*{6mm}
\includegraphics[width=0.22\linewidth]{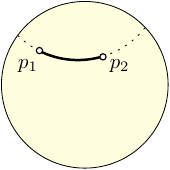}
    \caption{(Left) Upper half plane model. (Right) Poincar\'e disk model. }
\label{fig:hyperbolic}
\end{figure}

By the isometry invariance of hyperbolic space, we sometimes work in the Poincar\'e ball model where all points stay in an open unit ball $\mathbb{B}^{\dim}=\{x\in\mathbb R^{\dim}:\lVert x\rVert<1\}$, and geodesics are circular arcs orthogonal to the boundary of the ball. See \Cref{fig:hyperbolic} for an example.

\paragraph{Doubling Dimension}
For a metric $(X, d)$, the \emph{doubling dimension} of radius $r$ refers to $\log_2 M$ where $M$ is the minimum number of balls of radius $r/2$ to cover a ball of radius $r$. For $\dim$-dimensional Euclidean spaces, the doubling dimension is $\Theta(\dim)$, irrespective of the value of  $r$. For hyperbolic spaces, the doubling dimension of radius $r$ is in the order of $O(\dim+(\dim-1)r)$ and is only a constant when $r$ is a constant (i.e., at small distance scales). 
The proof of the following lemma and all other missing proofs can be found in \Cref{apd:pf}.

\begin{restatable}{lemma}{hyperbolicpacking}
\label{lem:hyperbolic-packing}
The $\dim$-dimensional hyperbolic space for a constant $\dim$ has a doubling dimension of $O(\dim+(\dim-1)r)$ at radius $r$.    
\end{restatable}

\begin{figure}
\centering\includegraphics[width=0.4\linewidth]{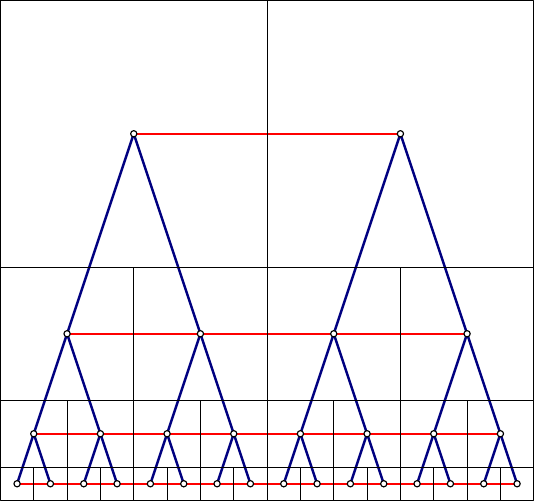} \hspace*{2mm}
\includegraphics[width=0.4\linewidth]{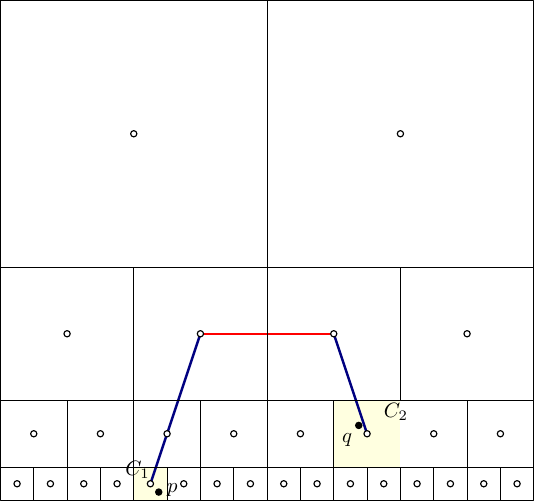}
    \caption{(Left) A binary tiling. The tree edges are marked in navy, and the bridge edges are marked in red. These edges are drawn by connecting the centers of the tiles (shown as hollow vertices). (Right) $d_H(C_1, C_2)$ is an approximation of $d_{\mathbb{H}}(p, q)$ with a constant additive error. Points in $X$ are drawn as solid vertices. }
\label{fig:binary-tiling}
\end{figure}

\paragraph{Binary Tiling} A \emph{binary tiling} of $\mathbb{H}^{\dim}$ uses hypercubes called \emph{tiles} to fill up the space. One of the tiles is the Euclidean unit cube with corners $(1, 0, \cdots, 0)$ and $(2, 1, \cdots, 1)$. And the other tiles are obtained by an isometric transformation $f_{b, 2^a}(z, x)=2^a \cdot (z, x+b)$ that translates $x$ by an integer vector $b\in \mathbb{R}^{\dim-1}$ and scales all coordinates by $2^a$ for an integer $a$. 
    Formally, a hyperbolic tile is a set of points $C(a, b_1, b_2, ...b_{\dim-1}) = \{(z, x_1, ... x_{\dim-1})\}$ satisfying:
    \begin{enumerate}
        \item $2^{a} \leq z< 2^{a + 1}$;
        \item $b_i2^{a} \leq x_i < (b_i +1)2^{a}$ for $1 \leq i \leq \dim-1$.
    \end{enumerate}
    We call $a$ the 
    \emph{level} of the tile, $\lvl(C)=a$. We call the Euclidean box $$[b_12^a, (b_1+1)2^a]\times[b_22^a, (b_2+1)2^a]\times \cdots [b_{\dim-1}2^a, (b_{\dim-1}+1)2^a]$$ the \emph{Euclidean range} of $C$.
Each tile $C(a, b_1, b_2, ...b_{\dim-1})$ is a Euclidean cube, and we will refer to the Euclidean center as the \emph{center} $M(C)$ of $C$ with coordinate $(3\cdot 2^{a-1}, (b_1+1/2)\cdot 2^a, \cdots, (b_{\dim-1}+1/2)\cdot 2^a)$. 
A point $(z, x)\in \mathbb{H}^d$ falls inside the tile $C(a,b_1, b_2, \cdots, b_{\dim-1})$ with $a=\lfloor \log_2 z \rfloor$ and $b_i=\lfloor x_i/2^a \rfloor$. 
Any two points in the same tile have distance at most $\mu_{\dim}=2\arcsinh (\sqrt{\dim-1}/2)$.
    A binary tiling of hyperbolic space is a set of tiles in $\mathbb{H}^d$, $\mathcal{C} = \{C(a, b_1, b_2, ... b_{d-1})\}$ for all $a, b_i \in \mathbb{Z}$.
See \Cref{fig:binary-tiling} for an example in $\mathbb{H}^{2}$.

Each tile $C(a, b_1, b_2, \cdots, b_{\dim-1})$ has exactly one adjacent tile $C'(a', b'_1, b'_2, \cdots, b'_{\dim-1})$ with a higher level ``on top of it'', called its parent: $a'=a+1$, $b'_i=\lfloor b_i /2\rfloor$. The bottom face of the parent tile contains the top face of its child tile. One tile has exactly $2^{\dim-1}$ child tiles.
Therefore, we have a natural tree/forest structure that preserves the parent-child relationship. Call this (infinite) forest $T$ and the edges in $T$ \emph{tree edges}. Using the tree edges, each tile has a well-defined \emph{ancestor chain} by walking up along the parent edges. Further, the descendant tiles of $C$ have Euclidean ranges contained in the Euclidean range of $C$, and we use $T(C)$ to denote the subtree of $C$, i.e., the set of descendants of $C$.

The \emph{horizontal neighbors} of $C$ are the tiles at the same level that share boundaries with $C$. Each tile has $3^{\dim-1}-1$ horizontal neighbors. We connect each tile with its horizontal neighbors by \emph{bridge} edges. We denote the (infinite) \emph{tiling graph} $H$ with vertices as all tiles in a binary tiling and the edges as the union of tree edges and bridge edges. 

The binary tiling and the tiling graph provide discrete approximations of the hyperbolic distance. 
\begin{definition}
    A \emph{hyperbolic tile path}, $P(C_1, C_2)$, between any two tiles, $C_1, C_2$, is the path in $H$ of minimum length 
    that begins with $C_1$ and ends with $C_2$, and contains at most one bridge edge. 
    We denote the hop length of this path as $d_H(C_1, C_2)$.
\end{definition}

If $C_1$ is the ancestor of $C_2$ (or vice versa), the tile path follows the ancestor path. Otherwise, the path is formed by walking upward from $C_1$ and $C_2$ until the lowest level when their ancestors are horizontal neighbors. 
Further, for fixed $\dim$ and assuming unit-cost coordinate operations, an implicit representation of the tile path and its hop length can be computed in $O(1)$ time. First, check whether the tiles are equal or one is an ancestor of the other using containment of their Euclidean ranges. In these cases no bridge is needed. Otherwise, let $p=M(C_1)$ and $q=M(C_2)$ be the tile centers. By Lemma~7 of~\cite{Park2025-ge}, the bridge level is $\ell=\lfloor\log_2\|x(p)-x(q)\|_{\infty}\rfloor$ or $\ell+1$. Choose the lowest candidate level at least $\max\{\lvl(C_1),\lvl(C_2)\}$ at which the ancestors of $C_1,C_2$ (possibly the tiles themselves) are horizontal neighbors.

As shown in~\cite{Park2025-ge}, $d_H$ is not a metric as the triangle inequality does not necessarily hold. Nevertheless, we sometimes still use the wording of $d_H$ distance in this paper. Consider a tile path between $C_1$ and $C_2$ and any tile $C'$ on this path. One can check that the tile path between $C'$ and $C_2$ is a sub-path of the tile path of $C_1$ and $C_2$. 
For points $p\in C_1$ and $q\in C_2$ we also abuse the notation a little bit and write $d_H(p, q)=d_H(C_1, C_2)$.

Now we state a useful property of the hyperbolic tile path and $d_H$.
\begin{theorem}[\cite{Park2025-ge}]\label{thm:cell_dist}
    For any two points in their respective hyperbolic tiles, $p \in C_1$, $q \in C_2$:
    \begin{equation}
        |d_{\mathbb{H}}(p, q) - \ln(2)\cdot d_H(C_1, C_2)| \leq \eta_{\dim}, \,\mbox{where }\eta_{\dim}=3 \ln (\dim) + O(1)
    \end{equation}
\end{theorem}

The $O(1)$ term above is unspecified in~\cite{Park2025-ge}; however, combining their Lemma~8, Lemma~13 and Theorem~6 shows that we may take the $O(1)$ term as a constant of $2+6\ln 2\approx 6.16$. 

\section{Navigable Graph in \texorpdfstring{$\mathbb{H}^{\dim}$}{H-d}}\label{sec:positive}
In this section, we will construct a navigable graph for a given set of points $X \subseteq \mathbb{H}^{\dim}$ with $|X| = n$. The construction can be separated in two parts: graph $G_{> D}$ for navigating to a destination of distance above a threshold $D$ and graph $G_{\le D}$ for navigating to a destination within $D$. We will call these \emph{global} and \emph{local} routing, respectively. A quick observation is that $G=G_{> D}\cup G_{\le D}$ is a navigable graph: for navigating from any $p\in X$ to any $q\in X$, graph $G_{> D}$ guarantees that if $d_{\mathbb{H}}(p, q)>D$ there is a neighbor $u$ of $p$ in  $G_{> D}$ that is closer to $q$; we take such neighbors repeatedly until (after a finite number of steps) we arrive at a node $x$ within distance $D$ from $q$; and then $G_{\le D}$ guarantees delivery.  We explain the construction of $G_{\le D}$ in \Cref{subsec:local}, then $G_{> D}$ with Steiner points in \Cref{subsec:global-steiner}, and finally $G_{> D}$ without Steiner points in \Cref{subsec:global-nosteiner}.

\subsection{Local Routing}
\label{subsec:local}

In local routing, we focus on navigating to a destination within hyperbolic distance at most $D$, where $D=O(1)$. As will be explained later, we set $D=3\eta_{\dim}$ with $\eta_{\dim}$ defined in~\Cref{thm:cell_dist}.

We give two constructions for local routing. The first is a warm-up, showing a direct extension of the Yao graph to hyperbolic space gives a navigable graph with $O_{\dim, D}(n)$ edges, but with no guarantee on hop bounds. We then give a randomized construction with $O_{\dim, D}(n\log n)$ edges that supports greedy routing in expected $O(\log^2 n)$ hops.

\subsubsection{A Linear-Size Local Navigable Graph}

\begin{figure}
\centering
\includegraphics[width=0.3\linewidth]{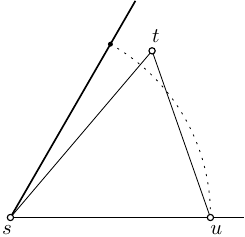}\hspace*{8mm}
\includegraphics[width=0.26\linewidth]{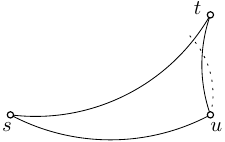}
    \caption{(Left) A Yao graph in the Euclidean space where each cone has an angle less than $\pi/3$ is a navigable graph. (Right) An analog for the hyperbolic setting as shown in \Cref{lem:yao-general}.}
    \label{fig:yao-euclidean}
\end{figure}

For $n$ points in the Euclidean plane, a Yao graph~\cite{Yao82constructing} of parameter $k$ on the points is defined in the following way. For each point $p$, place $k$ equally spaced rays such that they partition the plane into sectors of equal angles $\theta=2\pi/k$. Connect a directed edge from $p$ to the closest point inside each non-empty sector. As long as $k>6$, a Yao graph is a navigable graph. To see why, consider any pair $s$ and $t$. Choose the sector defined at $s$ that contains $t$, and consider the edge $su$ where $u$ is the closest point to $s$ among all points in this sector. If $u=t$ we are done. Otherwise, $\|st\|\geq \|su\|$. Notice that the angle $\angle ust\leq \theta<\pi/3$ is not the largest angle in $\triangle ust$. By sine law, $ut$ is not the longest edge in $\triangle stu$ and thus $\|ut\|<\|st\|$. This means moving to $u$ will be closer to $t$ than $s$. See \Cref{fig:yao-euclidean} for an example. This argument also works for points in $\dim$-dimensional Euclidean space, as long as any two rays inside one cone have an angle at most $\pi/3$. For a fixed dimension $\dim$, the total number of edges can be bounded by $O(c^{\dim} n)$ for some fixed constant $c>0$.

This observation can be extended to the hyperbolic space, but the required angle $\theta$ of two points inside the same cone depends on the distance $d_\mathbb{H}(s,u)$, the hyperbolic distance from $s$ to the closest point $u$ inside the cone. Here, for three points $t, u, s$, we can define the geodesic triangle and the angle $\angle_\mathbb{H} tsu$ is the angle between the tangent vectors of the geodesics $st$ and $su$ at point $s$. Notice that we use the subscript $_\mathbb{H}$ to differentiate from Euclidean triangles/angles.

\begin{restatable}{lemma}{yaogeneral}
\label{lem:yao-general}
    Consider two points $s$ and $u$ in $\mathbb{H}^{\dim}$ with $d_\mathbb{H}(s,u)= \xi>0$. If there is a third point $t$ such that $d_\mathbb{H}(s,t)>d_\mathbb{H}(s,u)$ and $\angle_\mathbb{H} tsu<\theta=\arccos(\frac{\cosh \xi}{\cosh \xi+1})$, then $d_\mathbb{H}(s,t)>d_\mathbb{H}(u,t)$.
\end{restatable}

The above Lemma suggests that an edge from $s$ to $u$ can help to cover other points $t$ whose direction from $s$ is sufficiently close. This immediately implies that we can use a Yao graph in the hyperbolic space for local routing. First, we need the following technical Lemma.

\begin{lemma}[\cite{Rogers1963,Dumer2006}]\label{lem:cone-cover}
    The surface of a unit $\dim$-dimensional Euclidean hypersphere can be covered by $O(1/\theta)^{\dim}$ spherical caps of angular radius $\theta/2$.
\end{lemma}

We use the Poincaré model. Without loss of generality, we assume point $p$ is at the origin of the Poincaré disk. We cover the hyperbolic space by \emph{cones} with apex at $p$ such that any two points inside the same cone form an angle at most
\[
\theta=\arccos{\frac{\cosh D}{\cosh D+1}}
=2\arcsin{\frac{1}{2\cosh (D/2)}}
\geq e^{-D/2}.
\]
Specifically, each (hyperbolic) cone in this family can be defined as the collection of points on geodesics $pt$, for all points $t$ inside a spherical cap of angular radius $\theta/2$.
By \Cref{lem:cone-cover} there are at most $O(e^{\dim D/2})$ cones. We connect $p$ to the closest point inside each cone. The graph has a total of $O_{\dim, D}(n)$ edges, where the big-O hides factors of functions of $\dim$ and $D$.
Applying \Cref{lem:yao-general} we immediately have,

\begin{theorem} \label{thm:yao}
For $n$ points in $\mathbb{H}^{\dim}$, the Yao graph with angle $\theta=\arccos{\frac{\cosh D}{\cosh D+1}}$ is a navigable graph of $O_{\dim, D}(n)$ edges for all pairs of distance at most $D$.
\end{theorem}

Theorem~\ref{thm:yao} gives a linear-size graph for local routing with possibly $\Omega(n)$ greedy routing hops. An example would be, for $n$ points on a line, the Yao graph can become a single path, and greedy routing requires $\Omega(n)$ hops. We next give a different local construction that uses a factor of $O(\log n)$ more edges to obtain an expected $O(\log^2 n)$ hop bound.

\subsubsection{A Bounded-Hop Local Navigable Graph}\label{subsubsec:local-boundedhop}

We now turn to the second local construction such that an accompanied greedy routing algorithm achieves bounded hops. The construction borrows the randomized-neighborhood graph idea of~\cite{arya93approximate}: after assigning the input points a random order, a vertex connects not only to its closest neighbor in a cone, but also to any point whose index is smaller than those of all closer points. On a path, these additional edges create shortcuts that can skip a constant fraction of the remaining vertices. There are two issues in transferring this idea to hyperbolic space. First, we need a bounded region in which a fixed collection of cones suffices, which we obtain by constructing a local graph around each tile of the binary tiling. Second, a useful randomized edge may be blocked by a closer point that does not make progress toward the destination.
To address the issues rigorously for the general case, there are additional technical challenges to show that an analog construction works for hyperbolic space with bounded distance $D$.

\paragraph{The local graph.}
The first issue is handled by localizing the construction. For each non-empty tile $C$ in the binary tiling, let $X_C$ consist of the input points lying in $C$ or in a tile within hyperbolic distance $D$ from $C$. We construct a \emph{local navigation graph} $G_C$ on $X_C$, and define $G_{\le D}\coloneq \bigcup_C G_C$,
where the union is over all non-empty tiles. Since $D=O(1)$, every input point belongs to at most $O_{\dim,D}(1)$ sets $X_C$. We describe the construction and routing analysis for one fixed $G_C$.

After global routing reaches a vertex $p$ with $d_{\mathbb H}(p,q)\le D$, let $C$ be the tile containing $p$. The remaining route to $q$ uses the edges of $G_C$.

Fix a tile $C$. Choose one point $v\in C$ and apply a hyperbolic isometry that maps $v$ to the origin $o$ of the Poincaré model. Then every point of $X_C$ lies within hyperbolic distance $r\coloneq D+2\mu_{\dim}=O(1)$ from $o$. The following lemma plays the same role as \Cref{lem:yao-general}, but its hypothesis is stated directly using the Euclidean angle in the Poincaré model. This allows us to organize the construction using Euclidean cones while measuring all distances in the hyperbolic metric. The proof is deferred to the Appendix~\ref{apd:pf}.

\begin{restatable}{lemma}{localeuc}
\label{lem:local-euclidean-angle}
Let \(r>0\). Suppose that \(s\in B(o,r)\) and
\(t,u\in B(s,r)\) satisfy $0<d_{\mathbb H}(s,u)<d_{\mathbb H}(s,t)$, if $\angle tsu\leq\arcsin\left(\frac{1}{2\cosh(r/2)}\right)$, where \(\angle tsu\) denotes the Euclidean angle in the Poincaré ball model, then $d_{\mathbb H}(t,u)<d_{\mathbb H}(s,t)$.
\end{restatable}

We use \emph{Euclidean cones} in the Poincaré model. This is only a choice of regions used to organize the neighbors: the vertices still lie in $\mathbb H^{\dim}$ and every distance in the construction is hyperbolic. The useful feature of Euclidean cones is that cones with the same direction vector at different apexes are translates of one another. This property will later give the nested sequence of cones used by the routing algorithm.

\begin{definition}[Euclidean cones]
Given a point $p$ in the Poincaré hyperbolic model which is an open unit ball
$\mathbb{B}^{\dim}=x\in\mathbb R^{\dim}:\lVert x\rVert<1$
and a unit vector $v$, the cone with apex $p$, direction $v$ and angular diameter $\theta$ is denoted as
\(      \mathcal{C}(p,v,\theta)
      =\left\{x\in \mathbb{B}^d:
         \angle xpv\leq\theta /2
      \right\}.\)
The \emph{core} of this cone $\mathcal{C}$ is $\mathcal{C}(p,v,\theta/3)$ and is denoted as $\hat{\mathcal{C}}$. We also define a \emph{bounded cone} with hyperbolic radius $r$ as
\(      \mathcal{C}(p,v,r,\theta)
      =\left\{x\in \mathcal{C}(p,v,\theta):
        0<d_{\mathbb{H}}(x,p) \leq r
      \right\}.\)
\end{definition}


We now specify the edges of $G_C$. At the origin $o$, choose a family of Euclidean cones of angular diameter $\theta/3$ whose directions cover the unit sphere, where the constant $\theta>0$ will be fixed later. Translate this family to every point $p\in X_C$. For each translated cone with direction $v$, use the larger cone $\mathcal C(p,v,\theta)$ when choosing neighbors; thus the original cone of diameter $\theta/3$ is the core of the larger cone.

Choose one random permutation of the input points, and call the rank of a point in this permutation its \emph{index}. The same permutation is used throughout the construction.

\begin{definition}[Randomized neighborhood graph]
For every point $p\in X_C$ and every cone $\mathcal{C}$ of Euclidean angular diameter $\theta$ at $p$, let $Y\subseteq X_C$ denote the points in the bounded cone with hyperbolic radius $r=D+2\mu_{\dim}$. We add a directed edge from $p$ to $x\in Y$ if $x$ has the smallest index among all points $y\in Y$ satisfying $d_{\mathbb H}(p,y)\le d_{\mathbb H}(p,x)$. Repeating this procedure for every $p\in X_C$ and every cone gives $G_C$.
\end{definition}

The construction is parameterized by the Euclidean angular diameter $\theta$, and we use the same value of $\theta$ for all local graphs $G_C$. For navigability, by \Cref{lem:local-euclidean-angle}, it is enough to require $\theta\leq \arcsin\left(\frac{1}{2\cosh(r/2)}\right)$.

The hop analysis introduces additional conditions needed to apply \Cref{lem:inner,lem:cap-q,lem:ratio}. Later, when we prove the hop bound, we show that these conditions and the navigability condition can be satisfied simultaneously by a single constant $\theta>0$. Since $r=D+2\mu_{\dim}$ depends only on $\dim$ and $D$, we denote this choice by $\theta=\theta(\dim,D)$ and use the same value throughout the construction.

The complete construction of $G_{\le D}$ is summarized in \Cref{alg:local-graph}.

\begin{algorithm}[htbp]
\caption{Construction of $G_{\le D}$}
\label{alg:local-graph}
\begin{algorithmic}[1]
\State Choose one random permutation of the input points and assign each point its index.
\State Fix the Euclidean angular diameter $\theta=\theta(\dim,D)$ as described above.
\For{each non-empty tile $C$}
\State Form $X_C$ from the points in $C$ and in the tiles within hyperbolic distance $D$ from $C$.
\State Map one point of $C$ to the origin of the Poincar\'e model.
\State Choose cones of angular diameter $\theta/3$ whose directions cover the unit sphere, and translate the same family to every $p\in X_C$.
\For{each $p\in X_C$ and each corresponding cone}
\State Expand the cone from angular diameter $\theta/3$ to $\theta$.
\State Consider the points of $X_C$ in this expanded cone within hyperbolic distance $r$ from $p$.
\State Order these points by increasing hyperbolic distance from $p$.
\State Connect $p$ to the first point and to every later point whose index is smaller than the indices of all earlier points.
\EndFor
\State Let the resulting edges form $G_C$.
\EndFor
\State Set $G_{\le D}=\bigcup_C G_C$.
\end{algorithmic}
\end{algorithm}

\begin{theorem}\label{thm:local-yao-hop}
For fixed $\dim$ and $D$, there exists a constant $\theta=\theta(\dim,D)>0$ such that the graph $G_{\le D}$ constructed in \Cref{alg:local-graph} has expected $O_{\dim,D}(n\log n)$ edges and is navigable for every pair of input points at hyperbolic distance at most $D$. Moreover, there exists a routing procedure that, for every such pair, uses the corresponding local graph $G_C$ and reaches the target in expected $O(\log^2 n)$ hops.
\end{theorem}

We first establish the edge bound and navigability in \Cref{thm:local-yao-hop}. Fix $p$ and one expanded cone, and order its points by increasing hyperbolic distance from $p$. The $i$-th point is connected to $p$ when its index is the smallest among the first $i$ points, which occurs with probability $1/i$. Thus, one cone contributes $\sum_{i=1}^{|Y|}1/i=O(\log n)$ outgoing edges in expectation. By \Cref{lem:cone-cover}, for a fixed $\theta$ there are $O_{\dim,\theta}(1)$ cones at each point. Since $\theta=\theta(\dim,D)$ and each input point belongs to only $O_{\dim,D}(1)$ sets $X_C$, we have
$$
    \mathbb E\!\left[|E(G_{\le D})|\right]
    =O_{\dim,D}(n\log n).
$$

The point of minimum hyperbolic distance from $p$ in every non-empty expanded cone is always among the randomized-neighborhood edges. For a target $q$, choose a cone whose core contains $q$. Since $\theta\leq\arcsin(1/(2\cosh(r/2)))$, \Cref{lem:local-euclidean-angle} implies that moving to such a closest point strictly decreases the hyperbolic distance to $q$. Hence $G_{\le D}$ is navigable for every source--target pair at hyperbolic distance at most $D$.

It remains to prove the $O(\log^2 n)$ expected hop bound and to verify that the choice $\theta=\theta(\dim,D)$ can satisfy all conditions required by the routing analysis. We show these below.

\paragraph{Proof of the Hop Bound in Theorem~\ref{thm:local-yao-hop}.}
We analyze the routing inside one fixed local graph $G_C$. Suppose that $p$ is the current vertex and $q$ is the target. To measure the progress toward $q$, let $\Cl(p,q)=B(q,d_{\mathbb H}(p,q))$, where $B(q,b)$ is the hyperbolic ball of radius $b$ centered at $q$. Whenever we write $|\Cl(p,q)|$ or $|B(q,b)|$, we count only the points of $X_C$ contained in the corresponding region. Thus, if we move from $p$ to a vertex $s$ closer to $q$, then $\Cl(s,q)\subsetneq\Cl(p,q)$, and the number of points in $\Cl(s,q)$ measures how much progress the move makes.

Now consider the point $x$ of smallest index in $\Cl(p,q)$. Since the indices come from a random permutation, $x$ is uniformly distributed among the points in $\Cl(p,q)$. Therefore, with probability at least $1/2$, $x$ lies among the half of these points closest to $q$, and hence $|\Cl(x,q)|\leq |\Cl(p,q)|/2$. If the graph contains the edge $px$, one hop therefore makes substantial progress toward $q$.

The difficulty is that $px$ may not be an edge of the randomized neighborhood graph. Although $x$ has the smallest index in $\Cl(p,q)$, there may be a point outside $\Cl(p,q)$ that lies in the same cone and is closer to $p$, preventing $x$ from being selected. The following definition captures this scenario.

\begin{definition}[Prunable vertex]
Consider a vertex $p\in X_C$, and a point $q\in X_C$. If there are two vertices $b,c\in X_C$ both inside a cone $\mathcal C$ with apex at $p$, with $c\in\Cl(p,q)$ and $b\notin\Cl(p,q)$, and $d_{\mathbb H}(p,c)\geq d_{\mathbb H}(p,b)$, we say $c$ is prunable.
\end{definition}

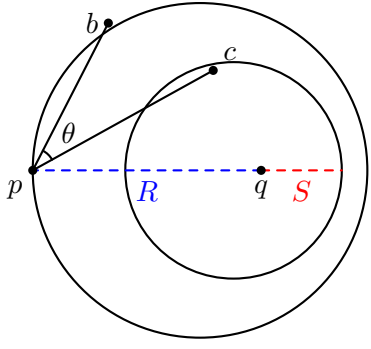
\begin{figure}[htbp]
\centering
\begin{tikzpicture}[scale=5, line cap=round, line join=round]

    \pgfmathsetmacro{\Rhyp}{1.4}
    \pgfmathsetmacro{\Shyp}{0.9}

    \pgfmathsetmacro{\a}{(exp(\Rhyp)-1)/(exp(\Rhyp)+1)} 


    \pgfmathsetmacro{\tR}{(exp(\Rhyp)-1)/(exp(\Rhyp)+1)}
    \pgfmathsetmacro{\centerR}{((1-\tR*\tR)*\a)/(1-\tR*\tR*\a*\a)}
    \pgfmathsetmacro{\radR}{((1-\a*\a)*\tR)/(1-\tR*\tR*\a*\a)}

    \pgfmathsetmacro{\tS}{(exp(\Shyp)-1)/(exp(\Shyp)+1)}
    \pgfmathsetmacro{\centerS}{((1-\tS*\tS)*\a)/(1-\tS*\tS*\a*\a)}
    \pgfmathsetmacro{\radS}{((1-\a*\a)*\tS)/(1-\tS*\tS*\a*\a)}

    \coordinate (p) at (0,0);
    \coordinate (q) at ({\a},0);

    \coordinate (OR) at ({\centerR},0);
    \coordinate (OS) at ({\centerS},0);

    \coordinate (spt) at ({\centerS+\radS},0);

    \coordinate (c) at ({\centerS - 0.20*\radS*cos(20)},
                       {2.70*\radS*sin(20)});

    \coordinate (b) at ({-0.4*cos(120)},
                       {0.45*sin(120)});


    \draw[thick] (OR) circle[radius=\radR];
    \draw[thick] (OS) circle[radius=\radS];

    \draw[dashed, thick, blue] (q) -- (p)
        node[midway, below, text=blue] {$R$};

    \draw[dashed, thick, red] (q) -- (spt)
        node[midway, below, text=red] {$S$};

    \draw[thick] (p) -- (b);
    \draw[thick] (p) -- (c);

    \fill (p) circle (0.012);
    \fill (q) circle (0.012);
    \fill (b) circle (0.012);
    \fill (c) circle (0.012);

    \node[below left] at (p) {$p$};
    \node[below] at (q) {$q$};
    \node[left] at (b) {$b$};
    \node[above right] at (c) {$c$};

\pic[
    draw,
    thick,
    "$\theta$",
    angle radius=8.32,
    angle eccentricity=2.35
] {angle = c--p--b};

\end{tikzpicture}
\caption{An illustration of Lemma~\ref{lem:inner} in the Poincaré ball model. We place \(p\) at the origin. The two circles represent the hyperbolic balls centered at \(q\) with radii \(R\) and \(S\), respectively, where \(R>S\). The angle \(\angle bpc\) is denoted by \(\theta\).}
\label{fig:lem_inner}
\end{figure}

See \Cref{fig:lem_inner} for an example. Specifically, for a prunable vertex $c$ there may not be an edge $p$ to $c$ even if $c$ has a very small index, as $p$ may have an edge to a vertex $b$ outside of $\Cl(p,q)$ which ``blocks'' the edge $pc$.

We handle this obstruction in two cases. At the beginning of a routing phase, let $R=d_{\mathbb H}(p,q)$, and let $S<R$ be a radius that will later be fixed as a constant fraction of $R$. If at least half of the remaining vertices lie in $B(q,S)$, \Cref{lem:inner} guarantees that a suitable vertex close to $q$ cannot be pruned, allowing a direct jump. Otherwise, we navigate through a region inside the cone that we call the cap. In this second case, \Cref{lem:cap-q} prevents the relevant cap vertices from being blocked, while \Cref{lem:ratio} guarantees that the translated cone continues to contain the target. The proofs of these geometric lemmas are deferred to Appendix~\ref{apd:pf}.

\begin{restatable}{lemma}{leminner}
\label{lem:inner}
Let \(0<S<R\), and let \(p,q\in \mathbb{B}^d\) satisfy $d_{\mathbb H}(p,q)=R$. Assume that $p,b,c\in B(0,r)$, and suppose that
$d_{\mathbb H}(q,b)\ge R$ and $d_{\mathbb H}(q,c)\le S$,
where $b\neq p$. Set $\rho=\tanh\left(\frac r2\right)$, and define
\[
\theta(R,S,r)
=
\frac{
2\arcsin\left(
\dfrac{\sinh\left(\frac{R-S}{2}\right)}
{\sinh(R+S)}
\right)}
{1+\rho\tanh\left(\frac{R+S}{2}\right)}.
\]
Then if $\angle bpc<\theta(R,S,r)$, we have
$d_{\mathbb H}(p,b)>d_{\mathbb H}(p,c)$.
\end{restatable}

Consider first the case $|B(q,S)|\geq|\Cl(p,q)|/2$. Let $x$ be the point of smallest index in $\Cl(p,q)$. With probability at least $1/2$, $x$ belongs to the half of the vertices in $\Cl(p,q)$ closest to $q$. Since $B(q,S)$ contains at least half of the vertices, in this event $x\in B(q,S)$ and
$|\Cl(x,q)|\leq|\Cl(p,q)|/2$.
By \Cref{lem:inner}, $x$ is not prunable, so the edge $px$ is present. Therefore, with probability at least $1/2$, a single direct jump reduces the number of remaining vertices by a factor of two.

It remains to handle the case $|B(q,S)|<|\Cl(p,q)|/2$. Here, most of the remaining vertices may lie near the boundary of $\Cl(p,q)$, and repeatedly taking short edges in the direction of $q$ may make little progress. We use the following region to control routing in this case.

\begin{definition}[$\CAP(p,\mathcal C)$]
Consider two points $p,q\in\mathbb H^d$ with $d_{\mathbb H}(p,q)=R$ and $S<R$. Take the cone $\mathcal C(p,v,\theta)$ with apex at $p$ whose core intersects the ball $B(q,S)$. We define the cap $\CAP(p,\mathcal C)$ as follows: for every Euclidean ray in $\mathcal C$ from $p$, the cap consists of points encountered before the ray enters the ball $B(q,S)$.
\end{definition}

To analyze this cap, we need to compare Euclidean angles in the Poincaré model with hyperbolic angles. We use the following lemma.

\begin{restatable}{lemma}{lemangle}
\label{lem:hyperbolic_angle}
For any Euclidean cone $\mathcal{C}(p,v,r,\theta)$ with $d_{\mathbb{H}}(p,o)\leq r$, and any two points $u,w\in\mathcal{C}(p,v,r,\theta)$, we have
\[
    \angle_{\mathbb H}upw
    \leq
    \delta
    \coloneq
    \theta+2\arcsin\!\left(\tanh^2(r/2)\right).
\]
\end{restatable}

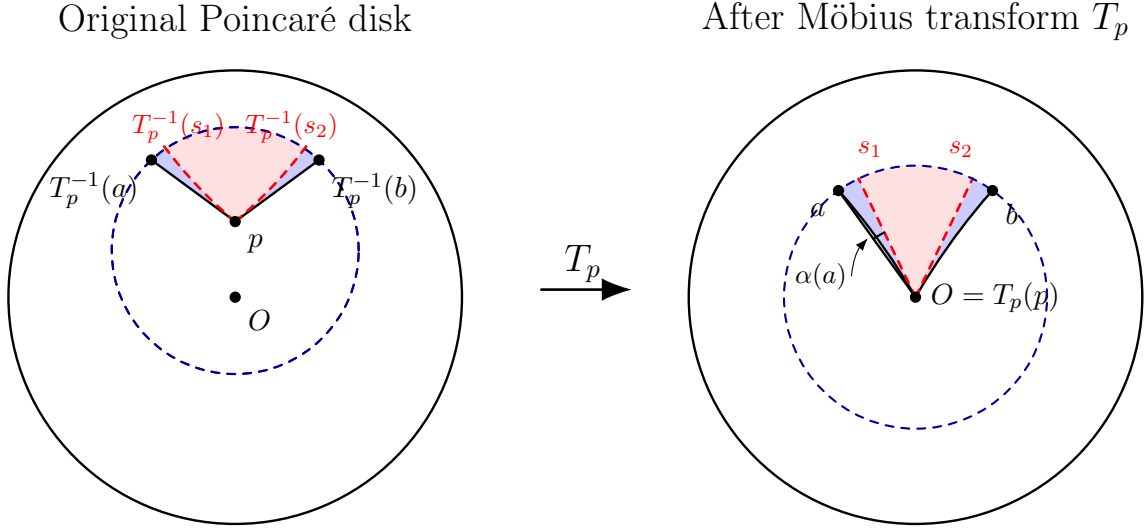
\begin{figure}[htbp]
\centering
\begin{tikzpicture}[
    line cap=round,
    line join=round,
    >=Latex,
    bluecone/.style={fill=blue!20, draw=none},
    redcone/.style={fill=red!12, draw=none},
    blackbd/.style={draw=black, line width=0.9pt},
    redgeo/.style={draw=red, dashed, line width=1pt},
    Lcircle/.style={draw=blue!60!black, dashed, line width=0.8pt},
    pt/.style={circle, fill=black, inner sep=1.5pt}
]

\def\R{3.0}
\def\sep{9.0}
\coordinate (Lcenter) at (0,0.616);
\def\Lrad{1.633}

\def\rhoD{1.740}

\begin{scope}
    \coordinate (O)  at (0,0);
    \coordinate (p)  at (0,1.00);

    \coordinate (A)  at (-1.85,2.36);
    \coordinate (B)  at ( 1.85,2.36);

    \coordinate (Ap) at (-1.30,2.70);
    \coordinate (Bp) at ( 1.30,2.70);

    \node[font=\Large] at (0,3.65) {Original Poincar\'e disk};

    \draw[blackbd] (O) circle (\R);

\path
    (p) arc[start angle=-130, end angle=-143.672, radius=5.940]
    coordinate (Gleft);

\path
    (p) arc[start angle=-50, end angle=-36.328, radius=5.940]
    coordinate (Gright);

\begin{scope}
    \clip (Lcenter) circle (\Lrad);
    \fill[bluecone]
        (p) -- (A)
        arc[start angle=128, end angle=52, radius=\R]
        -- cycle;
\end{scope}

\begin{scope}
    \clip (Lcenter) circle (\Lrad);
    \fill[redcone]
        (p)
        arc[start angle=-130, end angle=-143.672, radius=5.940]
        -- (-3.3,3.35)
        -- ( 3.3,3.35)
        -- (Gright)
        arc[start angle=-36.328, end angle=-50, radius=5.940]
        -- cycle;
\end{scope}

\draw[Lcircle] (Lcenter) circle (\Lrad);
    \draw[Lcircle] (Lcenter) circle (\Lrad);

\path[name path=Lcirclepath] (Lcenter) circle (\Lrad);

\path[name path=leftblackray]  (p) -- (A);
\path[name path=rightblackray] (p) -- (B);

\path[name path=leftredarc]
    (p) arc[start angle=-130, end angle=-143.672, radius=5.940];

\path[name path=rightredarc]
    (p) arc[start angle=-50, end angle=-36.328, radius=5.940];

\path[name intersections={of=leftblackray and Lcirclepath, by={a}}];
\path[name intersections={of=rightblackray and Lcirclepath, by={b}}];

\path[name intersections={of=leftredarc and Lcirclepath, by={sa}}];
\path[name intersections={of=rightredarc and Lcirclepath, by={sb}}];

\draw[blackbd] (p) -- (a);
\draw[blackbd] (p) -- (b);

\draw[redgeo]
    (p) arc[start angle=-130, end angle=-143.672, radius=5.940];

\draw[redgeo]
    (p) arc[start angle=-50, end angle=-36.328, radius=5.940];


    \node[pt] at (O) {};
    \node[below right=1pt] at (O) {$O$};

    \node[pt] at (p) {};
    \node[below right=1pt] at (p) {$p$};

    \node[pt] at (a) {};
    \node[below left=1pt] at (a) {$T_p^{-1}(a)$};

    \node[pt] at (b) {};
    \node[below right=1pt] at (b) {$T^{-1}_p(b)$};


    \node[red] at (-0.75,2.25) {\small{$T^{-1}_p(s_1)$}};
    \node[red] at ( 0.75,2.25) {\small{$T^{-1}_p(s_2)$}};
\end{scope}

\draw[-{Latex[length=4mm]}, line width=1pt]
    (4.05,0.10) -- (5.25,0.10);
\node[font=\Large] at (4.60,0.45) {$T_p$};

\begin{scope}[shift={(\sep,0)}]
    \coordinate (Oright) at (0,0);
    \coordinate (O)  at (0,0);

    \coordinate (A)  at (-1.85,2.36);
    \coordinate (B)  at ( 1.85,2.36);

    \coordinate (Ap) at (-1.30,2.70);
    \coordinate (Bp) at ( 1.30,2.70);

    \node[font=\Large] at (0,3.65) {After M\"obius transform $T_p$};

    \draw[blackbd] (O) circle (\R);

    \begin{scope}
        \clip (Oright) circle (\rhoD);
        \fill[bluecone]
            (O)
            .. controls (-0.55,0.95) and (-1.15,1.55) .. (A)
            arc[start angle=128, end angle=52, radius=\R]
            .. controls ( 1.15,1.55) and ( 0.55,0.95) .. (O)
            -- cycle;
    \end{scope}

    \begin{scope}
        \clip (Oright) circle (\rhoD);
        \fill[redcone]
            (O) -- (Ap)
            arc[start angle=116, end angle=64, radius=\R]
            -- cycle;
    \end{scope}

    \path[name path=TpL] (Oright) circle (\rhoD);

    \path[name path=leftcurve]
        (O) .. controls (-0.55,0.95) and (-1.15,1.55) .. (A);

    \path[name path=rightcurve]
        (O) .. controls ( 0.55,0.95) and ( 1.15,1.55) .. (B);

    \path[name path=leftred]  (O) -- (Ap);
    \path[name path=rightred] (O) -- (Bp);

    \path[name intersections={of=leftcurve and TpL, by={Ta}}];
    \path[name intersections={of=rightcurve and TpL, by={Tb}}];
    \path[name intersections={of=leftred and TpL, by={Sa}}];
    \path[name intersections={of=rightred and TpL, by={Sb}}];

    \draw[blackbd] (O) .. controls (-0.55,0.95) and (-1.15,1.55) .. (Ta);
    \draw[blackbd] (O) .. controls ( 0.55,0.95) and ( 1.15,1.55) .. (Tb);

    \draw[redgeo] (O) -- (Sa);
    \draw[redgeo] (O) -- (Sb);

    \node[pt] at (O) {};
    \node[right=2pt] at (O) {$O=T_p(p)$};

    \draw[Lcircle] (Oright) circle (\rhoD);


    \node[pt] at (Ta) {};
    \node[below left=1pt] at (Ta) {$a$};

    \node[pt] at (Tb) {};
    \node[below right=1pt] at (Tb) {$b$};

    \draw[blackbd] (O) -- (Ta);

\pic[
    draw,
    line width=0.8pt,
    angle radius=0.95cm
] {angle = Sa--O--Ta};

\node[
    fill=white,
    inner sep=1.5pt
] (alphaLabel) at (-1.25,0.28) {\small{$\alpha(a)$}};

\draw[->, thin]
    (alphaLabel.east)
    to[bend left=18]
    (-0.63,0.78);


    \node[red] at (-0.60,1.95) {\small{$s_1$}};
    \node[red] at ( 0.60,1.95) {\small{$s_2$}};
\end{scope}

\end{tikzpicture}
\caption{
Illustration of Lemma \ref{lem:hyperbolic_angle}.
Left: the black rays bound the Euclidean cone at $p$, the red curves are the
corresponding hyperbolic geodesics, and the blue dashed circle
bounds the ball centered at $p$ with hyperbolic radius $r$. Right: the same configuration after the
M\"obius map $T_p$ sending $p$ to the origin. Picking a point $a$ in the blue region, the angle $\alpha(a)$ measures
the deviation between the Euclidean radial direction and the corresponding
hyperbolic direction. Lemma \ref{lem:hyperbolic_angle} shows that this deviation is at
most $\arcsin(\tanh^2(r/2))$, and therefore the hyperbolic angular diameter of the cone is bounded by the Euclidean angular diameter plus twice this
error term.
}
\label{fig:lem_hyperbolic_angle}
\end{figure}

Informally, for a sufficiently small angular diameter $\theta$, \Cref{lem:cap-q} shows that every point in the cap is not prunable.

\begin{restatable}{lemma}{lemcap}
\label[lemma]{lem:cap-q}
Let $p\in B(o,r)$, where $o$ is the origin in the Poincaré ball model. We take $q\in\mathbb H^d$ such that $q$ lies in the core of bounded cone $\mathcal C(p,v,r,\theta)$. Suppose $d_\mathbb H(p,q)=R$, consider the cap $\CAP(p,\mathcal C)$ defined by $\mathcal C(p,v,r,\theta)$ and $B(q,S)$, where $S<R$. If
\[
    \theta
    <
    \arcsin\frac{\sinh S}{\sinh R}
    -
    \arcsin\left(\tanh^2\left(\frac r2\right)\right),
\]
every point $w\in\CAP(p,\mathcal C)$ and $u\in\mathcal C\setminus\Cl(p,q)$ satisfy
$d_{\mathbb H}(p,w)<d_{\mathbb H}(p,u)$.
\end{restatable}

\begin{figure}[htbp]
\centering
\begin{tikzpicture}[scale=8, line cap=round, line join=round]

    \pgfmathsetmacro{\eps}{0.45}
    \pgfmathsetmacro{\alp}{0.45}
    \pgfmathsetmacro{\rbase}{1.00}
    \pgfmathsetmacro{\Router}{(1+\eps)*\rbase}
    \pgfmathsetmacro{\Rinner}{(1+\alp*\eps)*\rbase}
    \def\coneRot{8} 

    \pgfmathsetmacro{\a}{tanh(\Router/2)}

    \pgfmathsetmacro{\tb}{tanh(\rbase/2)}
    \pgfmathsetmacro{\cb}{((1-\tb*\tb)*\a)/(1-\tb*\tb*\a*\a)}
    \pgfmathsetmacro{\rb}{((1-\a*\a)*\tb)/(1-\tb*\tb*\a*\a)}

    \pgfmathsetmacro{\ti}{tanh(\Rinner/2)}
    \pgfmathsetmacro{\ci}{((1-\ti*\ti)*\a)/(1-\ti*\ti*\a*\a)}
    \pgfmathsetmacro{\ri}{((1-\a*\a)*\ti)/(1-\ti*\ti*\a*\a)}

    \pgfmathsetmacro{\to}{tanh(\Router/2)}
    \pgfmathsetmacro{\co}{((1-\to*\to)*\a)/(1-\to*\to*\a*\a)}
    \pgfmathsetmacro{\ro}{((1-\a*\a)*\to)/(1-\to*\to*\a*\a)}

    \coordinate (p) at (0,0);
    \coordinate (q) at ({\a},0);
    \coordinate (Obase)  at ({\cb},0);
    \coordinate (Oinner) at ({\ci},0);
    \coordinate (Oouter) at ({\co},0);

\def\deltaA{18}
\def\coreA{6}
\def\vAng{4} 

\pgfmathsetmacro{\angA}{\vAng+\deltaA}
\pgfmathsetmacro{\angB}{\vAng-\deltaA}
\pgfmathsetmacro{\coreAup}{\vAng+\coreA}
\pgfmathsetmacro{\coreAdown}{\vAng-\coreA}

\pgfmathsetmacro{\tentA}{
  \ci*cos(\angA) - sqrt(\ri*\ri - \ci*\ci*sin(\angA)^2)
}
\pgfmathsetmacro{\tentB}{
  \ci*cos(\angB) - sqrt(\ri*\ri - \ci*\ci*sin(\angB)^2)
}

\coordinate (A) at ({\tentA*cos(\angA)}, {\tentA*sin(\angA)});
\coordinate (B) at ({\tentB*cos(\angB)}, {\tentB*sin(\angB)});

\pgfmathsetmacro{\phiA}{
  atan2(\tentA*sin(\angA), \tentA*cos(\angA)-\ci)
}
\pgfmathsetmacro{\phiBraw}{
  atan2(\tentB*sin(\angB), \tentB*cos(\angB)-\ci)
}
\pgfmathsetmacro{\phiB}{
  ifthenelse(\phiBraw < \phiA, \phiBraw + 360, \phiBraw)
}

    \coordinate (w) at ({0.095*cos(6)}, {0.095*sin(6)});   
    \coordinate (u) at ({0.92*cos(12)}, {0.92*sin(12)});   

    \fill[gray!50]
    (p) -- (A)
    arc[start angle=\phiA, end angle=\phiB, radius=\ri]
    -- cycle;

\draw[blue!70!black, thick] (p) -- (\angA:0.98);
\draw[blue!70!black, thick] (p) -- (\angB:0.98);

\draw[black, dashed, thick] (p) -- (\coreAup:0.98);
\draw[black, dashed, thick] (p) -- (\coreAdown:0.98);

\draw[->, blue!70!black, thick] (p) -- (\vAng:0.52)
    node[midway, below, text=blue!70!black] {$v$};

    \draw[red!75!black, thick]    (Oouter) circle[radius=\ro];
    \draw[orange!85!black, thick] (Oinner) circle[radius=\ri];


    \fill (p) circle (0.012);
    \fill (q) circle (0.012);
    \fill (w) circle (0.012);
    \fill (u) circle (0.012);

    \node[below left] at (p) {$p$};
    \node[below] at (q) {$q$};
    \node[above left] at (w) {$w$};
    \node[above right] at (u) {$u$};


    \node[gray] at (0.12,0.09) {$\operatorname{Cap}(p,\mathcal C)$};
    \node[blue!70!black] at (1.03,0.32) {$\mathcal C(p,v,r,\delta)$};
    \node[black] at (1.08, 0.12){$\mathcal C(p,v,r,\frac{\delta}{3})$};

    \draw[red!75!black, thick] (1.2,-0.18) -- (1.32,-0.18);
    \node[anchor=west, red!75!black] at (1.35,-0.18)
        {$B_{\mathbb H}(q,R)$};

    \draw[orange!85!black, thick] (1.2,-0.28) -- (1.32,-0.28);
    \node[anchor=west, orange!85!black] at (1.35,-0.28)
        {$B_{\mathbb H}(q,S)$};


\end{tikzpicture}
\caption{An illustration of Lemma~\ref{lem:cap-q} in the Poincaré ball model. The red and orange represent the balls of radius $R$ and $S$, respectively, all centered at \(q\). The solid blue rays bound the cone \(\mathcal C(p,v,r,\theta)\), while the dashed black rays indicate its core. The shaded region denotes the cap \(\operatorname{Cap}(p,\mathcal C)\).}
\label{fig:lem_cap}
\end{figure}
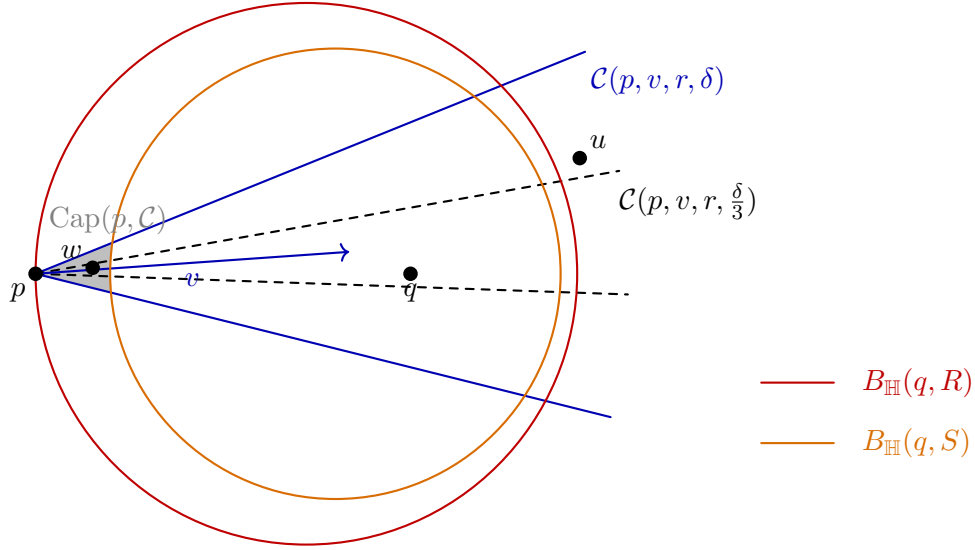

Next, we show that if, by greedy navigation, we move to a point $s$ inside the cap of a cone $\mathcal C(p,v,\theta)$, then the cone $\mathcal C(s,v,\theta)$ also contains $q$ and is inside $\mathcal C(p,v,\theta)$.

\begin{restatable}{lemma}{lemratio}
\label[lemma]{lem:ratio}
Fix a constant $r>0$. Let $o$ denote the origin of Poincaré model, $p$ satisfy $d_{\mathbb H}(p,o)\leq r$ and $v$ be a unit direction vector, and $\theta\le\arcsin(\frac{1}{2\cosh(r/2)})\leq \pi/6$. Suppose $s$ is in the cone $\mathcal C(p,v,r,\theta)$ whose core contains $q$. If
\[
    \beta
    =
    \frac{d_{\mathbb H}(p,s)}
         {d_{\mathbb H}(p,q)}
    \leq
    \frac{1}{3\cosh^2r},
\]
then $q\in\mathcal C(s,v,r,\theta)$.
\end{restatable}

\begin{figure}[htbp]
\centering
\begin{tikzpicture}[
    scale=5.0,
    line cap=round,
    line join=round,
    >=Latex,
    declare function={
        rho(\t)=(exp(\t)-1)/(exp(\t)+1);
        bcx(\ax,\ay,\R)=((1-rho(\R)*rho(\R))/(1-(\ax*\ax+\ay*\ay)*rho(\R)*rho(\R)))*\ax;
        bcy(\ax,\ay,\R)=((1-rho(\R)*rho(\R))/(1-(\ax*\ax+\ay*\ay)*rho(\R)*rho(\R)))*\ay;
        brad(\ax,\ay,\R)=((1-\ax*\ax-\ay*\ay)*rho(\R))/(1-(\ax*\ax+\ay*\ay)*rho(\R)*rho(\R));
    },
    rball/.style={draw=gray!65, dashed, line width=0.7pt},
    pcone/.style={fill=blue!20, fill opacity=0.45, draw=none},
    scone/.style={fill=red!20, fill opacity=0.30, draw=none},
    score/.style={fill=red!45, fill opacity=0.32, draw=none},
    conebd/.style={draw=blue!65!black, line width=0.9pt},
    sconebd/.style={draw=red!70!black, dashed, line width=0.9pt},
    scorebd/.style={draw=red!85!black, line width=0.8pt},
    axisline/.style={draw=black!70, line width=0.7pt},
    auxline/.style={draw=black!50, densely dotted, line width=0.55pt},
    pt/.style={circle, fill=black, inner sep=1.35pt},
    lab/.style={fill=white, fill opacity=0.86, text opacity=1, inner sep=1.5pt}
]

\def\R{2.20}             
\def\px{-0.50}           
\def\py{-0.34}
\def\vangle{2}           
\def\thetadeg{36}        
\pgfmathsetmacro{\half}{\thetadeg/2}
\pgfmathsetmacro{\corehalf}{\thetadeg/6} 
\def\L{1.90}             

\pgfmathsetmacro{\sx}{\px+0.10*cos(\vangle)}
\pgfmathsetmacro{\sy}{\py+0.10*sin(\vangle)}

\def\qx{0.23}
\def\qy{-0.31}

\pgfmathsetmacro{\rhoR}{rho(\R)}
\pgfmathsetmacro{\cpx}{bcx(\px,\py,\R)}
\pgfmathsetmacro{\cpy}{bcy(\px,\py,\R)}
\pgfmathsetmacro{\rp}{brad(\px,\py,\R)}
\pgfmathsetmacro{\csx}{bcx(\sx,\sy,\R)}
\pgfmathsetmacro{\csy}{bcy(\sx,\sy,\R)}
\pgfmathsetmacro{\rs}{brad(\sx,\sy,\R)}

\coordinate (o) at (0,0);
\coordinate (p) at (\px,\py);
\coordinate (s) at (\sx,\sy);
\coordinate (q) at (\qx,\qy);
\coordinate (Cp) at (\cpx,\cpy);
\coordinate (Cs) at (\csx,\csy);

\coordinate (pup)   at ($(p)+({\L*cos(\vangle+\half)},{\L*sin(\vangle+\half)})$);
\coordinate (plow)  at ($(p)+({\L*cos(\vangle-\half)},{\L*sin(\vangle-\half)})$);
\coordinate (paxis) at ($(p)+({\L*cos(\vangle)},{\L*sin(\vangle)})$);

\coordinate (sup)   at ($(s)+({\L*cos(\vangle+\half)},{\L*sin(\vangle+\half)})$);
\coordinate (slow)  at ($(s)+({\L*cos(\vangle-\half)},{\L*sin(\vangle-\half)})$);

\coordinate (scoreup)  at ($(s)+({\L*cos(\vangle+\corehalf)},{\L*sin(\vangle+\corehalf)})$);
\coordinate (scorelow) at ($(s)+({\L*cos(\vangle-\corehalf)},{\L*sin(\vangle-\corehalf)})$);

\draw[rball] (o) circle (\rhoR);
\node[gray!70,lab] at (0.00,0.77) {$B_{\mathbb H}(o,r)$};

\begin{scope}
    \clip (Cp) circle (\rp);
    \path[pcone] (p) -- (pup) -- (plow) -- cycle;
\end{scope}

\begin{scope}
    \clip (Cs) circle (\rs);
    \path[scone] (s) -- (sup) -- (slow) -- cycle;
\end{scope}

\begin{scope}
    \clip (Cs) circle (\rs);
    \path[score] (s) -- (scoreup) -- (scorelow) -- cycle;
\end{scope}

\begin{scope}
    \clip (Cp) circle (\rp);
    \draw[conebd] (p) -- (pup);
    \draw[conebd] (p) -- (plow);
    \draw[axisline] (p) -- (paxis);
\end{scope}
\begin{scope}
    \clip (p) -- (pup) -- (plow) -- cycle;
    \draw[conebd] (Cp) circle (\rp);
\end{scope}

\begin{scope}
    \clip (Cs) circle (\rs);
    \draw[sconebd] (s) -- (sup);
    \draw[sconebd] (s) -- (slow);
\end{scope}
\begin{scope}
    \clip (s) -- (sup) -- (slow) -- cycle;
    \draw[sconebd] (Cs) circle (\rs);
\end{scope}

\begin{scope}
    \clip (Cs) circle (\rs);
    \draw[scorebd] (s) -- (scoreup);
    \draw[scorebd] (s) -- (scorelow);
\end{scope}
\begin{scope}
    \clip (s) -- (scoreup) -- (scorelow) -- cycle;
    \draw[scorebd] (Cs) circle (\rs);
\end{scope}

\draw[blue!70!black]
    ($(p)+({0.11*cos(\vangle-\half)},{0.11*sin(\vangle-\half)})$)
    arc[start angle=\vangle-\half, end angle=\vangle+\half, radius=0.11];
\node[blue!70!black,lab] at ($(p)+(0.14,0.01)$) {$\theta$};

\draw[auxline] (p) -- (q);

\node[pt] at (o) {}; \node[above left=1pt] at (o) {$o$};
\node[pt] at (p) {}; \node[below left=1pt] at (p) {$p$};
\node[pt] at (s) {}; \node[below=2pt] at (s) {$s$};
\node[pt] at (q) {}; \node[above=3pt] at (q) {$q$};

\node[blue!70!black,lab] (CpLab) at (0.47,0.12) {$\mathcal C(p,v,r,\theta)$};
\draw[->, blue!70!black] (CpLab.west) to[out=180,in=45] (0.12,-0.18);

\node[red!70!black,lab] (CsLab) at (0.49,-0.58) {$\mathcal C(s,v,r,\theta)$};
\draw[->, red!70!black] (CsLab.west) to[out=180,in=-30] (0.10,-0.43);

\node[red!85!black,lab] (CoreLab) at (0.82,-0.12) {$\mathcal C(s,v,r,\theta/3)$};
\draw[->, red!85!black] (CoreLab.west) to[out=180,in=20] (0.06,-0.30);

\end{tikzpicture}
\caption{An illustration of Lemma \ref{lem:ratio} in a two-dimensional slice of the Poincaré model. The dashed circle represents the hyperbolic ball \(B_{\mathbb H}(o,r)\), with \(p\in B_{\mathbb H}(o,r)\). The blue region is the Euclidean bounded cone \(\mathcal C(p,v,r,\theta)\), and the red region is the Euclidean bounded cone \(\mathcal C(s,v,r,\theta)\). The darker red sector indicates the core \(\mathcal C(s,v,r,\theta/3)\) for reference; the lemma guarantees that \(q\) lies in the full translated cone \(\mathcal C(s,v,r,\theta)\). The sides of the cones are Euclidean rays, while the circular arcs indicate the hyperbolic-radius cutoff.}
\label{fig:lem_ratio}
\end{figure}
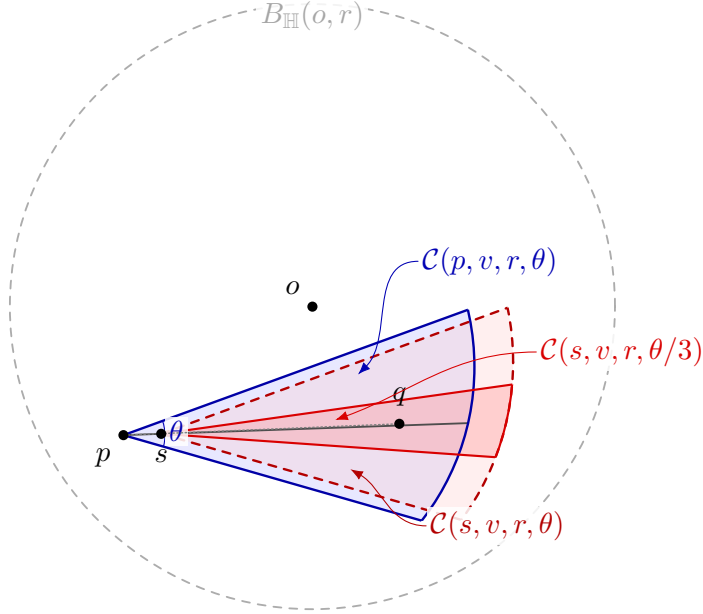

We now fix the parameters used in every phase. Recall that $r=D+2\mu_{\dim}$, and set
\[
    \kappa=1-\frac{1}{6\cosh^3 r},
    \qquad
    S=\kappa R.
\]
For this choice, there exists a sufficiently small constant $\theta=\theta(\dim,D)>0$ such that the navigability condition and the conditions needed to apply \Cref{lem:inner,lem:cap-q,lem:ratio} are all satisfied throughout the routing procedure. The details are deferred to Appendix~\ref{apd:angle_GN}.

We now define the routing procedure. It runs in phases. At the beginning of a phase, $p$ is the current vertex and $q$ is the target. We fix $R=d_{\mathbb H}(p,q)$, $S=\kappa R$, and $\Cl(p,q)$ throughout that phase. We denote by $\mathcal N(p)$ the neighbors of $p$ in the current local graph $G_C$, by $\mathcal N_{\mathcal C}(p)$ the neighbors lying in the current cone $\mathcal C$, and by $\low(U)$ the vertex of smallest index in a non-empty set $U$.

\begin{algorithm}
\caption{Greedy Navigation from $p$ towards $q$}
\begin{algorithmic}[1]
\Statex
\Function{GreedyNavigation}{$p,q$}
\State $S=\kappa \cdot d_{\mathbb{H}}(p,q)$
\State $x \gets \low\bigl(\mathcal N(p)\cap\Cl(p,q)\bigr)$
\If{$d_{\mathbb H}(x,q)<S$}
\State \Return $x$
\Else
\State Take direction $v$ such that the core of $\mathcal C=\mathcal C(p,v,\theta)$ contains $q$
\State $s\gets p$
\While{$\mathcal N_{\mathcal C}(s)\cap
\left(\CAP(p,\mathcal C)\cup B(q,S)\right)\neq\emptyset$}
\If{$\mathcal N_{\mathcal C}(s)\cap B(q,S)\neq\emptyset$}
\State $s\gets$ any point $\in\mathcal N_{\mathcal C}(s)\cap B(q,S)$
\State \Return $s$
\Else
\State $s\gets\low\bigl(\mathcal N(s)\cap\mathcal C\cap\CAP(p,\mathcal C)\bigr)$
\State $\mathcal C\gets\mathcal C(s,v,\theta)$
\EndIf
\EndWhile
\State \Return $s$
\EndIf
\EndFunction
\end{algorithmic}
\label{alg:q}
\end{algorithm}

The phase has two branches. It first checks the smallest-indexed, distance-improving neighbor $x$ of $p$. If $x\in B(q,S)$, it moves directly to $x$ and the phase ends. Otherwise, it chooses the cone whose core contains $q$ and navigates inside the cap. At a current vertex $s$, if a neighbor in the current cone lies in $B(q,S)$, the algorithm moves there and ends the phase. Otherwise, it moves to the smallest-indexed neighbor in both the current cone and the original cap, and translates the cone to the new vertex while keeping the same direction $v$.

\begin{lemma}\label{lem:navigation-in-cap}
The above procedure of navigating inside a cap $\CAP(p,\mathcal C)$ has $O(\log n)$ hops in expectation.
\end{lemma}

\begin{proof}
First, since all the cones are Euclidean translates with the same direction $v$, the cones at the vertices visited by the navigation algorithm form a nested chain. See \Cref{fig:chain}.

\begin{figure}[htbp]
\centering
\includegraphics[width=0.34\linewidth]{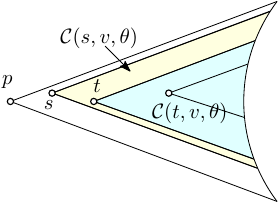}
\caption{Navigation within a cap $\CAP(p,\mathcal C)$: the cones of the vertices visited form a nested chain.}
\label{fig:chain}
\end{figure}

Suppose we move from a vertex $s$ to $t$, and let
\[
    k=|\mathcal C(s,v,\theta)\cap\CAP(p,\mathcal C)|.
\]
Build a directed graph on these $k$ vertices, where there is an edge from $x$ to $y$ if $\mathcal C(y,v,\theta)\subseteq\mathcal C(x,v,\theta)$. Between any two vertices there can be at most one directed edge, so this graph has at most $\binom{k}{2}$ edges and average out-degree at most $(k-1)/2$.

Conditioned on the routing history, the relative order of the remaining candidate vertices is still uniformly random. Hence $t$, the vertex of smallest index among the current candidates, is uniformly distributed among them. Therefore
\[
    \mathbb E\!\left[
    |\mathcal C(t,v,\theta)\cap\CAP(p,\mathcal C)|
    \,\middle|\, k
    \right]
    \leq \frac{k-1}{2}.
\]
Thus the expected number of remaining candidates decreases by a constant factor after every cap step, which gives an expected $O(\log n)$ hops before the cap navigation terminates.
\end{proof}

It remains to bound the number of phases. If $|B(q,S)|<|\Cl(p,q)|/2$, the cap branch reaches a vertex $s\in B(q,S)$, and hence
$|\Cl(s,q)|\leq |B(q,S)|<|\Cl(p,q)|/2$.
If $|B(q,S)|\geq|\Cl(p,q)|/2$, the argument following \Cref{lem:inner} shows that with probability at least $1/2$ the direct-jump branch moves to a vertex $x$ satisfying
$|\Cl(x,q)|\leq|\Cl(p,q)|/2$.
Therefore, after $O(1)$ phases in expectation, the number of remaining vertices decreases by a factor of two. There are at most $O(\log n)$ such decreases.

By \Cref{lem:navigation-in-cap}, each phase uses $O(\log n)$ hops in expectation. The total expected number of hops is therefore $O(\log^2 n)$, completing the proof of \Cref{thm:local-yao-hop}.

\subsection{Global Routing with Steiner Points}\label{subsec:global-steiner}

In global routing, we consider navigating to destinations at hyperbolic distance greater than $D$, using the binary tiling and its discrete distance $d_H$. We first retain the tile paths needed to connect the non-empty tiles and compress every maximal degree-two chain of empty tiles. The empty tiles that remain are assigned Steiner vertices, and every input or Steiner vertex is connected to the representative vertices of neighboring tiles in the compressed graph. This gives an auxiliary graph $G_H$ that supports greedy navigation measured by $d_H$. We then add a constant number of correction edges per vertex that jump farther along the same tile paths, enough to overcome the additive error between $d_H$ and $d_{\mathbb H}$; the resulting graph $G_{\mathbb H}$ supports greedy navigation measured by the actual hyperbolic distance. Finally, we optionally add shortcut edges to reduce the hop count to $O(\log^2 n)$.

\paragraph{Navigation using $d_H$.}
First, consider the binary tiling of hyperbolic space. Let $\mathcal{C}$ be the set of non-empty tiles. For each tile $C\in \mathcal{C}$, we designate one point of $X$ in the tile as the representative point, denoted as $v_C$.

Now consider all the tile paths in the tile graph $H$ between every pair of tiles of $\mathcal{C}$.
These paths $\mathcal{P}(\mathcal{C})$ are either ancestor chains or paths that connect two ancestor chains via a single bridge edge.

We compress the tile graph $H$ while preserving all paths in $\mathcal{P}(\mathcal{C})$. First, we remove tiles that do not appear on any such path. We then collapse every maximal ancestor chain of degree-two empty tiles into a single edge and remove the intermediate tiles. The new edge is assigned a weight equal to the number of edges in the chain before compression; all other edges have weight $1$.

\begin{figure}
\centering
\includegraphics[width=0.6\linewidth]{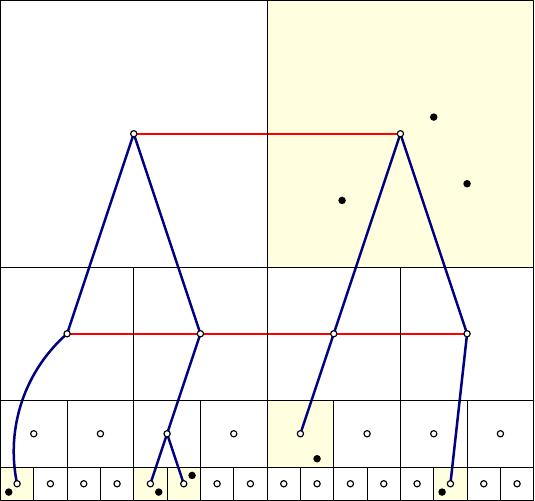} \hspace*{2mm}
\includegraphics[width=0.15\linewidth]{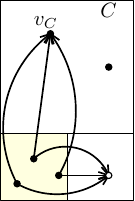}
\caption{(Left) The tile graph $H$ is compressed to $H(\mathcal{C})$ of size $O(n)$. The non-empty tiles in $\mathcal{C}$ are shaded. (Right) The edges in $G_H$ for points inside the tile are shaded to two neighboring tiles (with the horizontal neighbor being a Steiner vertex).}
\label{fig:steiner}
\end{figure}

Thus, all vertices of degree one (leaf vertices) are in $\mathcal{C}$. All vertices of degree two or higher that are empty tiles must be tiles where ancestor chains in $\mathcal{P}(\mathcal{C})$ meet for the first time or are endpoints of bridge edges on $\mathcal{P}(\mathcal{C})$. We denote these empty tiles as ``Steiner'' tiles $\mathcal{S}$. The compressed graph is denoted as $H(\mathcal{C})$ with vertex set $\mathcal{C}\cup \mathcal{S}$. The tile path between two tiles $C_1, C_2$ of $\mathcal{C}$, and its length (as the sum of weights on the edges), are both preserved in the compressed graph.
See \Cref{fig:steiner} (Left) for an example.

This graph $H(\mathcal{C})$ is exactly the same as the graph constructed in~\cite{Park2025-ge} in Theorem 16. Using the same argument, we can bound the size of $H(\mathcal{C})$.

\begin{lemma}\label{lem:steinergraph}
$|\mathcal{S}|=|S|=O_{\dim}(n)$, and each vertex of $\mathcal{C}\cup \mathcal{S}$ has degree of $O_{\dim}(1)$ in the compressed graph $H(\mathcal{C})$.
\end{lemma}

For each Steiner tile $C \in \mathcal{S}$, we pick an arbitrary point in $C$ as its representative point and a Steiner vertex. The set of Steiner vertices is denoted as $S$.

We now construct $G_H$ on the vertex set $X\cup S$ using the adjacencies of $H(\mathcal C)$. For every tile $C\in\mathcal C\cup\mathcal S$ and every neighboring tile $C'$ of $C$ in $H(\mathcal C)$, we add a directed edge from each vertex $p$ in $C$ to the representative vertex $v_{C'}$. Thus, the compressed graph specifies which tiles should be connected, while the representatives turn these tile adjacencies into edges of $G_H$. See \Cref{fig:steiner} (Right).

By \Cref{lem:steinergraph}, the out-degree of each vertex in $H(\mathcal{C})$ is a constant. Thus, we immediately have that the max out-degree is a constant and the total number of edges in $G_{H}$ is $O_{\dim}(n)$.

The reason for constructing $G_H$ this way is that it lets the routing algorithm follow the preserved tile path toward the destination. Suppose $p$ and $q$ lie in tiles $C_p$ and $C_q$, respectively. Let $C'$ be the next tile after $C_p$ on the tile path from $C_p$ to $C_q$ in $H(\mathcal C)$. By construction, $p$ has an edge to the representative $v_{C'}$, and the remaining tile path from $C'$ to $C_q$ is strictly shorter. Hence
$d_H(v_{C'},q)<d_H(p,q)$.
Repeating this argument gives a $d_H$-decreasing route until the current vertex is within hyperbolic distance $D$ from $q$.

\begin{lemma}\label{lem:greedyGH}
The greedy navigation algorithm $\mathcal{A}$ produces a path with strictly decreasing values of $d_H$ towards a destination $t\in X$ for any source $s\in X$ more than distance $D$ away from $t$.
\end{lemma}

Two points are worth noting. First, the argument above guarantees $d_H$-decreasing navigation from input points; a Steiner vertex need not support navigation to every destination unless it lies on the corresponding tile path. Second, once the current vertex is within distance $D$ from $q$, we switch to local routing. Therefore, in the Steiner construction, the Steiner vertices are also included in the local routing graph.

\paragraph{Navigation using $d_{\mathbb{H}}$.}

The graph $G_H$ guarantees progress with respect to $d_H$, but $d_H$ differs from the true hyperbolic distance by a constant additive error. Consequently, a single $d_H$-decreasing step need not decrease $d_{\mathbb H}$. To overcome this error, we augment $G_H$ with edges that jump sufficiently far along the same tile paths. The resulting graph is denoted by $G_{\mathbb H}$.

Specifically, for each tile $C\in \mathcal{C}\cup \mathcal{S}$ and each tile path $P\in \mathcal{P}(\mathcal{C})$ that $C$ stays on, we connect an edge from each point $p$ in $C$ to the representative vertex $v_{C'}$ in tile $C'$ on $P$ that meets the following conditions:
$d_H(C, C')>2\eta_{\dim}/\ln 2$ and the predecessor tile $C''$ along the tile path $P$ from $C$ to $C'$ has $d_H(C, C'')\leq 2\eta_{\dim}/\ln 2$.

Equivalently, along each relevant tile path through $C$, we add an edge to the representative of the first tile whose $d_H$ distance from $C$ exceeds $2\eta_{\dim}/\ln 2$. These are the correction edges of $G_{\mathbb H}$.

The graph $G_{\mathbb H}$ is the Steiner version of the global-routing graph denoted by $G_{>D}$ in our overall construction. We keep the notation $G_{\mathbb H}$ in this subsection to distinguish it from the auxiliary graph $G_H$ and from the later non-Steiner construction.

Similar to \Cref{lem:greedyGH}, we adapt the greedy algorithm $\mathcal{A}$. For a given pair $p, q\in X$ with distance greater than $3\eta_{\dim}$, $\mathcal{A}$ takes the next hop along the tile path towards $q$ with hyperbolic distance strictly smaller than the current node to $q$. The algorithm ends at a vertex with distance at most  $3\eta_{\dim}$ from $q$.

\begin{lemma}\label{lem:correction}
The greedy navigation algorithm $\mathcal{A}$ on graph $G_{\mathbb{H}}$ produces a path towards a destination $q\in X$ with strictly decreasing hyperbolic distance to $q$, for any source $p\in X$ of distance greater than $3\eta_{\dim}$.
$G_{\mathbb{H}}$ has $O_{\dim}(n)$ edges.
\end{lemma}

\begin{proof}
Suppose we want to navigate from $p$ to $q$, $d_{\mathbb{H}}(p, q)> 3\eta_{\dim}$. Suppose $p$ and $q$ stay in tiles $C_1$ and $C_2$, respectively.
By \Cref{thm:cell_dist} we have $d_H(C_1, C_2)\geq (d_{\mathbb{H}}(p, q)-\eta_{\dim})/\ln 2>2\eta_{\dim}/\ln 2$. Thus, we consider the neighbor $v_{C'}$ of $p$ in $G_{\mathbb{H}}$ such that $C'$ is on the tile path from $C_1$ to $C_2$ and $d_H(C_1, C')>2\eta_{\dim}/\ln 2$. Notice that $d_H(C_1, C_2)=d_H(C_1, C')+d_H(C', C_2)$. Now we have,
\begin{align*}
    d_{\mathbb{H}}(v_{C'}, q) & \leq \ln 2\cdot  d_H(C', C_2)+\eta_{\dim} \tag{By \Cref{thm:cell_dist}}\\
    & =\ln 2\cdot d_H(C_1, C_2) -\ln 2 \cdot d_H(C_1, C')+\eta_{\dim}\\
    &\leq  d_{\mathbb{H}}(p, q)+2\eta_{\dim}  -\ln 2 \cdot d_H(C_1, C') \tag{By \Cref{thm:cell_dist}}\\
    & <  d_{\mathbb{H}}(p, q)
\end{align*}
The number of edges in $G_{\mathbb{H}}$ is still linear in $n$. This is because each extra edge added to $p$ in $G_{\mathbb{H}}$ is charged to a neighbor of a tile within $d_H$ distance $2\eta_{\dim}/\ln 2$ of tile $C$ in $H(\mathcal{C})$.
We have a constant number of tiles within $d_H$ distance of $2\eta_{\dim}/\ln 2$ and the max degree in $H(\mathcal{C})$ is a constant.
\end{proof}

\paragraph{Reducing the hop count to $O(\log^2 n)$.}

The constructions above guarantee navigability but do not bound the number of hops: a long ancestor chain may still require $\Omega(n)$ hops. We therefore optionally augment the graph using the heavy-light path decomposition~\cite{Sleator1983-nk} applied to the tree edges of $H(\mathcal C)$.

The tree edges in the compressed graph $H(\mathcal{C})$ form a forest. For each tree $T$ in the forest, the heavy edges are identified as the edge from each node $v$ to the child whose subtree has the maximum number of tiles (ties are broken arbitrarily). These heavy edges form a set of node-disjoint paths. The remaining edges are called light edges. For any node in the tree, the ancestor chain to the root has at most $O(\log |T|)$ light edges -- this is because each light edge encountered means that the number of descendant tiles will at least double, and this cannot repeat more than $\log |T|$ times.

For each heavy path, we use a standard ``shortcutting'' technique to add additional edges. Specifically, given a heavy path $v_0, \cdots, v_{k}$ of length $k$, we connect an edge $(v_0, v_k)$. Then we connect both endpoints $v_0$ and $v_k$ with the middle vertex $v_{\lfloor k/2\rfloor}$, and recursively add shortcut edges on each sub-path. The total number of edges added is $O(k)$, and each node has degree $O(\log k)$. With the shortcut edges, for any node $v_i$ and $v_j$ there is a path of at most $O(\log k)$ hops between them. Now we are ready to shortcut the graphs.

\begin{itemize}
\item \emph{Shortcutting $G_H$.} For graph $G_H$, we apply the heavy-light decomposition to the tree edges of $H(\mathcal{C})$ and produces a graph $\tilde{G}_H$.
For each identified heavy path, we add shortcut edges between the representative points of the tiles. The total number of edges is still linear in $n$, $O_d(n)$, and the out-degree is $O_d(\log n)$. This immediately reduces the hop bound in greedy routing. Specifically, the properties of $G_H$ ensure that with one hop from $p$, we arrive at a representative vertex. We just need to show that greedy routing along an ancestor chain $P$ has a bounded hop length. Indeed, $P$ stays on $O(\log n)$ heavy paths; for each heavy path on $P$, the shortcut edges allow greedy navigation to use at most $O(\log n)$ hops. This gives a total of $O(\log^2 n)$ hops. Since greedy routing guided by $d_H$ uses at most two ancestor chains the hop bound follows.

\item \emph{Shortcutting $G_{\mathbb{H}}$.} The modification for $G_{\mathbb{H}}$ is similar. We will start with the graph $H(\mathcal{C})$ and add both the ``correction'' edges and shortcut edges.
Consider a heavy path $v_0, \cdots, v_{k}$ identified on the tree edges in $H(\mathcal{C})$, which is an ancestor chain in the tile graph $H(\mathcal{C})$. Each vertex $v_i$ is a representative vertex of a tile.
We first group the vertices along the heavy path as sequences of length $4\eta_{\dim}/\ln 2$ and call them $I_1, I_2, \cdots I_{\ell}$ with $\ell=\lceil k\ln 2/(4\eta_{\dim})\rceil$. For example, $I_1$ includes the first sequence of $4\eta_{\dim}/\ln 2$ vertices starting from $v_0$, and $I_2$ includes the next sequence of $4\eta_{\dim}/\ln 2$ vertices and so on.
We connect all vertices in $I_j$ with all vertices in the next interval $I_{j+1}$. By \Cref{lem:correction} we know that for each vertex in $I_j$, there is at least one vertex in $I_{j+1}$ whose $d_{\mathbb{H}}$ distance is strictly smaller towards the destination.
Now we add shortcut edges on top of the intervals -- we first add $O_{\dim}(1)$ edges between all vertices in $I_1$ to all vertices in $I_{\ell}$. Then we add edges connecting all vertices from $I_1$ and $I_{\ell}$ to the vertices in $I_{\lfloor \ell/2\rfloor}$, and then recurse in both parts.
We add these new edges on top of $G_{\mathbb{H}}$ and call this new graph $\tilde{G}_{\mathbb{H}}$.
Again the total number of extra edges is $O_{\dim}(n)$ and the out-degree is at most $O_{\dim}(\log n)$. The hop count reduction follows from the same argument. In one hop from $G_{\mathbb{H}}$ we can navigate to a representative vertex which is strictly closer to the destination. Then we navigate among representative vertices, for which a long ancestor chain can be now reduced to be of at most $O(\log^2 n)$ hops. Notice that for any navigation instance, there are at most two long ancestor chains. So the total number of hops is bounded by $O(\log^2 n)$.

\end{itemize}
For completeness, \Cref{alg:global-steiner} summarizes the entire Steiner construction above. 

\begin{algorithm}[htbp]
\caption{Global routing graph with Steiner points}
\label{alg:global-steiner}
\begin{algorithmic}[1]
\State Let $\mathcal C$ be the set of non-empty tiles and choose a representative $v_C$ for every $C\in\mathcal C$.
\State Let $\mathcal P(\mathcal C)$ be the collection of tile paths between pairs of tiles in $\mathcal C$.
\State Remove all tiles that do not occur on a path in $\mathcal P(\mathcal C)$.
\State Contract every maximal ancestor chain of degree-two empty tiles into one weighted edge.
\State Let $\mathcal S$ be the remaining empty tiles, and choose a Steiner representative $v_C$ for every $C\in\mathcal S$.
\For{each $C\in\mathcal C\cup\mathcal S$}
\For{each neighbor $C'$ of $C$ in $H(\mathcal C)$}
\State Add $p\to v_{C'}$ for every vertex $p$ in $C$.
\EndFor
\EndFor
\State Let $G_H$ be the graph consisting of these edges.
\For{each $C\in\mathcal C\cup\mathcal S$ and each $P\in\mathcal P(\mathcal C)$ containing $C$}
\State Let $C'$ satisfy
$d_H(C,C')>2\eta_{\dim}/\ln 2$, while its predecessor $C''$ on $P$ satisfies
$d_H(C,C'')\leq2\eta_{\dim}/\ln 2$.
\State Add $p\to v_{C'}$ for every vertex $p$ in $C$.
\EndFor
\State Let $G_{\mathbb H}$ be $G_H$ together with these correction edges.
\If{an $O(\log^2 n)$ hop bound is desired}
\State Apply the heavy-light decomposition to the tree edges of $H(\mathcal C)$.
\State Add the heavy-path shortcut edges described above to $G_{\mathbb H}$.
\EndIf
\State \Return the resulting graph.
\end{algorithmic}
\end{algorithm}

\subsection{Navigation Without Steiner Points}\label{subsec:global-nosteiner}

In the previous subsection, we used $O(n)$ Steiner vertices to represent empty tiles on the relevant tile paths. We now show that the Steiner vertices can be removed while keeping a graph $G$ of size $O_{\dim}(n)$ on the input set $X$ alone. For global routing, we use $d_H$ as the greedy routing measure, with node ID used to break ties between vertices with the same $d_H$ value. Once the route reaches a point within distance $D$ from the destination, we switch to local routing, which does not use Steiner points.

We use the same compressed graph $H(\mathcal C)$ and tile paths $\mathcal P(\mathcal C)$ as in the Steiner construction. The construction has three parts. First, inside each non-empty tile we choose one input point as its representative and add \emph{star edges} from the other points in the tile to this representative. Second, we add \emph{vertical edges} between representatives to support routing along the ancestor-chain portions of $H(\mathcal C)$. Finally, when a tile path uses a bridge whose endpoint is empty, there is no Steiner vertex through which we can route; we replace these missing vertices by \emph{crossing edges} between suitable input points.

\textbf{Star edges.}
For each non-empty tile $C\in\mathcal C$, let $v_C$ be the point of smallest ID in $C$, and use it as the representative of $C$. For every other point $p\in C$, add the directed edge $p\to v_C$. By the tie-breaking rule, this moves from $p$ to the preferred point in the same tile. All remaining edges can therefore be described using representative vertices.

\textbf{Vertical edges.}
For each non-empty tile $C\in\mathcal C$, if it has a non-empty ancestor tile, let $C'$ be the nearest such ancestor along its ancestor chain. We add the two directed edges $v_C\to v_{C'}$ and $v_{C'}\to v_C$. This gives $O(n)$ edges. By using these vertical edges, any point $p$ can navigate to a destination in an ancestor tile (or in a descendant tile) via greedy routing with $d_H$.

\textbf{Crossing edges.}
The vertical edges handle the ancestor-chain portions of a tile path. The remaining issue is a bridge $(P,Q)$ between horizontally adjacent tiles. When $P$ and $Q$ are non-empty, their representatives can play the same role as in the Steiner construction. When one or both are empty, we instead need input vertices that allow the route to move upward toward the bridge, cross it, and then move downward toward the destination.

Consider a bridge of two horizontally adjacent tiles $P$ and $Q$. If $P$ and $Q$ are non-empty tiles, then we can generally let all vertices in $T(P)$ first navigate to $v_P$, through an edge $(v_P,v_Q)$, and then route from $v_Q$ to the destination. However, when $P$ and $Q$ are empty tiles, and no Steiner points are allowed, we need to identify other pairs of vertices to take the place of the bridge $(P,Q)$.

Consider the tile $C_1$ that contains $p$ and tile $C_2$ that contains $q$. If the tile path from $C_1$ to $C_2$ uses a bridge edge $(P,Q)$, there are three cases:
\begin{itemize}
\item $C_1=P$ and $C_2\in T(Q)$, where $T(Q)$ is the subtree of $Q$.
\item $C_1\in T(P)$ and $C_2=Q$.
\item $C_1\in T(P_i)$ and $C_2\in T(Q_j)$, where $P_i$ is a child tile of $P$, $Q_j$ is a child tile of $Q$, and $P_i,Q_j$ are not horizontal neighbors. The last condition is because if $P_i,Q_j$ are horizontal neighbors, the tile path from the tiles in $T(P_i)$ to the tiles in $T(Q_j)$ will go through a lower bridge, e.g., $(P_i,Q_j)$, instead. Notice that for any tiles $U,U'\in T(P_i)$, their tile paths to the same destination tile $C_2\in T(Q_j)$ agree with each other starting from $P$. Thus, moving within $T(P_i)$ from $C_1$ to a tile closer to $P$ also brings it closer to $C_2$.
\end{itemize}

To realize this route without Steiner vertices, we define an exit vertex and an entrance vertex for each relevant tile. Exit vertices are used while moving upward toward the bridge, and entrance vertices are used while moving downward toward the destination.

\emph{Exit vertices and entrance vertices.}
For any tile $C$, define the \emph{exit vertex} $\hat{s}(C)$ as the representative vertex of a tile inside $T(C)$ that is closest to $C$ in $d_H$ distance.\footnote{There might be multiple tiles whose $d_H$ distances to $C$ are the same. In this case, we break the tie by the hyperbolic distance of their representative points.}
If $C$ is non-empty, then $\hat{s}(C)=v_C$.

We define the \emph{entrance vertex} $\tilde{t}(C)$ as the representative vertex of the tile closest to $C$ whose tile path to a descendant of $C$ passes through $C$. Equivalently, it is chosen among the tiles outside

$$
\bigcup_{C'\text{ child of }C}
\left(
T(C')
\cup
\bigcup_{C''\text{ horizontal neighbor of }C'}T(C'')
\right).
$$

If $C$ is non-empty, then $\tilde{t}(C)=v_C$.

\begin{figure}
\centering
\includegraphics[width=0.2\linewidth]{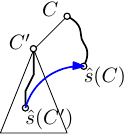} \hspace*{2mm}
\includegraphics[width=0.2\linewidth]{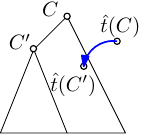}
\caption{(Left) The exit edge from $\hat{s}(C')$ to $\hat{s}(C)$. (Right) The entrance edge from $\tilde{t}(C)$ to $\tilde{t}(C')$.}
\label{fig:edges}
\end{figure}

\emph{Exit and entrance edges.}
For each downward tree edge $(C,C')$ of the compressed graph $H(\mathcal C)$, if $\hat{s}(C)\neq\hat{s}(C')$, then $\hat{s}(C)$ is closer to $C$ than $\hat{s}(C')$, since $\hat{s}(C')$ is a candidate in the definition of $\hat{s}(C)$. We add the directed edge
$\hat{s}(C')\to\hat{s}(C)$.
These are the \emph{exit edges}; intuitively, they move the route upward toward $C$. See \Cref{fig:edges} (Left).

Similarly, if $\tilde{t}(C)\neq\tilde{t}(C')$, we add the directed edge
$\tilde{t}(C)\to\tilde{t}(C')$.
For any destination tile inside $T(C')$ whose tile path passes through $C'$, this moves the route closer to the destination. These are the \emph{entrance edges}; intuitively, they move the route downward into the destination subtree. See \Cref{fig:edges} (Right).

\emph{Bridge edges.}
Consider a bridge $(P,Q)$ that occurs on a path in $\mathcal P(\mathcal C)$. If $P$ is non-empty, the source side of the bridge is represented by $v_P$; otherwise, for each child $P_i$ of $P$ whose subtree contains a non-empty tile, the source side is represented by $\hat{s}(P_i)$. Similarly, if $Q$ is non-empty, the destination side is represented by $v_Q$; otherwise, for each child $Q_j$ of $Q$ whose subtree contains a non-empty tile, the destination side is represented by $\tilde{t}(Q_j)$. For every relevant pair of source and destination sides, except when $P_i$ and $Q_j$ are themselves horizontal neighbors, we add the corresponding directed bridge edge. No edge is added when its two endpoints are the same.

\textbf{Navigation algorithm.}
We now show that these edges support greedy navigation using $d_H$. Let $p$ and $q$ lie in distinct tiles $C_1$ and $C_2$. We focus on the third case above, since the first two are obtained by omitting one side of the same argument. Suppose the tile path consists of the upward ancestor chain
$C_1=U_k,U_{k-1},\ldots,U_1,P$, followed by the bridge $(P,Q)$ and the downward chain
$Q,D_1,\ldots,D_\ell=C_2$.
The corresponding route in $G$ has three parts: it follows exit vertices upward toward the bridge, uses at most three crossing edges to pass the bridge, and then follows entrance vertices downward toward $C_2$. Initially, with at most one hop, $p$ moves to $\hat{s}(U_k)$.

\begin{itemize}
\item Suppose $u=\hat{s}(U_i)$. If $\hat{s}(U_{i-1})\neq\hat{s}(U_i)$, then $\hat{s}(U_{i-1})$ is closer to $C_2$ than $\hat{s}(U_i)$. By the construction of the exit edges, there is an edge from $u$ to $\hat{s}(U_{i-1})$. If $\hat{s}(U_{i-1})=\hat{s}(U_i)$, we continue along the tile path until either $u$ has a neighbor closer to $C_2$ or we reach $\hat{s}(U_1)$.

\item Suppose $u=\hat{s}(U_1)$. If $P$ and $Q$ are both non-empty, we move from $\hat{s}(U_1)$ to $v_P$, then from $v_P$ to $v_Q$, and then to $\tilde{t}(D_1)$. If $P$ is non-empty and $Q$ is empty, we skip $v_Q$ in this sequence. If $P$ is empty and $Q$ is non-empty, we move from $\hat{s}(U_1)$ to $v_Q$ and then to $\tilde{t}(D_1)$. If both $P$ and $Q$ are empty and $\hat{s}(U_1)\neq\tilde{t}(D_1)$, we move directly from $\hat{s}(U_1)$ to $\tilde{t}(D_1)$.

\item Suppose $u=\tilde{t}(D_i)$. As long as $\tilde{t}(D_{i+1})\neq\tilde{t}(D_i)$, moving from $\tilde{t}(D_i)$ to $\tilde{t}(D_{i+1})$ gets closer to $q$ by the definition of the entrance edge.
\end{itemize}

At $\tilde{t}(D_\ell)$, we switch to local routing.

\textbf{Size of $G$.}
The star edges and vertical edges each contribute $O(n)$ edges. For the crossing edges, every tree edge of $H(\mathcal C)$ contributes at most one exit edge and one entrance edge. For each relevant bridge $(P,Q)$, the number of children of $P$ and $Q$ is $O_{\dim}(1)$, so only $O_{\dim}(1)$ bridge edges are added. Since $H(\mathcal C)$ has $O_{\dim}(n)$ vertices and edges, the total number of crossing edges is $O_{\dim}(n)$. Therefore, $G$ has $O_{\dim}(n)$ edges. The maximum out-degree, however, need not be constant; in particular, many entrance edges may share the same tail endpoint.

For example, consider a binary tiling where the root has a vertex $r$ and $n$ vertices lie on the same level such that their pairwise distance is greater than their distance to $r$. In this case, $r$ keeps directed edges to all $n$ vertices.

Combined with the local routing with the Yao graph, we have the following.

\begin{theorem}\label{thm:nosteiner}
For $n$ points in $\mathbb{H}^{\dim}$, there is a navigable graph of size $O_{\dim}(n)$, where global greedy routing uses $d_H$ together with the stated tie-breaking rule, and local routing uses $d_{\mathbb H}$.
\end{theorem}

\textbf{Reducing the hop bound to $O(\log^2 n)$.}
The route above may still contain ancestor chains of $\Omega(n)$ hops. We shortcut the exit and entrance chains exactly as for $G_H$ in the Steiner construction: apply the heavy-light decomposition to the tree edges of $H(\mathcal C)$ and add the recursive midpoint shortcut edges along every heavy path. These shortcuts are placed between the corresponding exit vertices for an upward traversal and between the corresponding entrance vertices for a downward traversal. Since $d_H$ decreases monotonically along the relevant ancestor chain, these shortcut edges preserve greedy progress. A root-to-node chain meets $O(\log n)$ heavy paths, and each heavy path can be traversed in $O(\log n)$ hops. Hence global routing uses $O(\log^2 n)$ hops, while the shortcutting adds only $O_{\dim}(n)$ edges.

The global non-Steiner construction is summarized in \Cref{alg:global-nosteiner}.

\begin{algorithm}[htbp]
\caption{Global routing without Steiner points}
\label{alg:global-nosteiner}
\begin{algorithmic}[1]
\State Let $\mathcal C$ be the set of non-empty tiles.
\For{each $C\in\mathcal C$}
\State Let $v_C$ be the point of smallest ID in $C$.
\State Add the star edge $p\to v_C$ for every other point $p\in C$.
\EndFor
\For{each $C\in\mathcal C$ having a non-empty ancestor tile}
\State Let $C'$ be its nearest non-empty ancestor.
\State Add $v_C\to v_{C'}$ and $v_{C'}\to v_C$.
\EndFor
\For{each tile $C$ represented in $H(\mathcal C)$}
\State Compute $\hat{s}(C)$ and $\tilde{t}(C)$ as defined above.
\EndFor
\For{each downward tree edge $(C,C')$ of $H(\mathcal C)$}
\If{$\hat{s}(C)\neq\hat{s}(C')$}
\State Add $\hat{s}(C')\to\hat{s}(C)$.
\EndIf
\If{$\tilde{t}(C)\neq\tilde{t}(C')$}
\State Add $\tilde{t}(C)\to\tilde{t}(C')$.
\EndIf
\EndFor
\For{each bridge $(P,Q)$ occurring on a path in $\mathcal P(\mathcal C)$}
\State Let the source candidates be $v_P$ if $P$ is non-empty; otherwise use $\hat{s}(P_i)$ for every child $P_i$ whose subtree contains a non-empty tile.
\State Let the destination candidates be $v_Q$ if $Q$ is non-empty; otherwise use $\tilde{t}(Q_j)$ for every child $Q_j$ whose subtree contains a non-empty tile.
\State Add the corresponding directed bridge edges, except for horizontally adjacent pairs $P_i,Q_j$ and pairs whose two endpoints coincide.
\EndFor
\If{an $O(\log^2 n)$ global hop bound is desired}
\State Apply the heavy-light decomposition to the tree edges of $H(\mathcal C)$.
\State Add the recursive midpoint shortcuts between the corresponding exit vertices and entrance vertices.
\EndIf
\State \Return $G$.
\end{algorithmic}
\end{algorithm}

\subsection{\texorpdfstring{$(1+\eps)$}{(1+epsilon)}-Approximate Nearest Neighbor Search}\label{subsec:ANNS}

The navigable graph defined on $X$ allows reaching the target destination $q\in X$ in $O(\log^2 n)$ hops. In this subsection, we show that if the query $q\in \mathbb{H}^{\dim}$ is not in $X$, we can guarantee to reach a $(1+\eps)$-approximate nearest neighbor of $q$ in $O(\log^2 n)$ hops. Formally, $y\in X$ is a $(1+\eps)$-approximate nearest neighbor of $q$ if $d_{\mathbb{H}}(y, q)\leq (1+\eps)\cdot d_{\mathbb{H}}(x_q, q)$ where $x_q\in X$ is the nearest neighbor of $q$, $d_{\mathbb{H}}(x_q, q)\leq d_{\mathbb{H}}(x, q)$, $\forall x\in X$. For this, we need to modify our greedy navigation algorithms for both local and global routing regimes.

We first explain the high-level idea. For any query point $q\in \mathbb{H}^d$,  we start with global routing by using the approximate Voronoi diagram developed in~\cite{Park2025-ge}, which partitions the hyperbolic space into $O(n)$ Voronoi cells such that for each cell, an approximate nearest neighbor can be found with a constant additive error. 
Each Voronoi cell is a collection of tiles in the binary tiling. With a revised global routing phase, we can navigate to the Voronoi cell containing $q$ and identify a non-empty tile $X'_q$ such that $d_H(q, X'_q)$ is the smallest among all non-empty tiles. 
Notice that $d_H(q, X'_q)\leq d_H(q, x_q)$, where $x_q$ is the true nearest neighbor. But the tile $X'_q$ may not contain $x_q$.
Take an arbitrary point $x'_q$ in $X'_q$, we have by \Cref{thm:cell_dist},
$$d_{\mathbb{H}}(q, x'_q)\leq \ln 2\cdot d_H(q, x'_q)+\eta_{\dim} \leq \ln 2\cdot d_H(q, x_q)+\eta_{\dim} \leq d_{\mathbb{H}}(q, x_q)+2\eta_{\dim}.$$
If $2\eta\leq \eps \cdot d_{\mathbb{H}}(q, x_q)$, we are done and $x'_q$ is a $(1+\eps)$-approximate nearest neighbor. Otherwise, $2\eta> \eps \cdot d_{\mathbb{H}}(q, x_q)$. Thus,
    $$d_{\mathbb{H}}(x'_q, x_q)
    \leq d_{\mathbb{H}}(x'_q, q)+d_{\mathbb{H}}(q, x_q)\leq 2d_{\mathbb{H}}(q, x_q) +2\eta_{\dim}
    \leq 4\eta_{\dim}/\eps +2\eta_{\dim}=(4/\eps+2)\eta_{\dim}.$$
Now we run the local routing algorithm for ANNS at the tile $X'_q$ as well as all tiles that are within hyperbolic distance $(4/\eps+2)\eta_{\dim}$ from $X'_q$. There are at most $\dim^{O((\ln\dim)/\eps)}$ tiles to check. 
For each tile, we use local routing to find a $(1+\eps)$ approximate nearest neighbor. One of them will be successful.  We explain the two steps respectively.

\paragraph{Global routing for ANNS} Global routing for ANNS follows the design in~\cite{Park2025-ge} for the approximate Voronoi diagram. 
We have a few remarks. First we will use Steiner points for all non-empty tiles that will be identified below. Second, we always start the navigation from a specific ``root'' tile and use $d_H$ as guidance, instead of $d_{\mathbb{H}}$. 

For ease of explanation, we can assume that all points of $X$ have their $x$-coordinates to be within $[1/4, 1/2]^{\dim-1}$ and $z$-coordinates to be within $(0, 2]$-- note that we can transform all coordinates $(z, x)$ as $(\alpha \cdot z, \alpha \cdot x +\beta)$, with a factor $\alpha>0$ and translation vector $\beta \in \mathbb{R}^{\dim-1}$ and this does not change the hyperbolic distances. Now we build a compressed ``quadtree'' $\mathcal{Q}$ of all tiles within $[0, 2]\times [0, 1]^{\dim-1}$ with the root to be the tile $[1, 2]\times [0, 1]^{\dim-1}$. Our search for a query $q \in \mathbb{H}^d$ always starts from the root of the tree $\mathcal{Q}$, and navigates downward.  
The tiles in the binary tiling are partitioned into $O(n)$ Voronoi cells $\{V_i\}$. The navigation eventually arrives at the Voronoi cell that contains $q$. Each Voronoi cell $V_i$ is associated with a set of candidate tiles $\mathcal{R}(V_i)$, with $|\mathcal{R}(V_i)|=O_{\dim}(1)$, one of which is a non-empty tile with the smallest $d_H$ distance to $q$. The computation of these candidate tiles is described in~\cite{Park2025-ge}. Here we explain the quadtree and show how it can be navigated for ANNS. 

The quadtree $\mathcal{Q}$ is defined on the binary tiling, guided by quadtree $\mathcal{Q}_X$ of the points of $X$, projected on the $x$-coordinates. For each point $p\in X$ with $p=(z, x)$, its projection is $p'\in \mathbb{R}^{\dim-1}$ with coordinates $x$. 
Now we can build the (Euclidean) compressed quadtree $\mathcal{Q}_X$ of $X'$.
A node in $\mathcal{Q}_X$ is a \emph{regular} node when it has at least two non-empty children, or a \emph{leaf} node when it has at most one point inside, or a \emph{compressed node} with exactly one non-empty child (which is the smallest-sized quad). 
Each node is associated with an $x$-range. The $x$-range of a regular and a leaf node is its quad.  The $x$-range of a compressed node $\nu$ is the quad of $\nu$ excluding the quad of its non-empty child $\nu'$.
The quadtree $\mathcal{Q}$ on the hyperbolic tiling take the same tree structure as $\mathcal{Q}_X$, where a node in $\mathcal{Q}$ is a tile whose $x$-coordinate projection corresponds to the quad of the nodes in $\mathcal{Q}_X$. Now, the Voronoi diagram is defined by $\mathcal{Q}$. The Voronoi cell of a tile as a regular node in $\mathcal{Q}$ is itself. The Voronoi cell of a tile as a leaf node in $\mathcal{Q}$ is itself together with all descendant tiles. The Voronoi cell of a compressed tile $\nu$ (with non-empty child $\nu'$) is the collection of all descendant tiles of  $\nu$, with the descendants of $\nu'$ subtracted. The $x$-range of all leaf nodes and the compressed nodes form a partition of the domain $[0, 1]^{\dim-1}$. 
Further, for each tile $C$ in $\mathcal{Q}$ whose descendant tiles are not empty, we add to both $\mathcal{Q}$ and $\mathcal{Q}_X$ the nodes corresponding to the horizontal neighboring tiles (that might be empty of points of $X$) of $C$. 
Similarly, for a compressed node $\nu$ and its child $\nu'$, we add horizontal neighbors of both tiles corresponding to $\nu$ and $\nu'$. This will possibly introduce more compressed nodes.   

In summary, the nodes in the quadtree partition the space $[0, 1]^{\dim-1}\times (0, 2]$ into $O(n)$ Voronoi cells. The Voronoi cell of an ordinary node $\nu$ is the corresponding tile $C(\nu)$ itself; for a leaf node, it is the union of the corresponding cell and all descendants (i.e., all points with the same $x$-range and smaller $z$ coordinates); and for a compressed node the it is the set of points on or below $C(\nu)$ but now on or below $C(\nu')$ with $\nu'$ as $\nu$'s child. All nodes in $\mathcal{Q}$ that correspond to non-empty tiles are Steiner points. Edges are added following the quadtree edges to support top-down navigation. 
Greedy navigation using the quadtree $\mathcal{Q}$ walks down from the root. The next hop is the quadtree child tile whose subtree contains $q$ with the $d_H$ value to $q$ to be strictly decreasing. Similar to before, we can again use heavy-light decomposition and add shortcut edges to reduce the hop bound to $O(\log^2n)$ in the global navigation step. We omit the details here.

\paragraph{Local routing for ANNS} For approximate nearest neighbor search, we use exactly the same graph construction as in \Cref{subsubsec:local-boundedhop} but we need to modify the local routing algorithm because $q$ may not be in $X$. 
Similar to before, the algorithm proceeds in phases. Each phase is marked by a pair $(p, q)$.
Denote $R=d_{\mathbb{H}}(p, q)$, $L=R/(1+\eps)$, and $S=\kappa R$ with $1/(1+\eps)<\kappa<1$.
If $x_q\notin B(q, L)$, then $p$ is already a $(1+\eps)$ approximate nearest neighbor of $q$. In that case, the following algorithm (\Cref{alg:ANNS}) fails to find a next hop, and we are done. Otherwise, $x_q$ can be anywhere in $B(q, L)$; we consider all the cones $\mathcal{C}$ whose core intersects $B(q, L)$ and consider the cap  $\CAP(p, C)$ with respect to $\mathcal{C}$.
We define $\mathcal{N}_{\mathcal{C}}(s)$ as the set of neighbors of $s$ in the cone $\mathcal{C}$. Note that there are only $O(1)$ cones at each vertex.

\begin{algorithm}
\caption{Approximate Nearest Neighbor Search for $q\in \mathbb{H}^{\dim}$}
\begin{algorithmic}[1]
\Statex
\Function{ANNS}{$p,q$} 
\State 
$R=d_{\mathbb{H}}(p, q)$, $S=\kappa \cdot R$; $L=R/(1+\eps)$
\If {$\mathcal{N}(p)\cap \Cl(p,q)=\emptyset$} \Comment{$\Cl(p,q)$ is the ball centered at $q$ with radius $d_{\mathbb{H}}(p, q)$}
\State \Return $p$
\EndIf
    \State $x \gets$ $\low \bigl(\mathcal{N}(p) \cap \Cl(p,q)\bigr)$ 
    
    \If{$d_{\mathbb{H}}(x, q)<S$} \State \Return $x$
    \EndIf
    \For{every direction $v$ such that $\hat{\mathcal{C}}(p, v, \theta)\cap B(q, L)\neq \emptyset$} 
    \State $s\gets p$, $\mathcal{C}\gets \mathcal{C}(s, v, \theta)$
    \While{$\mathcal{N}_{\mathcal{C}}(s)\cap B(q, d_{\mathbb{H}(q, s)})\neq \emptyset$}
    \If{$\mathcal{N}_{\mathcal{C}}(s) \cap B(q, S) \neq \emptyset$}
                \State $t \gets$ any point $\in \mathcal{N}_{\mathcal{C}}(s) \cap B(q, S)$
                \State \Return $t$
            \Else
                \State $s \gets$ $\low\bigl(\mathcal{N}_{\mathcal{C}}(s)\cap \CAP(p, \mathcal{C}(p, v, \theta))\bigr)$
                \State $\mathcal{C}\gets\mathcal{C}(s,v,\theta)$
            \EndIf
        \EndWhile
        \EndFor
\EndFunction
\end{algorithmic}\label{alg:ANNS}
\end{algorithm}

In order to show that this modified local routing arrives at an approximate nearest neighbor, we need to re-examine the geometric structures in \Cref{lem:cap-q} and \Cref{lem:ratio}. 

Since $x_q$ can be anywhere in $B(q,L)$, we need to pick suitable $S=\kappa R$ and Euclidean angle $\theta$ such that the inequality in \Cref{lem:ratio} is true for any $x_q$ and corresponding points in the cap of a cone $\mathcal{C}(p,v,r,\theta)$. Here, we can pick $$S=\kappa R=\left(1-\frac{\eps^2}{6(1+\eps)(2+\eps)\cosh^2r\cosh^4(3r/2)}\right)R$$, and any $$\theta \leq \theta_0=\frac{3\eps}{32(2+\eps) \cosh^4(3r/2)}.$$ The detail why this makes the inequality in \Cref{lem:ratio} to be true are deferred to \Cref{lem:cone-cap-depth}. 

\begin{restatable}{lemma}{conecapdepth}
\label{lem:cone-cap-depth}
Consider the Poincar\'e ball $\mathbb B^d$, $\dim\geq2$, with origin $o$.
Fix $r>0$ and $p,q\in\mathbb B^{\dim}$ such that
$p\in\overline B_{\mathbb H}(o,r)$ and
\(
    R:=d_{\mathbb H}(p,q)<r, 0\leq L<R.
\)
Let $\mathcal{C}=\mathcal{C}(p,v,\widehat r,\theta)$ be the bounded Euclidean cone of
angular diameter $\theta$, and let
$\widehat C=C(p,v,\widehat r,\theta/3)$ be its core.  Assume that
\(
    \widehat C\cap\overline B_{\mathbb H}(q,L)\neq\varnothing.
\)
Set
\[
 a:=\frac{R-L}{3\cosh^2 r},\qquad
 \Lambda:=\cosh^2\!\left(\frac{3r}{2}\right),\qquad
 \sigma:=\frac{a(R-L)}{2\Lambda^2(R+L)},\qquad
 S:=R-\sigma,
\]
and
\[
 \theta_{\mathrm{cap}}:=\frac{3\sigma}{16a}
 =\frac{3(R-L)}{32\Lambda^2(R+L)}.
\]
Then $L<S<R$.  If $0<\theta\leq\theta_{\mathrm{cap}}$,
let $\operatorname{Cap}(p,\mathcal{C})$ denote the cap cut out by
$\overline B_{\mathbb H}(q,S)$,
then
\begin{equation}
 \operatorname{Cap}(p,\mathcal{C})
 \subset B_{\mathbb H}(p,a)
 \subset\overline B_{\mathbb H}\!\left(
 p,\frac{R-L}{3\cosh^2 r}\right).
 \label{eq:cone-cap-depth}
\end{equation}
\end{restatable}

Next, we present the following lemma, which shows that similar geometric structures also hold for the specific $S$.

\begin{restatable} {lemma} {annscap} \label{lem:ANNS-cap}
    Let $p\in B(o,r)$, where $o$ is the origin in ball model, $q\in \mathbb{H}^{\dim}$ with $d_{\mathbb{H}}(p,q)=R$, $L=\frac{R}{(1+\eps)}$, and $$S=\left(1-\frac{\eps^2}{6(1+\eps)(2+\eps)\cosh^2r\cosh^4(3r/2)}\right)R.$$  Assume that the core of $\mathcal{C}(p,v,r,\theta)$ intersects $B(q,L)$. Consider the cap defined by $\mathcal{C}(p,v,r,\theta)$ and $B(q,S)$, if \[\theta< \frac{2}{1+\tanh^2r}\arcsin\left(\frac{\sinh((R-S)/2)}{\sinh(\min(r,\frac{2+\eps}{1+\eps}R))}\right),\] every point $w \in \CAP(p,\mathcal{C})$ and $u\in \mathcal{C} \setminus \Cl(p, q)$, $d_{\mathbb{H}}(p, w)<d_{\mathbb{H}}(p, u)$. 
\end{restatable}

The proof of \Cref{lem:ANNS-cap} is deferred to the Appendix~\ref{apd:pf}. Then we can pick the Euclidean angle $\theta$ for ANNS to be a constant only depends on $\eps$ and $r$, and the detail are deferred to Appendix~\ref{apd:angle_ANNS}. 

\begin{theorem} 
For any $q\in \mathbb{H}^{\dim}$, we can design a greedy routing algorithm that takes guidance by $d_H$ (in global routing) or $d_{\mathbb{H}}$ (in local routing) on the navigable graph (with Steiner points) of size $O_{\dim}(n\log n)$ to arrive at a vertex $y\in X$ that is a $(1+\eps)$-approximate nearest neighbor of $q$, for any $\eps>0$. 
\end{theorem}

\section{Lower Bounds}\label{sec:lower}
In this section, we give lower bounds on the maximum degree and the number of edges of the navigable graphs for hyperbolic space. First, we give the lower bound of the number of edges for $\alpha$-navigable graph, $\alpha>1$, and the lower bound of maximum degree for 1-navigable graph in hyperbolic plane. Our construction is based on the construction of the following lemma. 

\begin{lemma}[\cite{Kisfaludi-Bak_vanWordragen_2025}] \label[lemma]{lem:lw-kw}
    For any $n \geq2$ and any $\epsilon \in \left(0,1\right]$, the point set $P(n,\epsilon) \subset \mathbb{H}^2$ of n points equally  placed on a circle of radius $r=\frac{1}{\epsilon}\ln{n}$ has the property that any distance between a pair of point is in $\left(2r(1-\epsilon), 2r\right]$. 
\end{lemma}

\begin{restatable}{theorem}{lowerboundalledges}
\label{thm:lw-alpha}
    For any $\alpha$>1, there is an $n$-point data set $X$ in $\mathbb{H}^2$ such that any $\alpha$-navigable graph for $X$ needs all $n(n-1)$ directed edges. 
\end{restatable}

\begin{theorem}
    The maximum out-degree of a $1$-navigable graph for points in the hyperbolic plane $\mathbb{H}^2$ may need to be $\Omega(n)$. 
\end{theorem}
\begin{proof}
    Let $\epsilon$ be $0.01$. We follow the construction in \Cref{thm:lw-alpha} to obtain the point set $Y$. The point set $Y'=Y\cup \{o\}$ achieves the lower bound, where $o$ denotes the center of the circle. 

    We claim that for 1-navigable graphs, the out-degree of $o$ is $\Omega(n)$. For any point $y_i$, $d_{\mathbb{H}}(o,y_i)\leq r+\tau$. By the proof of \Cref{thm:lw-alpha}, $d_{\mathbb{H}}(y_i, y_j)\geq 2r(1-\epsilon)-2\tau$. Since $\beta\leq 2\epsilon r$, substitute $\epsilon$ and $\tau$, we can verify that $d_{\mathbb{H}}(o,y_i) < d_{\mathbb{H}}(y_i, y_j)$. 

    The proof that there are no ties is similar to that of  \Cref{thm:lw-alpha}, so we omit it.
\end{proof}

Next, we give the lower bound on the number of edges for 1-navigability in high-dimensional hyperbolic space. Theorem 4 of \cite{diwan24navigable} states that for $n$ points distributed independently and uniformly in $\{-1,1\}^{\dim}$,
with high probability, there is no repeated point, and any 1-navigable graph under the Euclidean metric requires $\Omega(n^{3/2-\rho})$ edges, where $\rho=\max\{\frac{\log\log n}{\log n}, c\sqrt{\frac{\log n}{d}}\}$ for a universal constant $c$. So there exists a hard instance that achieves the lower bound in $\Omega(\log n) $- dimensional Euclidean space. We embed the hard instance from \cite{diwan24navigable} into hyperbolic space, and obtain the following results formally.

\begin{theorem}\label{thm:lb-high-d}
    There exists an $n$-point set in $\mathbb{H}^{\dim+1}$ for which the 1-navigable graph has number of edges $=\Omega(n^{3/2-\rho})$, where $\rho=\max\{\frac{\log\log n}{\log n}, c\sqrt{\frac{\log n}{\dim}}\}$ for a universal constant $c$. In particular, $d=\Theta(\delta^{-2}\log n)$ gives $\Omega(n^{3/2-\delta})$ edges, and $\dim=\Theta(\log^3n)$ gives $\Omega(n^{3/2}/\log n)$ edges. 
\end{theorem}
\begin{proof}
    Let $X=\{x_i\} \subset \{-1,1\}^{
    \dim}$ denote a hard instance in Euclidean space. For a point $x_i$, we define $p_i \in \mathbb{H}^{d+1}$ to be $(z, x_i)$, with any fixed $z>0$. Consider any two point $p_i$ and $p_j$, \[d_{\mathbb{H}}(p_i, p_j)=\arccosh \left (1+\frac{\|x_i-x_j\|^2}{2z^2}\right ).\] The function $\arccosh \left (1+\frac{a^2}{2z^2}\right )$ is strictly increasing for $a\geq 0$. Therefore, \[d_\mathbb{H}(p_i,p_j)< d_\mathbb{H}(p_k,p_j) \Leftrightarrow \|x_i-x_j\|<\|x_k-x_j\|.\] For a point $u$, define its relative ordering as the ordering of all other points according to their distances from $u$. Note that the requirements for constructing 1-navigable graphs depend only on these relative orderings. Therefore, if the relative ordering of every point remains unchanged, then the requirements for constructing the 1-navigable graph remain unchanged as well. \cite{diwan24navigable}'s lower bound therefore transfers unchanged to the hyperbolic space. 
\end{proof}



\section{Tree Metrics}
\begin{theorem}\label{thm:tree}
    Given a finite tree metric $M=(X,d)$ where each point $x\in X$ has distinct distances to all others in $X$, a minimum spanning tree (MST) of $X$ with each edge replaced by two bidirectional edges is a navigable graph. 
\end{theorem}
\begin{proof}
    We claim that for any two distinct vertices $p,q \in X$, the first vertex $u$ on the unique path from $p$ to $q$ in the MST satisfies $d(p,q) > d(u,q)$. 

    Since $M$ is a tree metric, there exists a tree $T$ with $X\subseteq T$ such that for every $p,q \in X$, $d(p,q)=d_T(p,q)$, where $d_T$ is the distance in the tree $T$. Given three points $p,q,u \in X$, there exists a unique intersection $c \in T$ of the paths $\overline{pu}, \overline{uq}$ and $\overline{pq}$ in the tree. Removing $c$, we get three subtrees $T_p, T_q$ and $T_u$.  
    
    Assume that $d(p,q)<d(u,q)$, we have \[d_T(p,c)+d_T(c,q)=d_T(p,q)=d(p,q)<d(u,q)=d_T(u,q)=d_T(u,c)+d_T(c,q),\] which implies that $d_T(p,c)<d_T(u,c)$. In the MST, if we remove the edge $(p,u)$, we obtain two connected components. Let $A$ be the part with $u$ and $q$. By the property of MSTs, $(p,u)$ is the shortest edge connecting $A$ and $X\setminus A$. In $A$, if we remove the first edge $(x,y)$ such that $x \notin T_q$ and $y \in T_q$, we yield two connected components. Let $A'$ be the part with $u$ and $x$. Again, by the property of MSTs, $(x,y)$ is the shortest edge connecting $A'$ and $X\setminus A'$. So $d(p,y)>d(x,y)$, which implies $d_T(p,c)>d_T(c,x)$. By triangle inequality, \[d_T(p,x) \leq d_T(p,c)+d_T(c,x)<2d_T(p,c).\] On the other hand, by the assumption, \[ d_T(p,u)=d_T(p,c)+d_T(c,u)>2d_T(p,c).\] So $d(p,u)=d_T(p,u)>d_T(p,x)=d(p,x)$, which contradicts with the fact that $(p,u)$ is the shortest edge connecting $A$ and $X\setminus A$. Therefore, $d(p,q)>d(u,q)$. 
\end{proof}

\section{Conclusion and Future Work}

One future direction is to see if one can build a linear size navigable graph for a discrete metric of $\delta$-hyperbolicity \footnote{Let $(X, d)$ be a metric space and let $\delta\geq 0$. For any four points $u, v, w, x \in X$, consider the three pairwise distance sums (called pairsums): $S_1 = d(u, v) + d(w, x)$, $S_2 = d(u, w) + d(v, x)$, $S_3 = d(u, x) + d(v, w)$. The space $X$ is $\delta$-hyperbolic if for every quadruple of points, the two largest sums among $S_1, S_2$, and $S_3$ differ by at most $2\delta$.}. First of all, we will need to assume a constant bounded doubling dimension at constant distance scale, as otherwise the example in \Cref{thm:lb-high-d} prevents a linear size navigable graph. 
If we do assume a constant bounded doubling dimension at constant distance scale, the local routing component can be accomplished by the methods in~\cite{Har-Peled2026-zm}\footnote{The work~\cite{Har-Peled2026-zm} provides a hop bound that depends on the spread of the input points.}. Thus the remaining element is whether one can design a graph supporting global routing routines. 
It is known that one can approximate a $\delta$-hyperbolic metric space of $n$ points by a tree metric with an additive error of $O(\delta \log n)$~\cite{Chepoi08diameters,Chepoi2012-pd}. This is not yet enough, since with a distance of $O(\delta \log n)$ there could be $\Omega(n)$ vertices. Instead, we need a structure similar to the binary tiling in a general $\delta$-hyperbolic setting to provide easy navigability with a constant additive error. This remains as future work. 
\newcommand{\etalchar}[1]{$^{#1}$}

\newpage
\appendix
\section{Missing Proofs in \texorpdfstring{\Cref{sec:pre} and \Cref{sec:positive}}{the Preliminaries and Navigable Graph Sections}} \label{apd:pf}

\hyperbolicpacking*
\begin{proof}
    Consider a ball $B$ of radius $r$ centered at $p$ and take a maximal set of points $P=\{p_i\}$ with pairwise distance at least $r/2$. Thus, every point is within distance $r/2$ from at least one point in $\{p_i\}$. All the balls of radius $r/2$ at $P$ collectively cover the ball $B$. And all balls of radius $r/4$ centered at points of $P$ are disjoint and fit within the ball at $p$ of radius $5r/4$. By volume packing, we have $$|P|\leq \frac{\volume(B(5r/4))}{\volume(B(r/4))}\leq 5\left (\frac{\sinh (5r/4)}{\sinh (r/4)}\right ) ^{d-1}$$
    The last inequality uses the volume formula $\frac{2\pi^{\dim/2}}{\Gamma(\dim/2)}\int_0^r\sinh^{\dim-1}(t)dt$ of the volume of a ball of radius $r$ in the hyperbolic space. 

    When $r$ is small, $\sinh(r)\approx r$ the bound approaches $5^{\dim}$; when $r$ is large, $\sinh(r)\approx e^r/2$ and the bound grows as $e^{(\dim-1)r}$. Since the doubling dimension is bounded by $\log_2 |P|$, it can be bounded by $O(\dim+(\dim-1)r)$ at radius $r$.
\end{proof}

\yaogeneral*
\begin{proof}
    We apply the hyperbolic law of cosines to the triangle $\triangle sut$:
    \begin{align*}
        \cosh d_\mathbb{H}(u,t)& =\cosh d_\mathbb{H}(s,u) \cosh d_\mathbb{H}(s,t)-\sinh d_\mathbb{H}(s,u) \sinh d_\mathbb{H}(s,t) \cos \angle_\mathbb{H} tsu\\
        & < \cosh \xi\cosh d_\mathbb{H}(s,t)-\sinh \xi \sinh d_\mathbb{H}(s,t) \frac{\cosh \xi}{\cosh \xi+1}
    \end{align*}
    Re-arrange and that $\frac{\sinh x}{\cosh x}$ is an increasing function of $x$, $d_\mathbb{H}(s,t)>d_\mathbb{H}(s,u)=\xi$, we have 
    \begin{align*}
        \frac{\cosh d_\mathbb{H}(u,t)}{\cosh d_\mathbb{H}(s,t)} & < \cosh \xi -\sinh \xi \cdot \frac{\sinh d_\mathbb{H}(s,t)}{\cosh d_\mathbb{H}(s,t)} \cdot\frac{\cosh \xi}{\cosh \xi+1}\\
        & <\cosh \xi -\sinh \xi \cdot \frac{\sinh \xi}{\cosh \xi} \cdot\frac{\cosh \xi}{\cosh \xi+1}\\
        &=\cosh \xi-\frac{\cosh^2 \xi-1}{\cosh \xi+1}=1
    \end{align*}
    The last equation uses the fact that $\sinh^2 \xi=\cosh^2 \xi-1$. This implies that $\cosh d_\mathbb{H}(u,t)< \cosh d_\mathbb{H}(s,t)$ and thus $d_\mathbb{H}(u,t)<  d_\mathbb{H}(s,t)$.
\end{proof}


\localeuc*
\begin{proof}
Set \(L=d_{\mathbb H}(s,t)\). Since \(t\in B(s,r)\), we have
\(0<L\leq r\). Moreover, the triangle inequality gives
\[
s,t,u\in B(o,2r).
\]

Suppose, to the contrary, that \(d_{\mathbb H}(t,u)\geq L\).
For \(0<\varepsilon<L\), let
\(\theta(L,\varepsilon,2r)\) be the threshold from
Lemma~\ref{lem:inner}. By its definition,
\[
\lim_{\varepsilon\downarrow0}\theta(L,\varepsilon,2r)
=
\frac{
2\arcsin\left(\dfrac{1}{2\cosh(L/2)}\right)
}{
1+\tanh(r)\tanh(L/2)
}.
\]
Because the denominator is strictly smaller than \(2\) and \(L\leq r\),
\[
\lim_{\varepsilon\downarrow0}\theta(L,\varepsilon,2r)
>
\arcsin\left(\frac{1}{2\cosh(L/2)}\right)
\geq
\arcsin\left(\frac{1}{2\cosh(r/2)}\right).
\]
Hence, for some sufficiently small \(\varepsilon>0\),
\[
\angle tsu<\theta(L,\varepsilon,2r).
\]

Apply Lemma~\ref{lem:inner} with
\[
p=s,\qquad q=t,\qquad b=u,\qquad c=t,\qquad
R=L,\qquad S=\varepsilon.
\]
Indeed,
\[
d_{\mathbb H}(q,b)=d_{\mathbb H}(t,u)\geq L,
\qquad
d_{\mathbb H}(q,c)=0\leq\varepsilon,
\]
and \(b\neq p\). The lemma therefore yields
\[
d_{\mathbb H}(s,u)>d_{\mathbb H}(s,t),
\]
contradicting the hypothesis.
\end{proof}

\lemangle*
\begin{proof}
It suffices to compare, for each point $x\in \mathcal{C}(p,v,r,\theta)$,
the Euclidean direction of the segment $px$ with the tangent direction at
$p$ of the hyperbolic geodesic from $p$ to $x$.

Fix such a point $x\neq p$. The relevant geometry is contained in the
two-dimensional Euclidean plane spanned by $p$ and $x$, whose intersection
with $\mathbb{B}^d$ is a totally geodesic Poincar\'e disk. Thus we may work
in the Poincar\'e disk $\mathbb{B}^2$. After applying a rotation, assume
that
\[
    p=\rho\in [0,1)
\]
is real, and write
\[
    x=p+\ell e^{i\varphi}
\]
for some $\ell>0$ and $\varphi\in \mathbb{R}$.

Consider the disk automorphism
\[
    T_p(z)=\frac{z-p}{1-\overline p z},
\]
which sends $p$ to $0$. Since \(p=\rho\) is real, we have
\[
    T_p'(p)=\frac{1}{1-\rho^2}>0.
\]
Hence, infinitesimally at \(p\), the Euclidean direction \(e^{i\varphi}\)
is mapped to the same direction. On the other hand, \(T_p\) maps the
hyperbolic geodesic from \(p\) to \(x\) to the Euclidean diameter from
\(0\) to \(T_p(x)\). Therefore, if \(\alpha(x)\) denotes the angle at \(p\)
between the Euclidean segment \(px\) and the hyperbolic geodesic from \(p\)
to \(x\), then
\[
    \alpha(x)
    =
    \left|\arg T_p(x)-\varphi\right|.
\]
Now
\[
    T_p(x)
    =
    \frac{\ell e^{i\varphi}}
    {1-\rho^2-\rho\ell e^{i\varphi}}.
\]
Thus, setting
\[
    A=1-\rho^2-\rho\ell e^{i\varphi},
\]
we obtain
\[
    \alpha(x)=|\arg A|.
\]
Since \(x\in \mathbb{B}^2\), we have
\[
    \operatorname{Re} A
    =
    \operatorname{Re}(1-\rho x)
    \geq 1-\rho |x|
    >0.
\]
Hence \(0\leq \alpha(x)<\pi/2\), so the sine function is monotone on the
relevant interval.

Let
\[
    h=d_{\mathbb H}(p,x).
\]
Since \(T_p\) sends \(p\) to \(0\), the standard distance formula in the
Poincar\'e disk gives
\[
    |T_p(x)|=\tanh(h/2).
\]
But also
\[
    |T_p(x)|
    =
    \frac{\ell}{|A|}.
\]
Therefore
\[
    |A|=\frac{\ell}{\tanh(h/2)}.
\]
Consequently,
\[
    \sin \alpha(x)
    =
    \frac{|\operatorname{Im} A|}{|A|}
    =
    \frac{\rho\ell |\sin\varphi|}{|A|}
    =
    \rho\,\tanh(h/2)\,|\sin\varphi|.
\]
Since \(d_{\mathbb H}(p,o)\leq r\), we have
\[
    \rho=|p|
    =
    \tanh\!\left(\frac{d_{\mathbb H}(p,o)}{2}\right)
    \leq
    \tanh(r/2).
\]
Also \(h=d_{\mathbb H}(p,x)\leq r\), because the cone is truncated at
hyperbolic radius \(r\). Hence
\[
    \sin\alpha(x)
    \leq
    \tanh(r/2)\tanh(r/2)
    =
    \tanh^2(r/2).
\]
Since \(0\leq \alpha(x)<\pi/2\), this implies
\[
    \alpha(x)
    \leq
    \arcsin\!\left(\tanh^2(r/2)\right).
\]

Now take \(u,w\in \mathcal{C}(p,v,r,\theta)\), and let \(e_u,e_w\) denote
the Euclidean directions of the segments \(pu,pw\). Let \(h_u,h_w\) denote
the tangent directions at \(p\) of the hyperbolic geodesics from \(p\) to
\(u,w\), respectively. By the Euclidean cone condition,
\[
    \angle(e_u,e_w)\leq \theta.
\]
By the estimate just proved,
\[
    \angle(h_u,e_u)
    \leq
    \arcsin\!\left(\tanh^2(r/2)\right),
    \qquad
    \angle(h_w,e_w)
    \leq
    \arcsin\!\left(\tanh^2(r/2)\right).
\]
Therefore, using the triangle inequality for angles,
\[
    \angle_{\mathbb H} upw
    =
    \angle(h_u,h_w)
    \leq
    \angle(h_u,e_u)+\angle(e_u,e_w)+\angle(e_w,h_w).
\]
Thus
\[
    \angle_{\mathbb H} upw
    \leq
    \theta
    +
    2\arcsin\!\left(\tanh^2(r/2)\right),
\]
as claimed.
\end{proof}

\leminner*
\begin{proof}
Set
\(
\psi=\angle bpc.
\)
Suppose, for contradiction, that
\(
d_{\mathbb{H}}(p,b)\leq d_{\mathbb{H}}(p,c).
\)
Define
\(
t=d_{\mathbb{H}}(p,c),
\)
since
\(
d_\mathbb{H}(p,c)\neq d_\mathbb{H}(q,c)
\)
we have \(c\neq p\), and consequently \(t>0\).

We first replace \(b\) by a point on the same Euclidean ray from
\(p\). Let
\(
v=\frac{b-p}{\lVert b-p\rVert}
\)
and consider the Euclidean ray
\[
\ell=\{p+sv:s\geq 0\}\cap\mathbb{B}^{n}.
\]

In the Poincaré ball model, every hyperbolic ball is also a Euclidean
ball. In particular, \(B_{\mathbb{H}}(p,t)\) is a Euclidean ball
containing \(p\) in its interior. Hence the ray \(\ell\) meets the
hyperbolic sphere
\(
\partial B_{\mathbb{H}}(p,t)
\)
at exactly one point other than \(p\). Denote this point by \(b^{*}\),
where \(
\angle b^{*}pc
=
\angle bpc
=
\psi
\) and
\(
d_{\mathbb{H}}(p,b^{*})=t.
\)

We next show that
\(
d_{\mathbb{H}}(q,b^{*})\geq R.
\)
The intersection of \(B_{\mathbb{H}}(q,R)\) with the Euclidean
ray \(\ell\) is either empty after \(p\), or is an initial segment of
the ray starting at \(p\).
Since \(b\neq p\) and
\(
d_{\mathbb{H}}(q,b)\geq R,
\)
the point \(b\) does not lie in the interior of
\(B_{\mathbb{H}}(q,R)\). Every point on \(\ell\) lying beyond \(b\)
is therefore also outside the interior of \(B_{\mathbb{H}}(q,R)\).
Since \(b^{*}\) lies at least as far along the ray as \(b\), it follows
that
\(
d_{\mathbb{H}}(q,b^{*})\geq R.
\)

Let
\(
\phi=\angle_{\mathbb{H}} b^{*}pc
\)
be the hyperbolic angle at \(p\) between the geodesic segments
\(pb^{*}\) and \(pc\). By the triangle inequality,
\(
\begin{aligned}
d_{\mathbb{H}}(b^{*},c)
\geq
d_{\mathbb{H}}(q,b^{*})-d_{\mathbb{H}}(q,c)\geq R-S.
\end{aligned}
\)
Meanwhile,
\(
d_{\mathbb{H}}(p,b^{*})
=
d_{\mathbb{H}}(p,c)
=
t.
\)
Applying the hyperbolic law of cosines to the triangle
\(pb^{*}c\), we obtain
\[
\cosh d_{\mathbb{H}}(b^{*},c)
=
\cosh^{2}t-\sinh^{2}t\cos\phi.
\]
Since
\(
\cosh^{2}t-\sinh^{2}t=1,
\)
this can be rewritten as
\[
\cosh d_{\mathbb{H}}(b^{*},c)-1
=
\sinh^{2}t\,(1-\cos\phi).
\]
Using the identities
\(
\cosh x-1=2\sinh^{2}\left(\frac{x}{2}\right)
\)
and
\(
1-\cos\phi=2\sin^{2}\left(\frac{\phi}{2}\right),
\)
we get
\[
\sinh\left(
\frac{d_{\mathbb{H}}(b^{*},c)}{2}
\right)
=
\sinh t\,
\sin\left(\frac{\phi}{2}\right).
\]
Therefore,
\[
\sin\left(\frac{\phi}{2}\right)
=
\frac{
\sinh\left(
\frac{d_{\mathbb{H}}(b^{*},c)}{2}
\right)
}{
\sinh t
}.
\]
Since \(d_{\mathbb{H}}(b^{*},c)\geq R-S\), we have
\(
\sin\left(\frac{\phi}{2}\right)
\geq
\frac{
\sinh\left(\frac{R-S}{2}\right)
}{
\sinh t
}.
\)
Furthermore, the triangle inequality
give
\(
t\leq R+S.
\)
Consequently,
\[
\sin\left(\frac{\phi}{2}\right)
\geq
\frac{
\sinh\left(\frac{R-S}{2}\right)
}{
\sinh(R+S)
}.
\]
Since \(0\leq\phi\leq\pi\), the function
\(\sin(\phi/2)\) is increasing in \(\phi\). Hence
\[
\phi
\geq
2\arcsin\left(
\frac{
\sinh\left(\frac{R-S}{2}\right)
}{
\sinh(R+S)
}
\right).
\tag{1}
\]

We now compare the hyperbolic angle \(\phi\) with the Euclidean chord
angle \(\psi\).
Let
\(
\tau=\tanh\left(\frac{t}{2}\right),
\)
and for \(p,y\in\mathbb{B}^{n}\),
define the M\"obius transformation by
\[
L_{p}(y)= \frac{
\left(1+2\langle p,y\rangle+\lVert y\rVert^{2}\right)p
+
\left(1-\lVert p\rVert^{2}\right)y
}{
1+2\langle p,y\rangle
+\lVert p\rVert^{2}\lVert y\rVert^{2}
},
\]
which is a hyperbolic isometry of the Poincaré ball satisfying
\(L_{p}(0)=p\). Since
\(
d_{\mathbb{H}}(p,b^{*})
=
d_{\mathbb{H}}(p,c)
=
t,
\)
and at the origin of the Poincaré ball we have the explicit distance formula \( d_\mathbb{H}(o,y) = 2\arctanh\|y\|
\), therefore there exist unit vectors
\(
n_{b},n_{c}\in\mathbb{S}^{n-1}
\)
such that
\(
b^{*}=L_p(\tau n_{b}),
c=L_p(\tau n_{c}).
\)
Because \(L_{p}\) is conformal and its differential at the origin is
a positive scalar multiple of the identity, we have
\[
\phi=\angle_{\mathrm{E}}(n_{b},n_{c}).
\tag{2}
\]

A direct calculation from the definition of M\"obius addition gives
\[
L_p(\tau n)-p
=
\frac{
(1-\lVert p\rVert^{2})\tau
}{
1+2\tau\langle p,n\rangle
+\tau^{2}\lVert p\rVert^{2}
}
\left(n+\tau p\right)
\]
for every \(n\in\mathbb{S}^{n-1}\). The scalar coefficient in this
formula is positive. Thus the Euclidean unit vector pointing from
\(p\) toward \(L_p(\tau n)\) is
\[
F_{a}(n)=\frac{n+a}{\lVert n+a\rVert},
\qquad
a=\tau p.
\]
It follows that
\[
\psi
=
\angle_{\mathrm{E}}\bigl(F_{a}(n_{b}),F_{a}(n_{c})\bigr).
\tag{3}
\]

We claim that
\[
\angle_{\mathrm{E}}(n_{b},n_{c})
\leq
(1+\lVert a\rVert)
\angle_{\mathrm{E}}
\bigl(F_{a}(n_{b}),F_{a}(n_{c})\bigr).
\tag{4}
\]

To prove this claim, observe first that \(\lVert a\rVert<1\), and the
map
\[
F_{a}(n)=\frac{n+a}{\lVert n+a\rVert}
\]
is a diffeomorphism of \(\mathbb{S}^{n-1}\). Its inverse is explicitly
given by
\[
F_{a}^{-1}(u)
=
\left(
\langle a,u\rangle
+
\sqrt{
1-\lVert a\rVert^{2}
+\langle a,u\rangle^{2}
}
\right)u-a.
\]

Let \(v\in T_{n}\mathbb{S}^{n-1}\), so that
\(\langle v,n\rangle=0\), and write
\(
u=F_{a}(n).
\)
Differentiating \(F_{a}\) gives
\[
D F_{a}(n)[v]
=
\frac{
v-\langle u,v\rangle u
}{
\lVert n+a\rVert
}.
\]
Therefore,
\[
\left\lVert D F_{a}(n)[v]\right\rVert^{2}
=
\frac{
\lVert v\rVert^{2}\lVert n+a\rVert^{2}
-\langle a,v\rangle^{2}
}{
\lVert n+a\rVert^{4}
}.
\]
Because \(v\perp n\),
\[
\langle a,v\rangle^{2}
\leq
\left(
\lVert a\rVert^{2}-\langle a,n\rangle^{2}
\right)\lVert v\rVert^{2}.
\]
It follows that
\[
\left\lVert D F_{a}(n)[v]\right\rVert^{2}
\geq
\frac{
\left(1+\langle a,n\rangle\right)^{2}
}{
\lVert n+a\rVert^{4}
}
\lVert v\rVert^{2}.
\]
Thus the operator norm of $D(F_{a}^{-1})_{F_{a}(n)}$ satisfies
\[
\left\lVert
D(F_{a}^{-1})_{F_{a}(n)}
\right\rVert_{\mathrm{op}}
\leq
\frac{
\lVert n+a\rVert^{2}
}{
1+\langle a,n\rangle
}.
\]
If
\(
A=\lVert a\rVert,
z=\langle a,n\rangle,
\)
then
\(
\frac{
\lVert n+a\rVert^{2}
}{
1+\langle a,n\rangle
}
=
\frac{1+2z+A^{2}}{1+z}.
\)
Moreover,
\[
(1+A)(1+z)-(1+2z+A^{2})
=
(1-A)(A-z)\geq 0,
\]
because \(z\leq A<1\). Hence
\[
\left\lVert
D(F_{a}^{-1})_{F_{a}(n)}
\right\rVert_{\mathrm{op}}
\leq 1+\lVert a\rVert.
\tag{5}
\]

Let \(\gamma\) be a shortest spherical geodesic joining
\(F_{a}(n_{b})\) to \(F_{a}(n_{c})\). Its length is \(\psi\). By
applying \(F_{a}^{-1}\) to \(\gamma\) and using (5), we obtain a curve
from \(n_{b}\) to \(n_{c}\) whose length is at most
\[
(1+\lVert a\rVert)\psi. 
\]
The spherical distance between \(n_{b}\) and \(n_{c}\) is no greater
than the length of this curve. This proves (4).

Combining (2), (3), and (4), we obtain
\(
\phi\leq(1+\lVert a\rVert)\psi.
\)
Since \(a=\tau p\), we have
\(
\phi\leq(1+\tau\lVert p\rVert)\psi.
\)
The assumption \(p\in B_{\mathbb{H}}(0,r)\) implies
\(
\lVert p\rVert
\leq
\tanh\left(\frac{r}{2}\right)
=
\rho.
\)
Also, since \(t\leq R+S\),
\[
\tau=\tanh\left(\frac{t}{2}\right)
\leq
\tanh\left(\frac{R+S}{2}\right).
\]
It follows that
\[
\phi
\leq
\left(
1+\rho\tanh\left(\frac{R+S}{2}\right)
\right)\psi.
\tag{6}
\]

Combining (1) and (6), we obtain
\[
2\arcsin\left(
\frac{
\sinh\left(\frac{R-S}{2}\right)
}{
\sinh(R+S)
}
\right)
\leq
\left(
1+\rho\tanh\left(\frac{R+S}{2}\right)
\right)\psi.
\]
Therefore,
\[
\psi
\geq
\frac{
2\arcsin\left(
\frac{
\sinh\left(\frac{R-S}{2}\right)
}{
\sinh(R+S)
}
\right)
}{
1+\rho\tanh\left(\frac{R+S}{2}\right)
}
=
\theta(R,S,r).
\]
This contradicts the assumption
\[
\psi=\angle_{\mathrm{E}}bpc<\theta(R,S,r).
\]
Thus
\[
d_{\mathbb{H}}(p,b)>d_{\mathbb{H}}(p,c),
\]
as required.
\end{proof}

\lemcap*
\begin{proof} 
First, we define the hyperbolic cone $\mathcal{C}_\mathbb{H}$ at $p$ with Euclidean direction vector $v=\Vec{pq}$, cone angle $2A$ and diameter $r$ by 
\[
\mathcal{C}_\mathbb{H}(p,v,r,2A)=\{x\in\mathbb{H}^d:d_\mathbb{H}(x,p)\leq r, \angle_\mathbb{H}xpv\leq A\}.
\]
Assume that each boundary ray of \(\mathcal C\) intersects \(B_{\mathbb H}(q,S)\),
equivalently \[ A\leq \arcsin\frac{\sinh S}{\sinh R}. \]
The hyperbolic cap $\CAP(p,\mathcal{C}_\mathbb{H})$ is the set of points on hyperbolic rays from $p$ inside  $\mathcal{C}_\mathbb{H}(p,v,r,2A)$ until the first intersection with $B_\mathbb{H}(q,S)$ for each ray. Our first goal is to show for every $\hat{w}\in\CAP(p,\mathcal{C}_\mathbb{H})$ and $\hat{u}\in\mathcal{C}_\mathbb{H}\setminus \Cl(p,q)$, we have $d_\mathbb{H}(p,\hat{w})<d_\mathbb{H}(p,\hat{u})$.




Let \(\xi_0\) be the unit tangent vector pointing from \(p\) to \(q\). Every geodesic ray in \(\mathcal C_\mathbb{H}\) has the form
\[ \gamma_\xi(t)=\exp_p(t\xi), \qquad t\geq 0, \]
where $\exp_p(t\xi)=\cosh(t)p+\sinh(t)\xi$, \(\xi\) is unit and 
\[ \alpha:=\angle(\xi,\xi_0)\leq A. \]
For such a ray, set 
\[ F_\alpha(t)=d_{\mathbb H}(q,\gamma_\xi(t)). \]
By the hyperbolic cosine law, 
\[ \cosh F_\alpha(t) = \cosh R\cosh t-\sinh R\sinh t\cos\alpha. \]
The assumption $ A\leq \arcsin\frac{\sinh S}{\sinh R} $ is precisely the condition that every ray with \(\alpha\leq A\) intersects \(B_{\mathbb H}(q,S)\). For \(\alpha\leq A\), let \(t_s^-(\alpha)\) be the first time at which \(\gamma_\xi\) meets \(\partial B_{\mathbb H}(q,S)\). Similarly, let \(t_R^+(\alpha)\) be the second time at which \(\gamma_\xi\) meets \(\partial B_{\mathbb H}(q,R)\). The first time for radius \(R\) is \(t=0\), because \(d_{\mathbb H}(p,q)=R\). For fixed \(t>0\), the quantity \(F_\alpha(t)\) is increasing in \(\alpha\), since \(\cos\alpha\) decreases as \(\alpha\) increases. Hence \[ t_s^-(\alpha)\leq t_s^-(A) \] for all \(\alpha\leq A\): the most tilted ray enters \(B_{\mathbb H}(q,s)\) latest. Similarly, 
\[ t_R^+(\alpha)\geq t_R^+(A) \] 
for all \(\alpha\leq A\): the most tilted ray exits \(B_{\mathbb H}(q,R)\) earliest. Since \(S<R\), along the boundary ray \(\alpha=A\) the first intersection with \(\partial B_{\mathbb H}(q,S)\) occurs strictly before the far-side intersection with \(\partial B_{\mathbb H}(q,R)\). Thus \[ t_s^-(A)<t_R^+(A). \] Therefore, for all \(\alpha,\beta\leq A\), \[ t_s^-(\alpha) \leq t_s^-(A) < t_R^+(A) \leq t_R^+(\beta). \] 
Now let \(\hat{w}\in \operatorname{CAP}(p,\mathcal C_\mathbb{H})\). Then \(\hat{w}=\gamma_\alpha(t_{\hat{w}})\) for some \(\alpha\leq A\), and since \(\hat{w}\) lies in the hyperbolic cap, \[ t_{\hat{w}}<t_s^-(\alpha). \] Let \({\hat{u}}\in \mathcal C_\mathbb{H}\setminus B_{\mathbb H}(q,R)\), then \({\hat{u}}=\gamma_\beta(t_{\hat{u}})\) for some \(\beta\leq A\), and \[ t_{\hat{u}}>t_R^+(\beta). \] Combining the inequalities gives \[ t_{\hat{w}}<t_s^-(\alpha)\leq t_s^-(A)<t_R^+(A)\leq t_R^+(\beta)<t_{\hat{u}}. \] Since the ray parameter \(t\) is hyperbolic distance from \(p\), we have \[ t_{\hat{w}}=d_{\mathbb H}(p,{\hat{w}}), \qquad t_{\hat{u}}=d_{\mathbb H}(p,{\hat{u}}). \] Hence \[ d_{\mathbb H}(p,{\hat{w}})<d_{\mathbb H}(p,{\hat{u}}). \]

Now we shrink the hyperbolic cone angle $A$ so that the conclusion still holds under the Euclidean cone setting. 
Let $\theta_1=\theta_2(\leq A\leq\arcsin\frac{\sinh S}{\sinh R})$ be the hyperbolic angles between boundary of $\mathcal{C}_\mathbb{H}$ and Euclidean direction $\Vec{pq}$, we shrink $\theta_1$ to $\hat{\theta_1}=\theta_1-\arcsin\left(\tanh^2(r/2)\right)$ and same thing to $\theta_2$. Since $R-S$ can be arbitrary small and $r$ is bounded, we may choose $\epsilon$ small enough such that $ \arcsin\left(\tanh^2(r/2)\right)<\frac{\sinh S}{\sinh R}<1$. Then angle $\theta_0 = 2\left(\arcsin\frac{\sinh S}{\sinh R}-\arcsin(\tanh^2(r/2))\right)>0$ for some $\epsilon>0$.

then by lemma \ref{lem:hyperbolic_angle}, for the Euclidean cone $\mathcal{C}(p,\Vec{pq},r,\theta_0)$ in a bounded domain, we have
\[
\CAP(p,\mathcal{C})\subset\CAP_\mathbb{H}(p,\mathcal{C}), \quad \mathcal{C}\setminus\Cl(p,q)\subset \mathcal{C}_{\mathbb{H}}\setminus\Cl(p,q).
\] 
Hence \( d_{\mathbb H}(p,{w})<d_{\mathbb H}(p,{u}) \) still holds
for any $w\in \CAP(p,\mathcal{C})$ and $u\in \mathcal{C}\setminus\Cl(p,q)$.

Since the direction vector $v$ is not equal to $\Vec{pq}$ in general but $q$ should lie in the cone of $\mathcal{C}(p,v,r,\theta)$, we have 
\[
\theta\leq \frac{1}{2}\theta_0,
\]
so $\theta$ should satisfy 
\[
\theta\leq \left(\arcsin\frac{\sinh S}{\sinh R}-\arcsin(\tanh^2(r/2))\right).
\]

\end{proof}

\lemratio*
\begin{proof}
\label[proof]{pf:ratio}
    If $s=p$, the conclusion follows from the core assumption. We therefore assume $s\neq p$ below.

    Since $d_{\mathbb{H}}(o,p)\leq r$ and $q,s \in \mathcal{C}(p,v,r,\theta)$, $d_{\mathbb{H}}(o,s)\leq 2r$ and $d_{\mathbb{H}}(o,q)\leq 2r$.  Let $B_{\mathbb{H}}(o,r')$ denote the hyperbolic ball at $o$ with hyperbolic radius $r'$. Then, $p,q,s \in B_{\mathbb{H}}(o,2r)$. In Poincaré model, $d_{\mathbb{H}}(o,x)=2\arctanh \|x\|$. Hence $x \in B_{\mathbb{H}}(o,2r)$ implies $\|x\|\leq \tanh r$. Set $\rho=\tanh r$. On the Euclidean ball $\|x\|\leq \rho$, the hyperbolic metric satisfies: \[ds_{\mathbb H} = \frac{2\|dx\|}{1-\|x\|^2}.\] Therefore, for $x,y \in B_{\mathbb{H}}(o,2r)$, \[2\|x-y\|\le d_{\mathbb H}(x,y)\le\frac{2}{1-\rho^2}\|x-y\|.\] Plug in $\rho=\tanh r$, \[2\|x-y\| \le d_{\mathbb H}(x,y)\le2\cosh^2r\,\|x-y\|.\] For $s$ and $p$, we have \[2\|s-p\|\leq d_{\mathbb{H}}(s,p).\] For $p$ and $q$, we have \[d_{\mathbb{H}}(p,q)\leq 2\cosh^2 r\|p-q\|.\] Since $\beta = d_{\mathbb{H}}(p,s)/d_{\mathbb{H}}(p,q)$, \[\|s-p\|\leq \beta \cosh^2r\|p-q\|. \] By assumption $\beta \leq \frac{1}{3\cosh^2 r}$, \[\|s-p\|\leq \frac 13\|p-q\|.\] Let $\eta=\frac{\|s-p\|}{\|p-q\|}$, then $\eta\in(0,1/3]$. Decompose vectors into components parallel and perpendicular to $v$. we can write \[\quad q-p=(q-p)_\parallel v+(q-p)_\perp. \] Since $\angle{qpv}\leq \theta/6$, \[\|(q-p)_\perp\|\le \|q-p\|\sin (\theta/6), \quad \|(q-p)_\parallel\|\ge \|q-p\|\cos (\theta/6).\] Since $\angle{spv}\leq \theta/2$,  \[\|(s-p)_\perp\|\le \|s-p\|\sin (\theta/2), \quad \|(s-p)_\parallel\|\leq  \|s-p\|.\] Now consider $q-s=(q-p)-(s-p)$. Its perpendicular component satisfies
    \[
    \begin{aligned}
    \|(q-s)_\perp\|
    &\le \|(q-p)_\perp\|+\|(s-p)_\perp\|\\
    &\le \|q-p\|\sin(\theta/6)+\|s-p\|\sin(\theta/2)\\
    &\leq \|p-q\|\bigl(\sin(\theta/6)+\eta\sin(\theta/2)\bigr).
    \end{aligned}
    \]
    For the parallel component, we have $(q-s)_\parallel=(q-p)_\parallel-(s-p)_\parallel.$ Hence \[\|(q-s)_\parallel\|\ge \|p-q\|\cos (\theta/6)-\|s-p\|=\|p-q\|(\cos(\theta/6)-\eta).\] Let $\alpha=\angle{qsv}$. Then \[\tan\alpha=\frac{\|(q-s)_\perp\|}{\|(q-s)_\parallel\|}\le\frac{\sin(\theta/6)+\eta\sin(\theta/2)}{\cos(\theta/6)-\eta}.\] We now show that $\alpha\leq \theta/2$, since $\theta/2<\pi/2$ and $\eta\leq1/3$, it suffices to prove \[\frac{\sin(\theta/6)+\frac13\sin(\theta/2)}{\cos(\theta/6)-\frac13}\le\tan(\theta/2).\] Equivalently, \[\left(\sin(\theta/6)+\frac13\sin (\theta/2)\right)\cos (\theta/2)\le\left(\cos(\theta/6)-\frac13\right)\sin (\theta/2).\] Simplify, \[\frac13\sin(\theta/2)(1+\cos(\theta/2))\le \sin (\theta/3).\] Since $1+\cos(\theta/2)\leq 2$, it is enough to show that $2\sin(\theta/2)\leq3\sin(\theta/3)$, which is equivalent to $2\sin(\theta/6)(4\cos^2(\theta/6)-1)\le 6\sin (\theta/6)\cos (\theta/6).$ Since $\sin (\theta/6)>0$ and $0<\cos(\theta/6)<1$, this is equivalent to $4\cos^2(\theta/6)-3\cos(\theta/6)-1=(4\cos(\theta/6)+1)(\cos(\theta/6)-1)\le 0.$ It is always true, so we have $\alpha=\angle{qsv}\leq \theta/2$.  

    Since $\theta\leq \arcsin(\frac{1}{2\cosh(r/2)})$, by \Cref{lem:local-euclidean-angle}, $d_{\mathbb{H}}(s,q)<d_{\mathbb{H}}(p,q)\leq r$. So $q\in\mathcal C(s,v,r,\theta)$. 

\end{proof}

\conecapdepth*
\begin{proof}
    Since
\(
 \frac{\sigma}{R-L}=
 \frac{a}{2\Lambda^2(R+L)}<1,
\)
we have $L<S<R$. Choose $t\in\widehat C\cap\overline B_{\mathbb H}(q,L)$ and put
$K_b:=\overline B_{\mathbb H}(q,b)$. Since $K_R\subset B_{\mathbb H}(o,3r)$,
the Poincar\'e metric and the Euclidean convexity of $K_R$ give
\begin{equation}
 2\lVert x-y\rVert\leq d_{\mathbb H}(x,y)
 \leq2\Lambda\lVert x-y\rVert,
 \qquad x,y\in K_R.
 \label{eq:cap-metric-comparison}
\end{equation}

We have $d_{\mathbb H}(p,t)\geq R-L>a$.  Let
$x_t=(1-\lambda)p+\lambda t$ be the unique point of $pt$ satisfying
$d_{\mathbb H}(p,x_t)=a$.  Equation
\eqref{eq:cap-metric-comparison} and
$d_{\mathbb H}(p,t)\leq R+L$ imply
\begin{equation}
    \lambda\geq\frac{a}{\Lambda(R+L)}.
 \label{eq:cap-lambda}
\end{equation}
Write $K_b=\overline B_{\mathrm E}(c_b,\rho_b)$, where $B_{\mathrm E}(c_b,\rho_b)$ denotes the Euclidean ball centered at $c_b$ with radius $\rho_b$ and set
$\operatorname{dep}_b(x):=\rho_b-\lVert x-c_b\rVert$.  If
$y\in\partial K_R$ is closest to $t$ in Euclidean distance, then
\[
 d_{\mathbb H}(t,y)\geq R-L,
 \qquad
 \operatorname{dep}_R(t)\geq\frac{R-L}{2\Lambda}.
\]
Since Euclidean depth in a ball is concave and $p\in\partial K_R$,
\eqref{eq:cap-lambda} yields
\begin{equation}
    \operatorname{dep}_R(x_t)\geq\sigma.
 \label{eq:cap-depth-R}
\end{equation}

Every point of $K_R$ is at hyperbolic distance at most
$R-S=\sigma$ from $K_S$: move it toward $q$ until it reaches $K_S$.
Hence its Euclidean distance from $K_S$ is at most $\sigma/2$.  Since
$K_S\subset K_R$ are Euclidean balls, their Euclidean distance satisfies
\[
 \lVert c_R-c_S\rVert+\rho_R-\rho_S\leq\frac{\sigma}{2}.
\]
Together with \eqref{eq:cap-depth-R}, this gives
\begin{equation}
    \operatorname{dep}_S(x_t)\geq\frac{\sigma}{2}.
 \label{eq:cap-depth-S}
\end{equation}

For $\xi\in\mathbb S^{d-1}$, let $X_a(\xi)$ be the intersection of the
ray $p+\mathbb R_{>0}\xi$ with $\partial B_{\mathbb H}(p,a)$.  This map
is $a$-Lipschitz in spherical angular distance.  Indeed, if
$B_{\mathbb H}(p,a)=B_{\mathrm E}(c,\rho)$, then
\[
 \rho=\frac{(1-\lVert p\rVert^2)\tanh(a/2)}
 {1-\lVert p\rVert^2\tanh^2(a/2)}
 \leq\tanh(a/2)\leq\frac a2;
\]
and differentiating the radial exit map of this Euclidean ball gives
\[
 \lVert DX_a\rVert_{\mathrm{op}}
 \leq\rho+\lVert p-c\rVert<2\rho\leq a.
\]

The direction of $pt$ is within $\theta/6$ of $v$, whereas every
direction $\xi$ of $C$ is within $\theta/2$ of $v$.  Therefore
\[
 \lVert X_a(\xi)-x_t\rVert
 \leq\frac{2a\theta}{3}\leq\frac{\sigma}{8}.
\]
By \eqref{eq:cap-depth-S}, every $X_a(\xi)$ lies in the interior of
$K_S$.  The witness condition also gives
$\widehat r\geq d_{\mathbb H}(p,t)>a$, so these points occur before the
bounded-cone cutoff.  Since $p\notin K_S$, every cone ray enters $K_S$
strictly before its radius-$a$ endpoint.  This proves
\eqref{eq:cone-cap-depth}.

Both $K_R$ and $B_{\mathbb H}(p,a)$ are Euclidean balls.  Thus the
initial segment of every cone ray up to its radius-$a$ endpoint lies in
$K_R$, while every later point of that ray lies outside
$B_{\mathbb H}(p,a)$.
\end{proof}

\annscap*
\begin{proof}
    Choose any $z\in \hat{\mathcal{C}}\cap B_{\mathbb H}(q,L)$
    and set $ t_0=d_{\mathbb H}(p,z)$.
    Since \(z\in \mathcal{C}\), we have \(t_0\leq r\). Furthermore,
    \[
    t_0
    \leq d_{\mathbb H}(p,q)+d_{\mathbb H}(q,z)
    <R+L=\frac{2+\eps}{1+\eps}R.
    \]
    Consequently,
    $ t_0\leq
    \min\left(
    r,\frac{2+\eps}{1+\eps}R
    \right)$.
    
     We first estimate equal-radius cross-sections of the cone. Let \(x\) and \(y\) lie on two Euclidean rays from \(p\), let \(\beta\) be the
    Euclidean angle between those rays, and suppose that \[d_{\mathbb H}(p,x)=d_{\mathbb H}(p,y)=t\leq t_0.\] Let $o$ denote the origin in the ball model, we have $\angle_{\mathbb{H}}uov=\angle_{\mathbb{E}}uov$ for any points $u$ and $v$. 
    Using this fact and the same arguments as (2)-(6) in the proof of \Cref{lem:inner}, let $\alpha=\angle_{\mathbb{H}}xpy$ and $\beta=\angle_{\mathbb{E}}xpy$, we have \[\alpha\leq (1+\tanh(t/2)\|p\|)\beta\leq (1+\tanh^2(r/2))\beta.\] The hyperbolic cosine rule therefore yields
    \[
    \begin{aligned}
    d_{\mathbb H}(x,y)
    &=
    2\operatorname{arsinh}
    \left(\sinh t\sin(\alpha/2)\right)\\
    &\leq
    2\operatorname{arsinh}
    \left(
    \sinh t_0
    \sin\left(\frac{(1+\tanh^2(r/2))\beta}{2}\right)
    \right).
    \end{aligned}
    \] Since $\beta\leq \theta$, \begin{equation}
        d_{\mathbb{H}}(x,y)<R-S. \label{eq:cross-section-unified}
    \end{equation}

    We next prove that every ray of \(\mathcal C\) intersects
\(B_{\mathbb H}(q,S)\). Let \(\ell\) be any ray of \(\mathcal C\), and
let \(z_\ell\in\ell\) satisfy
$d_{\mathbb H}(p,z_\ell)=t_0$.
The ray \(pz\) belongs to the core and therefore makes an angle at most
\(\theta/6\) with \(v\), while \(\ell\) makes an angle at most
\(\theta/2\) with \(v\). Thus $ 
\angle_{\mathbb{E}} zpz_\ell\leq2\theta/3.$ Since $ \angle_{\mathbb{E}} zpz_\ell<\theta,$ by \eqref{eq:cross-section-unified}, \[
d_{\mathbb H}(z,z_\ell)<R-S<S-L.\] 
Since \(d_{\mathbb H}(q,z)<L\), it follows that
\[
d_{\mathbb H}(q,z_\ell)
\leq d_{\mathbb H}(q,z)+d_{\mathbb H}(z,z_\ell)
<S.
\]
Therefore every ray of \(\mathcal C\) enters
\(B_{\mathbb H}(q,S)\), and its first entry occurs at \(p\)-distance
at most \(t_0\).

Let
$
w\in\CAP(p,\mathcal C)
$ and $
u\in\mathcal C\setminus B_{\mathbb H}(q,R).
$ Let \(x\) be the first point at which the ray containing \(w\) enters
\(B_{\mathbb H}(q,S)\). Then
$
d_{\mathbb H}(q,x)=S
$
and
$
d_{\mathbb H}(p,w)
\leq d_{\mathbb H}(p,x)
\leq t_0.
$

Suppose, for contradiction, that
$
d_{\mathbb H}(p,w)\geq d_{\mathbb H}(p,u).
$
Set
$
t_x =d_{\mathbb H}(p,x),
$
and let \(y\) be the point on the ray containing \(u\) such that
$
d_{\mathbb H}(p,y)=t_x.
$
Then
\[
d_{\mathbb H}(p,u)\leq t_x=d_{\mathbb H}(p,y).
\]

The hyperbolic ball \(B_{\mathbb H}(q,R)\) is a Euclidean ball in the
Poincaré model, and
$
p\in\partial B_{\mathbb H}(q,R).
$
The ray containing \(u\) intersects
$
B_{\mathbb H}(q,S)\subset B_{\mathbb H}(q,R),
$
so its intersection with \(B_{\mathbb H}(q,R)\) is an interval
starting immediately after \(p\). Since
$
u\notin B_{\mathbb H}(q,R),
$
the point \(u\) lies at or beyond the exit point of this interval.
Hyperbolic distance from \(p\) is strictly increasing along a Euclidean
ray from \(p\), so \(y\) lies at or beyond \(u\). Hence
\[
d_{\mathbb H}(q,y)\geq R.
\]

The points \(x\) and \(y\) have the same \(p\)-distance
\(t_x\leq t_0\), and the angle between their rays is at most \(\theta\).
Thus \eqref{eq:cross-section-unified} gives
\[
d_{\mathbb H}(x,y)<R-S.
\]
On the other hand, the reverse triangle inequality gives
\[
d_{\mathbb H}(x,y)
\geq
d_{\mathbb H}(q,y)-d_{\mathbb H}(q,x)
\geq
R-S,
\]
which is a contradiction. Therefore
$d_{\mathbb H}(p,w)<d_{\mathbb H}(p,u).$
\end{proof}

\section{Euclidean Angle Diameter for Greedy Navigation} \label{apd:angle_GN}
    First, we pick $S=(1-1/(6\cosh^3 r))R$ and we have $r>R$. Now, we need to show that there exists a positive angle $\theta$ which only depends on $r$ and satisfies three geometric lemmas.  
\begin{itemize}
    \item In Lemma~\ref{lem:inner}, 
    we can pick any angle less than \[\frac{2}{1+\tanh\left(\frac r2\right)\tanh\left(\frac{R+S}{2}\right)}\arcsin\left(\frac{\sinh\left((R-S)/2\right)}{\sinh(R+S)}\right).\] Since $1+\tanh(\frac{r}{2})\tanh(\frac{R+S}{2})<2$, we can pick any $$\theta\leq\arcsin\left(\frac{\sinh\left((R-S)/2\right)}{\sinh(R+S)}\right).$$
    The upper bound of $\theta$ 
is decreasing in $R>0$. Since $\arcsin(x)$ is increasing in $(0,1]$, after substituting $S$ into the function, we can pick \[\theta_1=
\arcsin\left(
\frac{\sinh\left(\frac{r}{12\cosh^3r}\right)}
     {\sinh\left(\left(2-\frac{1}{6\cosh^3r}\right)r\right)}
\right)>0.\]
    \item In Lemma~\ref{lem:cap-q}, we can pick any angle less than \[ \arcsin(\frac{\sinh S}{\sinh R})-\arcsin(\tanh^2(r/2)).\] Since $\cosh t<\cosh R$ for $S\le t\le R$, we have \[\sinh R-\sinh S=\int_S^R\cosh t\,dt<(R-S)\cosh R=\frac{1}{6\cosh^3r}R\cosh R.\] Re-arrange, we have $$\frac{\sinh S}{\sinh R}>1-\frac{1}{6\cosh^3r}R\coth R.$$ Since $\sinh R>R$ and $R<r$, we have \[R\coth R=\frac{R\cosh R}{\sinh R} <\cosh R<\cosh r.\] Plug in the previous inequality, we have $$\frac{\sinh S}{\sinh R}>1-\frac{1}{6\cosh^2 r}. $$ Since the $\arcsin$ is strictly increasing on $[0,1]$, we can pick \[\theta_2=\left(\arcsin(1-\frac{1}{6\cosh^2 r})-\arcsin(\tanh^2(r/2))\right)>0.\]
    \item In Lemma~\ref{lem:ratio}, we can pick \[\theta_3=\min\left(\arcsin\left(\frac{1}{2\cosh(r/2)}\right),\frac{1}{2\cosh r}\right).\] With such choice of $\theta$, every ray in the cone points sufficiently toward \(q\) that, after
traveling hyperbolic distance \(T=R/(3\cosh^2 r)\), its distance to \(q\) has decreased
by at least \(T/(2\cosh r)\). Hence, every cone ray enters
\(B_{\mathbb H}(q,S)\) before reaching distance \(T\) from \(p\).
Consequently, every point \(x\) in the cap satisfies
\(
d_{\mathbb H}(p,x)\leq T
\)
Since \(R=d_{\mathbb H}(p,q)\), it follows that
\[
\frac{d_{\mathbb H}(p,x)}{d_{\mathbb H}(p,q)}
\leq \frac{1}{3\cosh^2 r}.
\]
\end{itemize}
In all, we should pick $\theta=\min\{\theta_1,\theta_2,\theta_3\}$.

\section{Euclidean Angle Diameter for ANNS} \label{apd:angle_ANNS}
Fix $r>0$ and $0<\eps<1$, and let $0<R<r$ and $L=R/(1+\eps)$.
Set
\[
 B=\frac{\eps^2}
 {6(1+\eps)(2+\eps)\cosh^2r\cosh^4(3r/2)},
 \qquad S=(1-B)R.
\]
Since $0<B<\eps/(1+\eps)$, we have $L<S<R$.
We choose a positive angular diameter depending only on $\eps$ and $r$
that satisfies the three geometric conditions.
\begin{itemize}
 \item For Lemma~\ref{lem:inner}, the admissible angular bound is
 \[
 \frac{2}{1+\tanh(r/2)\tanh((R+S)/2)}
 \arcsin\left(\frac{\sinh((R-S)/2)}{\sinh(R+S)}\right).
 \]
 The prefactor is greater than one, so it is enough to bound
 \[
 A(R)=\arcsin\left(\frac{\sinh(BR/2)}{\sinh((2-B)R)}\right).
 \]
 For $0<a<b$, the function $\sinh(aR)/\sinh(bR)$ is strictly
 decreasing for $R>0$: its logarithmic derivative is
 $a\coth(aR)-b\coth(bR)<0$, because $x\coth x$ is strictly
 increasing on $(0,\infty)$.
 Taking $a=B/2$ and $b=2-B$, we obtain $A(R)\geq A(r)$ and may set
 \[
 \theta_1=\arcsin\left(\frac{\sinh(Br/2)}{\sinh((2-B)r)}\right)>0.
 \]

 \item For Lemma~\ref{lem:ANNS-cap}, write
 \[
 c=\frac{2+\eps}{1+\eps},\qquad
 K=\frac{2}{1+\tanh^2r},\qquad R_* = \frac{r}{c}.
 \]
 The required bound is $\theta<g(R)$, where
 \[
 g(R)=K\arcsin\left(\frac{\sinh(BR/2)}{\sinh(\min(r,cR))}\right).
 \]
 If $0<R\leq R_*$, then $\min(r,cR)=cR$.
 Since $B/2<c$, the same monotonicity calculation gives
 \[
 g(R)\geq K\arcsin\left(\frac{\sinh(BR_*/2)}{\sinh(cR_*)}\right).
 \]
 If $R_*\leq R<r$, then $\min(r,cR)=r$ and the numerator
 $\sinh(BR/2)$ is increasing in $R$. Hence
 \[
 g(R)\geq K\arcsin\left(\frac{\sinh(BR_*/2)}{\sinh r}\right).
 \]
 As $cR_*=r$, both cases yield the common lower bound
 \[
 \theta_2=\frac{2}{1+\tanh^2r}
 \arcsin\left(
 \frac{\sinh\left(\frac{B(1+\eps)r}{2(2+\eps)}\right)}{\sinh r}
 \right)>0.
 \]

 \item Lemma~\ref{lem:cone-cap-depth}, with $L=R/(1+\eps)$,
 gives the above value of $S$ and permits
 \[
 \theta_{\mathrm{cap}}=
 \frac{3\eps}{32(2+\eps)\cosh^4(3r/2)}.
 \]
 Every cap point $w$ then satisfies
 \[
 d_{\mathbb H}(p,w)<\frac{R-L}{3\cosh^2r}.
 \]
 For a cone whose core contains a nearest neighbor
 $x_q\in B_{\mathbb H}(q,L)$, the reverse triangle inequality gives
 $d_{\mathbb H}(p,x_q)\geq R-L$.
 Thus
 \[
 \frac{d_{\mathbb H}(p,w)}{d_{\mathbb H}(p,x_q)}
 <\frac{1}{3\cosh^2r},
 \]
 as required in Lemma~\ref{lem:ratio} with target $x_q$.
 To satisfy its angular condition as well, set
 \[
 \theta_3=\min\left\{
 \arcsin\left(\frac{1}{2\cosh(r/2)}\right),
 \frac{3\eps}{32(2+\eps)\cosh^4(3r/2)}
 \right\}.
 \]
\end{itemize}
Any $0<\theta<\min\{\theta_1,\theta_2,\theta_3\}$ satisfies all three
conditions. This choice depends only on $\eps$ and $r$, not on $R$.
\section{Lower Bound Proofs}

\lowerboundalledges*

\begin{proof}
    Let $\epsilon = \frac{\alpha-1}{2\alpha}$, we construct the point set $P(n,\epsilon)$ as \Cref{lem:lw-kw}. In order to break tie, we modify the construction. Let $\beta$ be minimum positive gap between distinct distance values in $P(n, \epsilon)$. $\beta < 2\epsilon r$. Take $\tau=\beta/10$. For point $x_i \in P(n,\epsilon)$, let $y_i$ be the point obtained by moving $x_i$ along the radial direction to distance $r+d_i$ from the center of circle, where $d_i=i\tau/n$. The point set $Y=\{y_i\}$ meets the conditions in the claim.  

    First, we claim that for $i\neq j, j\neq k \in [n]$ and $\alpha > 1$, $d_{\mathbb{H}}(y_i,y_j)\leq \alpha \cdot d_{\mathbb{H}}(y_k,y_j)$. By triangle inequality and \Cref{lem:lw-kw}, \[d_{\mathbb{H}}(y_i,y_j)\leq d_{\mathbb{H}}(x_i,x_j)+d_{\mathbb{H}}(y_i,x_i)+d_{\mathbb{H}}(x_j,y_j)\leq 2r+2\tau,\] and \[d_{\mathbb{H}}(y_k,y_j)\geq d_{\mathbb{H}}(x_k,x_j)-d_{\mathbb{H}}(x_k,y_k)-d_{\mathbb{H}}(x_j,y_j)\geq 2r(1-\epsilon)-2\tau.\] Substitute $\epsilon$ and $\tau$, we have $\beta (\alpha+1)/5 \leq r(\alpha-1)$. Since $\beta < 2\epsilon r$, we have $4\alpha^2-5\alpha+1\geq 0$, which is always true when $\alpha > 1$.   

    Second, we show that there is no tie. Let $o$ denote the center of the circle. For a point $y_j \in Y$, we do a case study. 
    \begin{enumerate}
        \item For two point $y_i, y_k \in Y$ such that $\angle y_ioy_j\neq \angle y_joy_k$, we have $|d(x_j,x_i)-d(x_j,x_k)|\geq \beta$. without loss of generality, we assume that $d(x_j,x_k) - d(x_j,x_i)\geq \beta$. By triangle inequality, \[d_{\mathbb{H}}(y_i,y_j)\leq d_{\mathbb{H}}(x_i,x_j)+d_{\mathbb{H}}(y_i,x_i)+d_{\mathbb{H}}(x_j,y_j)\leq d_{\mathbb{H}}(x_i,x_j)+2\tau,\] and \[d_{\mathbb{H}}(y_k,y_j)\geq d_{\mathbb{H}}(x_k,x_j)-d_{\mathbb{H}}(x_k,y_k)-d_{\mathbb{H}}(x_j,y_j)\geq d_{\mathbb{H}}(x_k,x_j)-2\tau.\] So $d_{\mathbb{H}}(y_k,y_j)-d_{\mathbb{H}}(y_i,y_j)\geq \beta - 4\tau>0.$
        \item If $\angle y_ioy_j = \angle y_joy_k=\theta$, let $a=d_\mathbb{H}(y_i, y_j)$ and $b=d_\mathbb{H}(y_k,y_j)$, by the Hyperbolic law of cosines, \[ \cosh(a) = \cosh(r+d_j)\cosh(r+d_i)-\sinh(r+d_j)\sinh(r+d_i)\cos{\theta}, \] and 
        \[ \cosh(b) = \cosh(r+d_j)\cosh(r+d_k)-\sinh(r+d_j)\sinh(r+d_k)\cos{\theta}.\] Compare by subtracting and then simplifying, let $M=r+\frac{d_i+d_k}{2}$ and $R_j=r+d_j$, \[\cosh(b)-\cosh(a) = 2 \sinh(\frac{d_k-d_i}{2})\cosh M\cosh R_j(\tanh M-\cos\theta \tanh R_j).\] Since $\theta \geq 2\pi/n$, $r=\log n/\eps$ and $\eps<\frac{1}{2}$, $\tanh M>\tanh r>\cos(2\pi/n)>\cos \theta \tanh R_j$. With $d_k\neq d_i$, the right-hand side is nonzero. Hence $\cosh(b)-\cosh(a)\neq 0$, which implies that $a\neq b$. 
    \end{enumerate}
\end{proof}

\end{document}